\RequirePackage{snapshot}

\documentclass[journal=jactcce,manuscript=article]{achemso}
\setkeys{acs}{articletitle = true}
\usepackage[T1]{fontenc} 
\usepackage{amsmath,amsthm,amssymb,amsfonts}
\usepackage{xcolor}
\usepackage{hhline}
\usepackage{multirow}
\usepackage{xr} 
\usepackage{cleveref}
\usepackage{siunitx}
\usepackage{xargs}                      
\usepackage{tikz}   
\usepackage{xfrac}  

\newcommand{\angstrom}{\mbox{\normalfont\AA}}
\usetikzlibrary{calc,math,arrows,shapes,shapes.misc,shapes.geometric,decorations.pathmorphing,positioning,fit,decorations.markings,patterns}

\crefformat{figure}{Figure S#2#1#3}

\author{Alina Wettstein}
\email{alina.wettstein@gmail.com}
\affiliation[IPC]{\footnotesize Institut f\"ur physikalische Chemie, Westf\"alische Wilhelms-Universit\"at M\"unster, Corrensstra{\ss}e 28/30, D-48149 M\"unster, Germany}
\author{Diddo Diddens}
\author{Andreas Heuer}
\email{andheuer@wwu.de} 
\affiliation[HIMS]{\footnotesize Institut f\"ur Energie- und Klimaforschung, Ionics in Energy Storage, Helmholtz Institut M\"unster, Forschungszentrum J\"ulich, Corrensstra{\ss}e 46, 48149 M\"unster, Germany}
\alsoaffiliation[IPC]{\footnotesize Institut f\"ur physikalische Chemie, Westf\"alische Wilhelms-Universit\"at M\"unster, Corrensstra{\ss}e 28/30, D-48149 M\"unster, Germany}

\title[An \textsf{achemso} demo]{Polymer electrolytes in strong external electric fields: Modification of structure and dynamics}


\begin{document}

\begin{tocentry}

\includegraphics[width=1\textwidth]{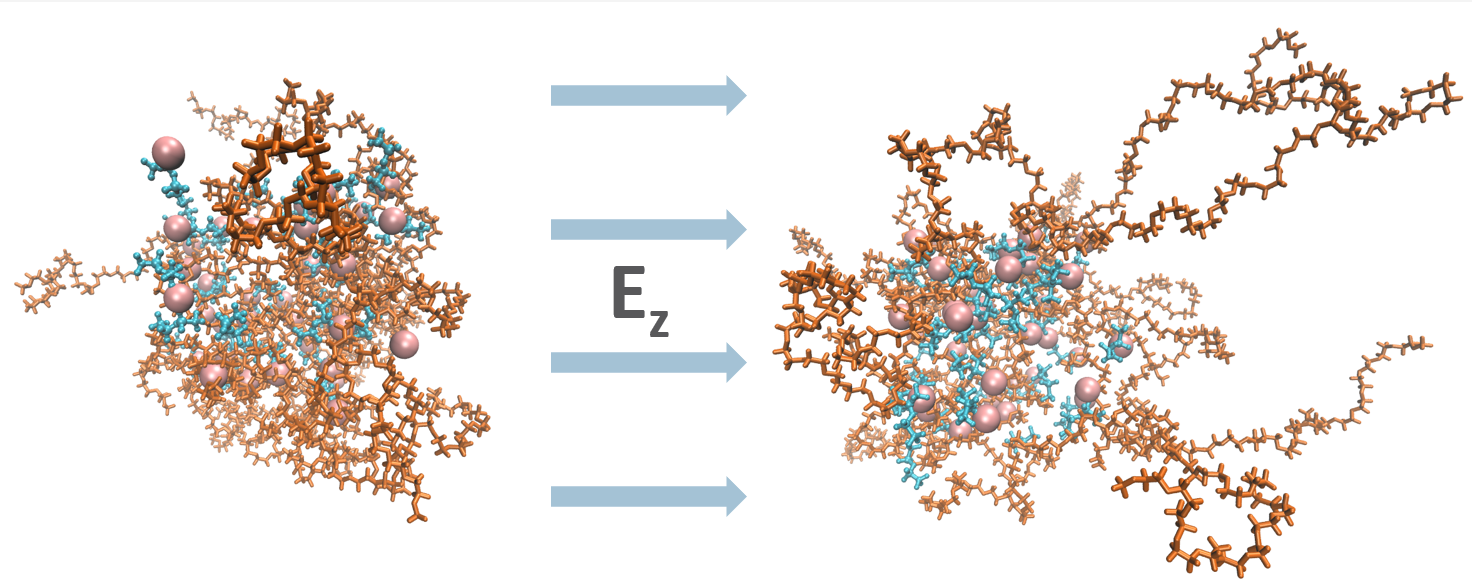}

\end{tocentry}

\begin{abstract}

We present the results from an extensive atomistic molecular dynamics simulation study of poly(ethylene oxide) (PEO) doped with various amounts of lithium-bis(trifluoromethane)sulfonimide (LiTFSI) salt under the influence of external electric field strengths up to $1\,$V/nm. 
The motivation stems from recent experimental reports on the nonlinear response of mobilities to the application of an electric field in such electrolyte systems and arising speculations on field-induced alignment of the polymer chains, creating channel-like structures that facilitate ion passage.
Hence, we systematically examine the electric field impact on the lithium coordination environment, polymer structure as well as ionic transport properties and further present a procedure to quantify the susceptibility of both structural and dynamical observables to the external field.
Our investigation reveals indeed a coiled-to-stretched transformation of the PEO strands along with a concurrent nonlinear behavior of the dynamic properties.  
However, from studying the temporal response of the unperturbed electrolyte system to field application we are able to exclude a structurally conditioned enhancement of ion transport and surprisingly observe a slowing down. A microscopic understanding is supplied.

\end{abstract}



To meet the ever increasing need for electrochemical energy storage systems in the context of progressive orientation on a sustainable and mobile energy supply, scientific focus remains on the improvement of lithium-ion battery technology \cite{Tarascon2010}. As an alternative to the state-of-the-art liquid electrolytes, setting the standard for ionic conductivities but bearing safety risks associated amongst others with electrolyte decomposition \cite{Scrosati2010,Tarascon2001,Xu2004}, solid polymer electrolytes provide inherently robust mechanical properties, which are required for the diverse ambient conditions that batteries are deployed in. The most investigated solid polymer electrolyte material is an amorphous poly(ethylene) oxide (PEO) matrix dissolving a lithium salt, for example  lithium-bis(trifluoromethane)sulfonimide (LiTFSI) \cite{Armand1986,Bruce1993,Fenton1973,Quartarone1998,Gadjourova2001}. 
However, the contemporary achievable ionic conductivity at ambient temperatures is still too low to compete in technological application, which hence further fuels the necessity to fully understand the transport behavior in these systems \cite{Quartarone1998,Xu2014,ManuelStephan2006}. 

\begin{figure}[H]
	\centering
	\includegraphics[width=1.0\textwidth]{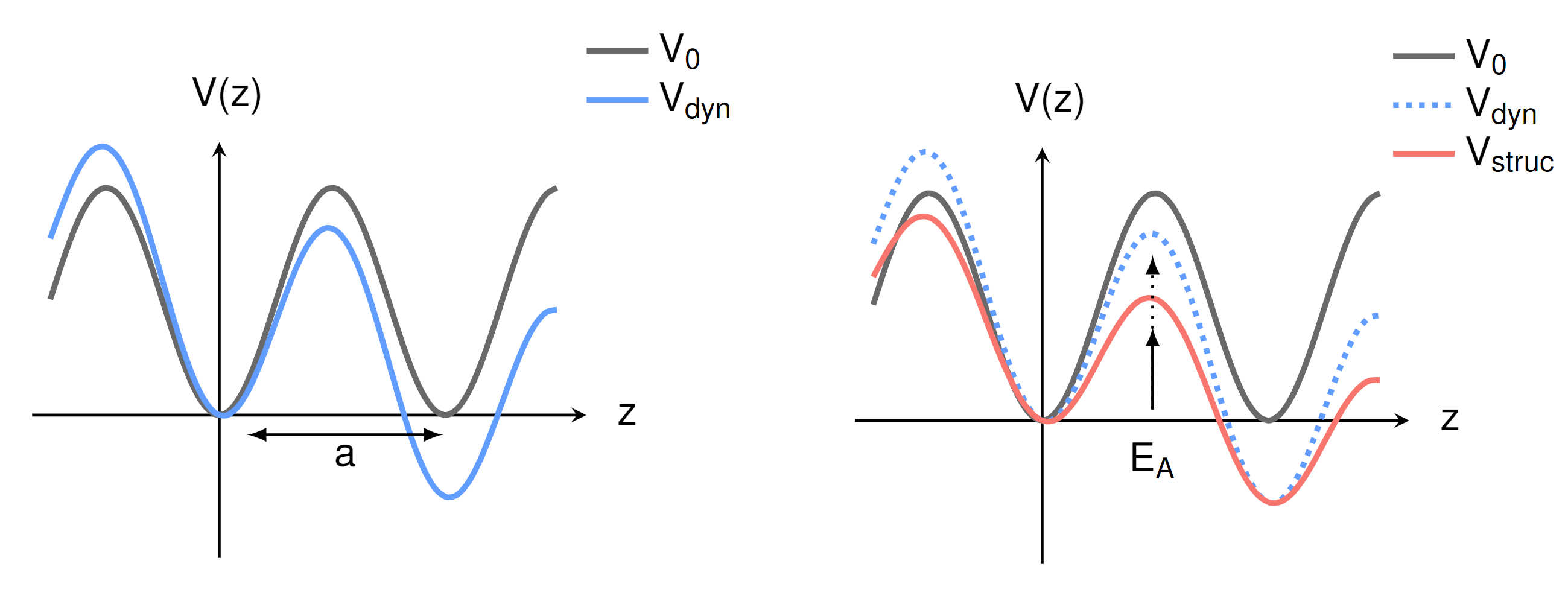}
	\caption{Sketch of one-dimensional potential energy profile V(z) in direction of the electric field. Left: Effective lowering of activation barriers $\text{E}_{\text{A}}$ due to tilting of energy surface in field direction. Right: Additional decrease of barrier heights due to structural modifications.} 
	\label{fig:exemplary_energy_landscape_tilt_structure}
\end{figure}

Studying the impact of electric fields on polymer electrolyte systems emerges as highly relevant in regard to the electrode/electrolyte interfaces where ionic concentration gradients evoke local potential drops that correspond to local field strengths in the order of volts per nanometer \cite{Jorn2013,Matse2020,Kreuzer2002,Stuve2011}. Strong electric fields may further be induced by dendritic lithium structures that form upon cycling the battery and pierce through the polymer host matrix \cite{Harry2014,Jaeckle2019,Linlin2018}.

Both experimental \cite{Rosenwinkel2019,Wang2012,SunithaV.R.2015} and theoretical \cite{Huang2017} research on ion transport in polymeric systems in the presence of an external electric field have indicated a nonlinear enhancement of ion dynamics. 
Based on the observation of field-dependent mobilities in electrophoretic NMR measurements carried out at electric field strengths $\mathcal{O}\left(10\,\text{V/cm}\right)$ Rosenwinkel et al. put forward the hypothesis of evolving polymer channel structures that guide ion transport efficiently \cite{Rosenwinkel2019}.
Even though this line of reasoning comes along as very intuitive, a causal understanding of such nonlinear effects can turn out much more complex. 
For instance, experimental and theoretical work on inorganic lithium silicate glasses detected a nonlinear response of lithium ion dynamics to an external field, yet the systems experienced no structural modifications\cite{Heuer2017,Kunow2006,Schroer2015}. 
From the perspective of the potential energy surface two entirely independent effects may give rise to a field-induced enhancement of the dynamics. In Figure \ref{fig:exemplary_energy_landscape_tilt_structure} we show a sketch of a potential energy profile in one dimension which is pictured for reasons of simplicity via a sinusoidal shape. In absence of an external field the ion vibrates in a potential well, that is a local minimum of the energy landscape. To migrate over a further distance, i.e. the width $a$ of the potential well, the ion requires an activation energy $\text{E}_{\text{A}}$ to overcome the potential barrier. While thermal fluctuations of the environment may enable such a crossing occasionally, the application of an external field effects an additional energy gain in field direction (see Figure \ref{fig:exemplary_energy_landscape_tilt_structure} left). As a consequence, the energy profile is tilted such that the activation barriers for migration in field direction are lowered (respectively raised for backward jumps) which hence causes an increasingly directed ion motion.
Mathematically, one can capture the biasing effect of the field on forward and backward jumping rates $\Gamma_{\text{f/b}}$ through an additional energy contribution in the Boltzmann factor \cite{Roling2008,Genreith2016}:
\begin{equation}
\Gamma_{\text{f/b}} \qquad = \qquad \Gamma_0 \cdot \exp\left(\pm \dfrac{q\text{E}a}{2\text{k}_{\text{B}}T}\right),
\end{equation} 
where $\Gamma_0$ denotes the undirected jumping rate without field and $q$ the charge of the particle which is forced by the external field $\text{E}$. A Taylor expansion of the thus stated hyperbolic sine dependence of the ion drift velocity $\nu$ on the external field naturally yields a nonlinear transport behavior for sufficiently high fields \cite{Huang2017,Roling2008,Genreith2016}:
\begin{equation}
\langle \nu \rangle\,=\,\dfrac{d}{dt} \langle z \rangle \,=\,a\,\left(\Gamma_{\text{f}}-\Gamma_{\text{b}}\right)\,=\,a\,\Gamma_0 \sinh\left(\dfrac{q\text{E}a}{2\text{k}_{\text{B}}T}\right) \,=\,t_1\cdot\text{E}\,+\,t_3\cdot\text{E}^3\,+\,...
\label{eq:field_dependence_mu}
\end{equation}
If the field furthermore induces structural changes that cause an additional decrease of the potential barriers  (see Figure \ref{fig:exemplary_energy_landscape_tilt_structure} right), the nonlinear response may either be complementary enhanced or even dominated by such structural modifications.

In this work, we address three questions by means of atomistic molecular dynamics (MD) simulations covering a broad range of electric field strengths: 
Does the field alter structural and/or dynamical properties of the salt-in-polymer electrolyte? If so, at which field strengths and how pronounced do these effects occur? Is the nonlinear increase of transport properties conditioned by such conformation changes in this soft matter system?

The paper is organized as follows: First, we present our results on the ion dynamics as a function of field strength. Then, we discuss the lithium coordination environment and the polymer chain structure under field application. Next, we bring forward an explanatory approach to rationalize the structuring of the polymer host matrix and further develop a concept to classify the field susceptibility of both structural and dynamic observables. Lastly, we discuss the correlation between structure and dynamics on the basis of field-switch simulations and close with an overall conclusion. 

\textbf{Simulated systems}
 
The MD simulations were performed with the GROMACS \cite{Berendsen1995,VanDerSpoel2005,Pall2015,Abraham2015} software package relying on the classic all-atom, non-polarizable  OPLS-AA and OPLS-AA-based CL\&P force field \cite{WilliamL.Jorgensen1996,CanongiaLopes2012,Lopes2004,Lopes*2004,Shimizu2010} at a temperature of $423\,$K. The details of the structure generation and simulation protocol are given in the \textit{Supporting Information} section. 
Inspired by the experimental study of Rosenwinkel et al. the integral part of our study covers three salt-in-PEO electrolyte systems with salt-to-monomer ratios r = [Li]/[EO] = 0.06, 0.10 and 0.15.
Each system contains 20 coiled PEO chains composed of N = 27 monomer units and the corresponding amount of LiTFSI ion pairs (36, 54 and 81 Li/TFSI molecules).
Beyond that a variety of slightly modified setups was simulated to answer questions arising in the course of this investigation. 
These include a plain polymer melt (r = 0.0) as well as an artificial scenario (r = 0.06) where only lithium and TFSI experience an external electric force to capture the purely ionic response to the field.
To determine if the occurrence of field effects is limited by the molecular weight of the polymer chains, we investigated an r = 0.06 electrolyte mixture containing 10 coiled PEO chains double in length, i.e. N = 54.
Furthermore, we performed a series of 100 short (5 ns) simulations where an electric field of 1\,V/nm is applied on $\text{E}_{\text{z}}$\,=\,0.0\,V/nm structures of the r = 0.06 mixture to trace how structural and dynamical observables equilibrate to their steady state value.
All simulation trajectories were run for at least $300\,$ns and cover up to $2\,\mu$s to ensure that the diffusive regime has been reached.


\textbf{Dynamical properties.} \quad We investigate the impact of the electric field on ion dynamics on the basis of the electrophoretic mobilities $\mu$, as measured in the electrophoretic NMR experiments \cite{Rosenwinkel2019}, and diffusion coefficients in field direction $D_{\parallel}$, i.e. the diffusion within the moving coordinate frame of reference that is constituted by the drift motion. 
We extract
\begin{equation}
\mu = \dfrac{\langle \nu \rangle}{E} = \lim\limits_{t \rightarrow \infty} \dfrac{z(t) - z(0)}{E\cdot t} 
\end{equation}
from the ion's stationary drift velocity $\nu$. In order to assess the diffusion coefficient in field direction, we analyze the parallel component of the mean-square displacement (MSD) in the moving coordinate frame \cite{Heuer2005,Blickle2007}:
\begin{equation}
D_{\parallel} = \lim \limits_{t\rightarrow\infty} \dfrac{\text{MSD}_{\text{parallel}} - \langle \nu \rangle ^2 \cdot t^2}{2 \cdot t}.
\end{equation}
For small electric fields, i.e. small perturbations from equilibrium, it is expected that neither $\mu$ (see equation \eqref{eq:field_dependence_mu}) nor $D_{\parallel}$ depend on $\text{E}_{\text{z}}$. Thus any deviation from the linear response can be directly interpreted as a nonlinear effect. 
Since the Einstein relation connecting diffusion constant and mobility has been found to no longer hold beyond the linear response regime even in systems without ion correlations\cite{Blickle2007}, the influence of the electric field on $D_{\parallel}$ is a priori unknown. Therefore, investigation of $D_{\parallel}$ provides novel information on the complexity of the nonlinear dynamic effects. 

\begin{figure}[H]
\centering
\includegraphics[width=0.8\textwidth]{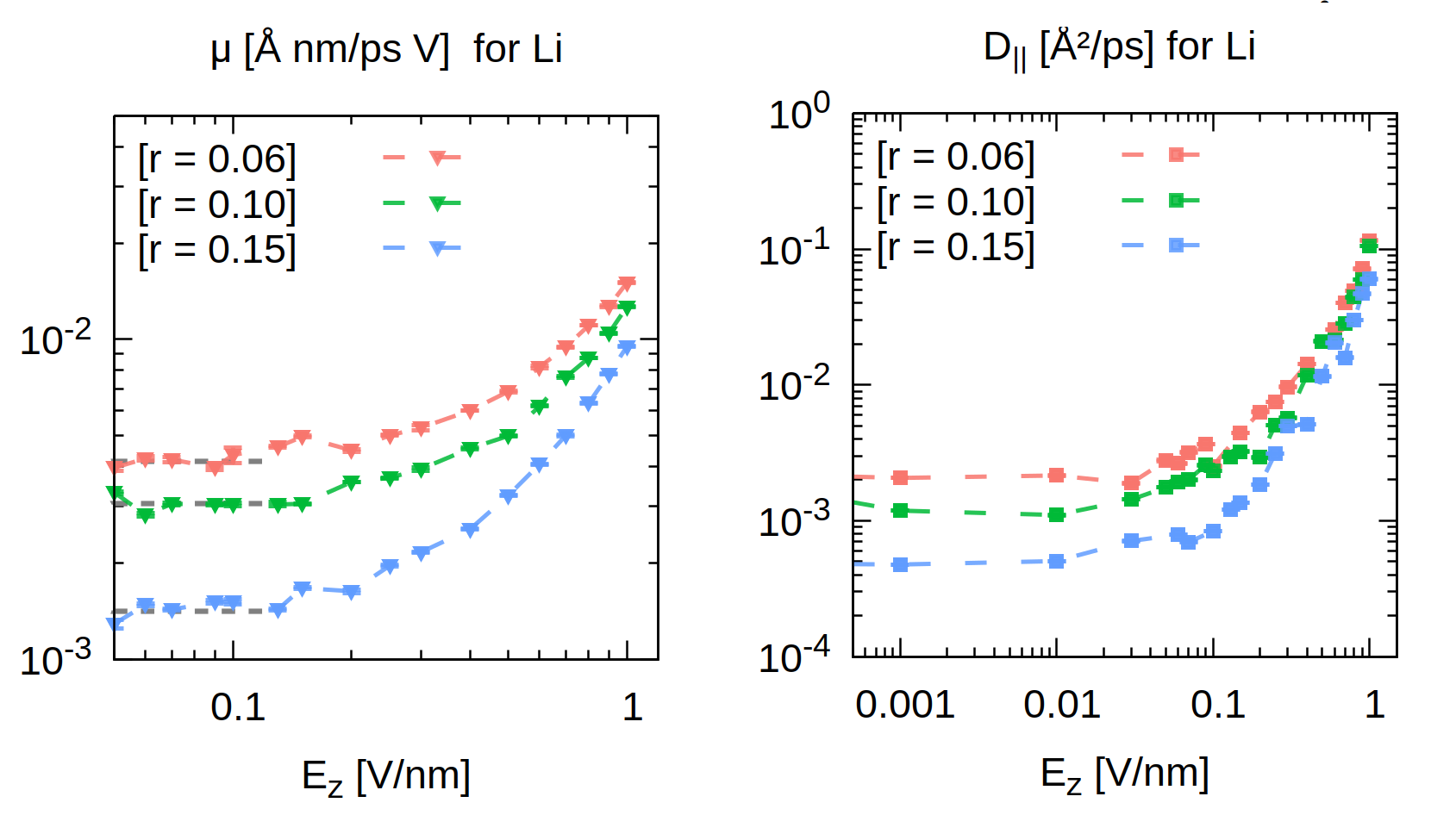}
\caption{Lithium mobilities $\mu$ and parallel diffusion constants $\text{D}_{\parallel}$ as a function of electric field strength for three different salt concentrations. The dashed grey horizontal lines mark the linear response mobilities at the respective concentration, which were employed to evaluate $\text{D}_{\parallel}$ for $\text{E}_{\text{z}}$\,< 0.03\,V/nm.}
\label{fig:nonlinear_dynamics_mu_d_para_li}
\end{figure}

The resulting mobilities and diffusion coefficients for lithium are shown in Figure  \ref{fig:nonlinear_dynamics_mu_d_para_li} (see \Cref{supp-fig:nonlinear_dynamics_mu_d_para_tfsi} for TFSI dynamics). 
For application of an electric field we observe a substantial enhancement of $\mu$  and $\text{D}_{\parallel}$ for both ion species, whereas the impact is more pronounced for $\text{D}_{\parallel}$, which is increased by nearly two orders in magnitude.
Interestingly, ranking the intensity of the enhancements of $\mu$ and $\text{D}_{\parallel}$ in this soft matter system matches the observations that have been made previously on lithium silicate glasses \cite{Kunow2006,Heuer2017,Schroer2015}. Demonstrating a strong similarity to such inorganic glasses, the threshold electric field strengths to observe nonlinear effects in our simulation, i.e. $\text{E}_{\text{z}} \gtrapprox $ 0.2\,V/nm for $\mu_{\text{Li}}$ and $\text{E}_{\text{z}} \gtrapprox$ 0.05\,V/nm for $\text{D}_{\parallel  \text{Li}}$,  lie considerably above the electric field strengths $\mathcal{O}\left(10\,\text{V/cm}\right)$ employed in the eNMR measurements.

We note that in absence of an external field the diffusion coefficients for both lithium and TFSI for the r\,=\,0.06 mixture are in accordance with simulations of an amorphous LiTFSI-PEO melt (r\,=\,0.06, T\,=\,416\,K) employing a many-body polarizable force field \cite{OlegBorodin*2006b}. Furthermore, the transport properties decrease in magnitude for increasing salt concentration as expected from the increasing viscosity \cite{Brooks2018,Timachova2015}. Due to the strong interactions between lithium and the polymer ether oxygens lithium is less mobile and as a result less diffusive than  TFSI despite its smaller size \cite{Brooks2018,Timachova2015,Rosenwinkel2019}.

To ensure within the realms of simulation work that our observations of enhanced transport properties induced by the electric field are not limited by the molecular weight of the simulated chains, we performed additional simulations for the r\,=\,0.06 electrolyte employing chains double in length. We show in the Supplementary Information section that nonlinear dynamics are observed for this system as well. A closer discussion of the chain length dependence is, however, beyond the scope of this work.

 
\begin{figure}[H]
	\centering
	\includegraphics[width=1.0\textwidth]{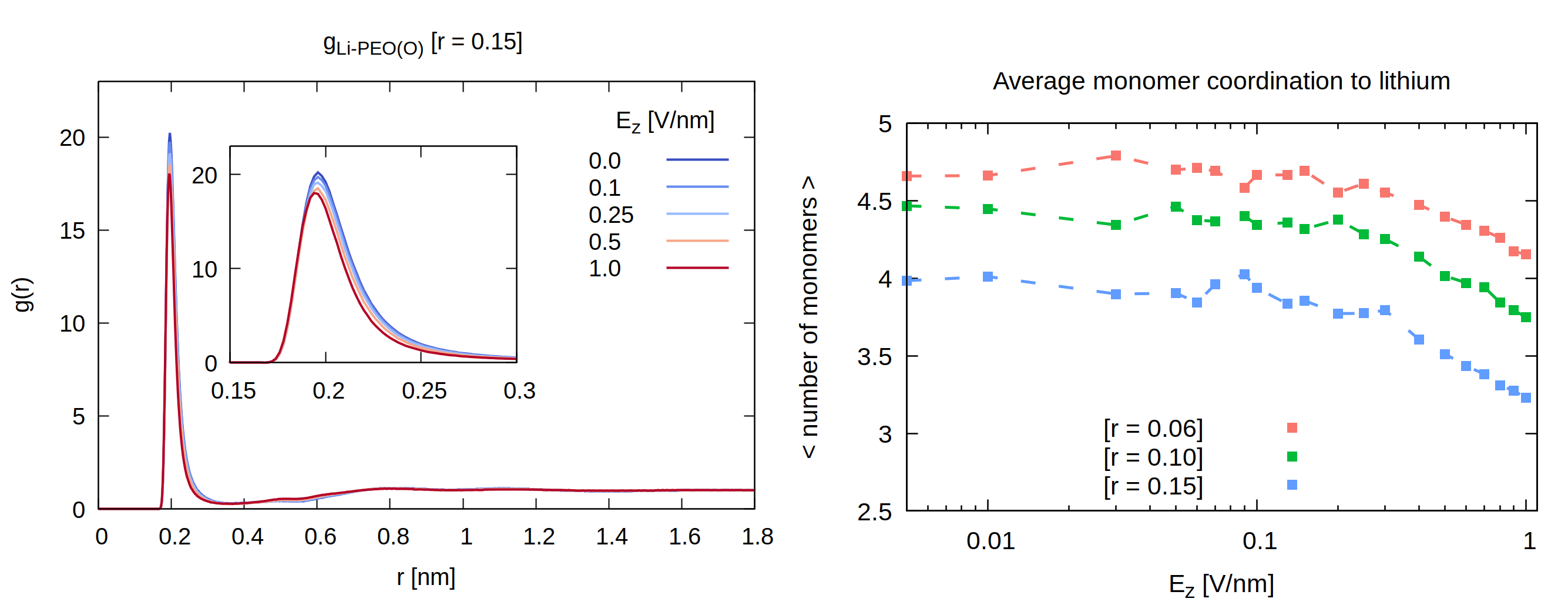}
	\caption{Left: Exemplary lithium-ether oxygen radial distribution function for a variety of electric field strengths $E_z$ for the Li[TFSI]-PEO mixture with a monomer-lithium ratio r$\,=\,0.15$. Right: Average lithium-monomer coordination number as a function of electric field strength. } 
	\label{fig:RDF_CN_LI_PEO_OE}
\end{figure}
\textbf{Lithium coordination environment.} \quad To ascertain whether the electric field modifies the structural properties of the lithium salt-polymer system, we investigate the lithium coordination environment. 
The radial distribution functions $g_{\text{Li-PEO(O)}}(r)$ of polymer oxygens around lithium as a reference are shown in Figure \ref{fig:RDF_CN_LI_PEO_OE} exemplary for the r\,=\,0.15 system for various  $\text{E}_{\text{z}}$. 
In agreement with previous simulation studies the first sharp coordination peak for the lithium - ether oxygen bonds emerges at a distance of $2.0$~\si{\angstrom} while the second coordination sphere appears to be smeared at a distance of $4.5$~ \si{\angstrom} \cite{Costa2015,Diddens2014}. 
We observe that the peak heights decrease with increasing field strength which is indicative of a liberation of the lithium ions from the polymer backbone.
Integration of $g_{\text{Li-PEO(O)}}(r)$ up to the first minimum yields the average number of monomers coordinating to lithium as shown Figure \ref{fig:RDF_CN_LI_PEO_OE}, which we find for all salt concentrations to decrease accordingly.

\begin{table} [hbtp]
	\centering
	\def\arraystretch{1.5}
	\begin{tabular}{l| l l l | r  }
		\hline
		{$\text{E}_{\text{z}}$ [V/nm] }  &  {free [\%] } & {1 chain [\%] }&  {2 chains [\%] }&  {$\tau$ [ns] } \\
				\hline
		0.0 & 0.5 & 96.9 & 2.6  & 263.4\\ 
		1.0 & 3.0 & 81.3 & 15.7 & 2.8  \\
		\hline
	\end{tabular}
	
	\caption{Relative frequencies of chain coordination motifs in absence of electric field and a maximum electric field strength of $\text{E}_{\text{z}}\,=\,1.0$~V/nm exemplary for r\,=\,[Li]/[EO]\,=\,0.06 (see Table S1 for higher salt contents). The label 'free' means that lithium is not coordinated by the polymer. $\tau$ denotes the mean residence time of lithium on a polymer chain.
}\label{tab:ratio_coordination_chains}

\end{table}

\begin{figure}[H]
	\centering
	\includegraphics[width=0.7\textwidth]{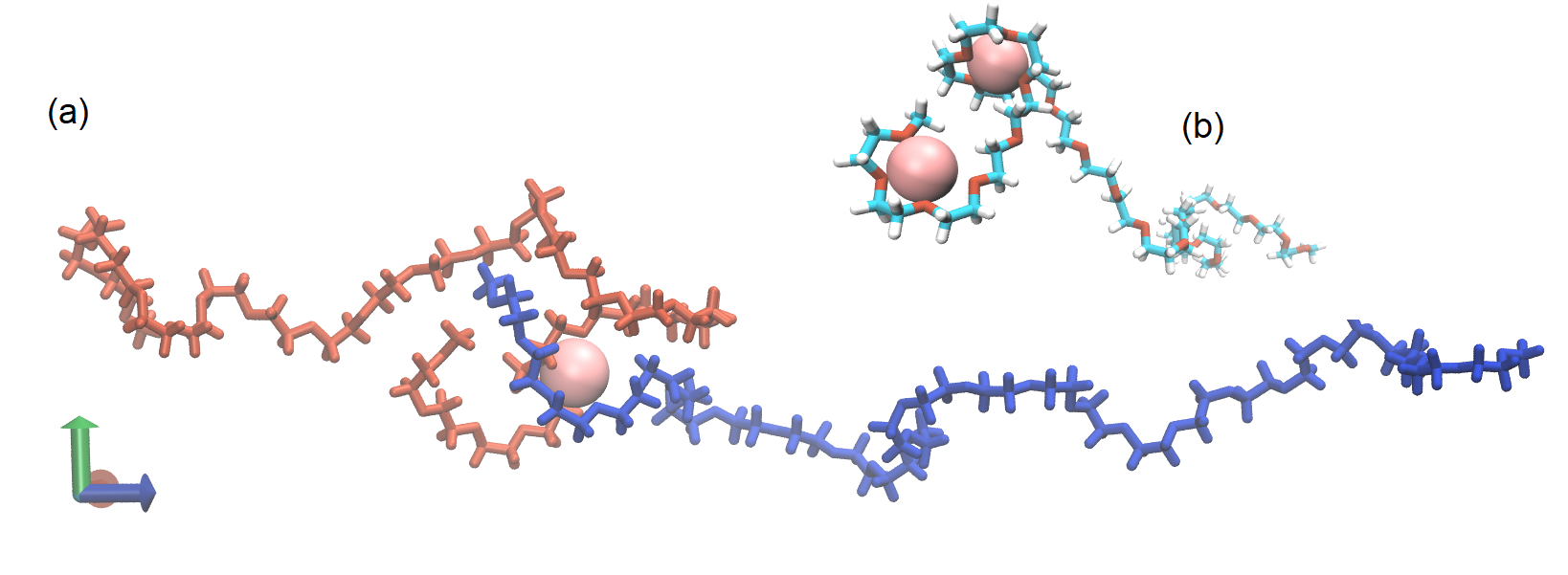}
	\caption{Top right: Snapshot of typical crown ether coordination environment of lithium via a single polymer chain. Bottom: Simulation snapshot of lithium coordinating to two polymer chains for $\text{E}_{\text{z}}$\,=\,0.9\,V/nm. The blue axis points in z-direction.}
	\label{fig:examplary_snapshot_two_chains}
\end{figure}

Analysis of the lithium-polymer coordination data as a function of chain identity is provided in Table \ref{tab:ratio_coordination_chains} and confirms primary coordination by a single polymer chain in the absence of an external field as ascertained in other simulation studies \cite{Molinari2018a,MullerPlathe1995,OlegBorodin*2006b}. 
We make the observation that under influence of an external field the coordinating monomers are increasingly provided by two different polymer chains (see simulation snapshot in Figure \ref{fig:examplary_snapshot_two_chains}) and further, that a growing percentage of the lithium ions is structurally decoupled completely from the polymer at maximum field strength.
We find further indication of such weakening attachment of lithium to the polymer in drastically declining mean residence times $\tau$ of lithium on the chain, which drop approximately two orders in magnitude (see Table \ref{tab:ratio_coordination_chains}). 
One might tentatively postulate that decreasing interaction of lithium and polymer, i.e. the decreasing coordination numbers as well as a breaking up of the typical crown-ether coordination cage that is provided by one polymer chain wrapping around the lithium ion, allows for more frequent  lithium transfers between the polymer chains and, eventually, accounts for the large increase of lithium mobility and diffusion.

We note that $g_{\text{Li-Li}}$ and $g_{\text{Li-TFSI(O)}}$ respond differently to field application as both probability distributions increase at smaller distances and thus yield lithium-TFSI and lithium-lithium coordination numbers increasing with field strength (see Figures S5-S8). We assess this as remarkable evidence of a structural reorganization induced by the electric field.

\begin{figure}[H]
\centering
\includegraphics[width=0.6\textwidth]{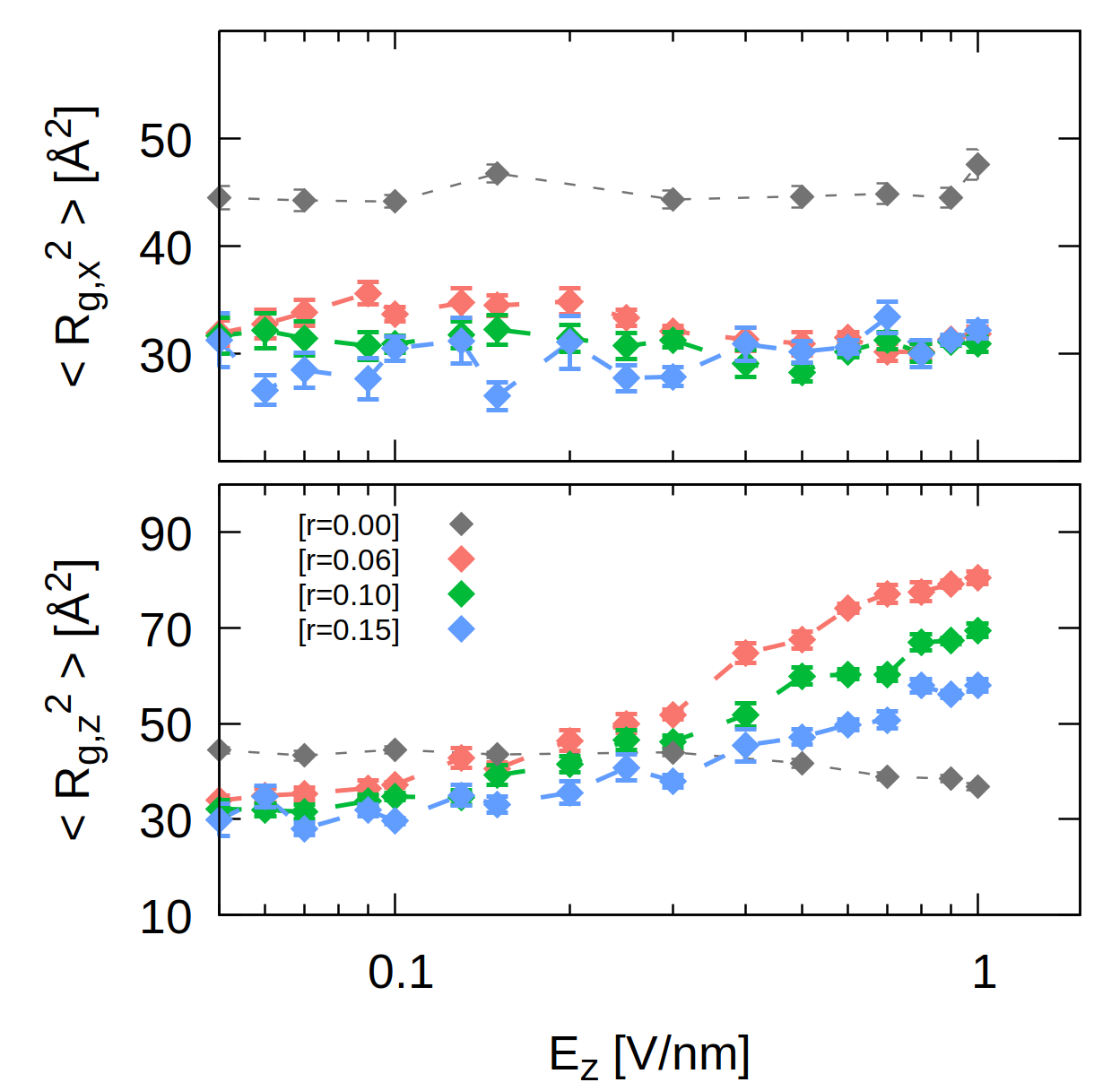}
\caption{Polymer elongation as a function of electric field strength for various lithium salt concentrations. Mean squared gyration radii $\text{R}_{\text{g,x}}^2$ (top) and  $\text{R}_{\text{g,z}}^2$ (bottom).}
\label{fig:polymer_structure}
\end{figure}

\textbf{Polymer structure.} \quad We analyze the conformational properties of the polymer chains in terms of the mean squared radius of gyration parallel ($\text{R}_{\text{g,z}}^2$) and orthogonal ($\text{R}_{\text{g,x}}^2$) to field direction. 
The results are depicted in Figure \ref{fig:polymer_structure}. 
Whereas the electric field hardly affects the chain elongation in x-direction, we observe that for exceeding a field strength $\text{E}_{\text{z}}$\,=\,0.1\,V/nm the chains start to expand in field direction in the lithium salt containing systems. 
We find that $\text{R}_{\text{g,z}}^2$ is increased more efficiently for decreasing salt content.

The ordering effect levels off at elevated field strengths and a maximum capacity of chain stretching can be observed. The final plateau is even more pronounced in the reference simulations containing polymer chains of double length as shown in \Cref{supp-fig:chain_length_comparison_Rgz2}. The polymer chains exhibit a finite length, which naturally calls for a maximum  elongation achievable within the respective conditions.
The plain polymer does not display such a chain stretching behavior, but at most a slight trend of compaction in field direction. 

We provide a microscopic understanding of this observation in the Supplementary Information section S4, which can be briefly summarized as follows: As a consequence of counteracting orientation of the polymer monomers, which constitute a local dipole due to unequally distributed partial charges on oxygen and carbon atoms, the polymer chains are contracted in field direction. This effect counteracts the global chain alignment and may contribute to the decelerating trend of chain ordering at strong field strengths.

To clarify if the chain stretching originates from the polymer itself in the salt containing systems, we performed additional simulations for the r\,=\,0.06 mixture, in which only lithium and TFSI experienced the electric field but not the partial charges of the polymer backbone. \Cref{supp-fig:R_g_field_acceleration} shows that the chains are equally stretched in this artificial scenario except for deviations at high electric field strengths close to 1\,V/nm. We speculate that for these high fields the previously mentioned chain contraction originating from the polymer itself comes into effect and causes this discrepancy. Yet, we can conclude that the lithium salt causes the chain unfolding.

In the Supplementary Information section S6, we present a comprehensive discussion of the the coiled-to-stretched transformation of the polymer conformation, which we find to be induced by lithium, closely attached to the ether oxygens, pulling the polymer chain into the elongated shape. 
We note that the stretching is most efficient for an asymmetric distribution of the lithium ions along the polymer backbone, which hence serves as an explanation of a more effective chain alignment in field direction for decreasing salt content as observed in Figure \ref{fig:polymer_structure}. The probability for an asymmetric positioning of lithium on the polymer chains declines with increasing salt content (see \Cref{supp-fig:s_asym_statistics_r_0_15}) and so does consequently the ability of the lithium ions to stretch the chains.

\begin{figure}[H]
\centering
\includegraphics[width=1.0\textwidth]{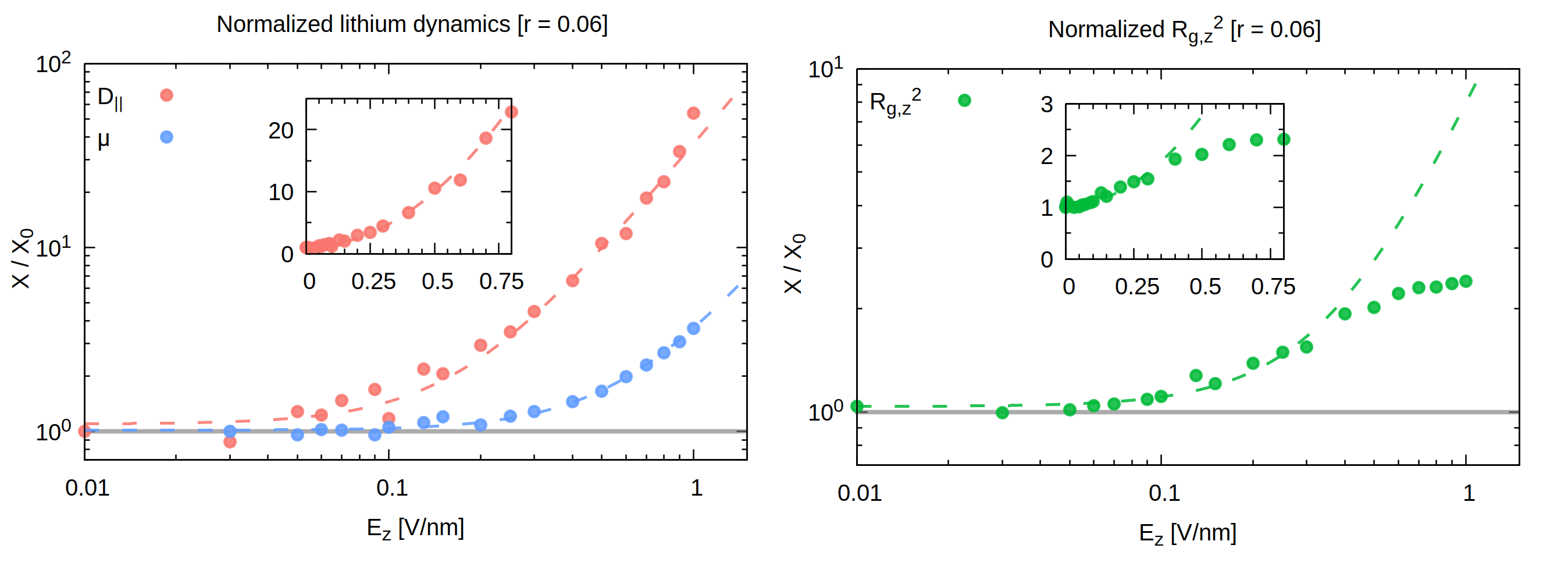}
\caption{Deviation from the equilibrium value of lithium diffusion $\text{D}_{0}$ and polymer $\text{R}_{\text{g,z,0}}^2$ in absence of an electric field, respectively deviation from the force independent plateau mobility $\mu_0$ in the linear response regime of the drift velocities. The dashed lines are the quadratic fits. Whereas $\text{D}_{\parallel}$(E) and $\mu$(E) are fitted over the entire field range, $\text{R}_{\text{g,z}}^2$(E) is only fitted up to $\text{E}_{\text{z}}\,=\,$0.4V/nm once the effect of saturating chain order sets in.}
\label{fig:normalized_dynamics_lithium_field}
\end{figure}

\textbf{Overall picture of field dependence.} \quad Since the response of either structural or dynamical properties $X(E)$ must be independent of the direction of the electric field (a reversal of the field causes solely  a reversal of the drift direction), the field dependence can be captured by an even polynomial series:
\begin{equation}
X(E) =   X_0 + \chi \cdot E^2 + \mathcal{O} \left(E^4\right),
\end{equation}
i.e. in first approximation by a parabola, which recovers in the limit of small electric fields $E \rightarrow 0$ the quiescent value $X_0$. We note in passing that for special systems also non-analytical terms, e.g. $a_1|E|$ may occur \cite{Roling2008}.   
We show in Figure \ref{fig:normalized_dynamics_lithium_field} exemplary for lithium that the dependencies of $\mu$, $\text{D}_{\parallel}$ and $\text{R}_{\text{g,z}}^2$ on $\text{E}_{\text{z}}$ can be described by a quadratic fit. For better comparability the observables are normalized to their respective value in absence of an external field. We point out that the plateauing trend of $\text{R}_{\text{g,z}}^2$ is not reflected in a deceleration of the dynamic enhancements.

In order to enable a comparison between inherently different properties, such as the response of the gyration radius or parallel diffusion coefficients to the external field, we define a critical field strength $E_c$, which denotes a doubling of the observable in absence of a field, that is $X_0$. This critical field strength thus may be extracted from the quadratic fit parameters (for example fits see dashed lines in Figure \ref{fig:normalized_dynamics_lithium_field}):
\begin{equation}
2 \cdot X_0 = X_0 + \chi \cdot E_c^2 \quad \Rightarrow \quad E_c = \sqrt{X_0 / \chi}.
\label{eq:critical_field_strength}
\end{equation}

\begin{figure}[H]
\centering
\includegraphics[width=0.7\textwidth]{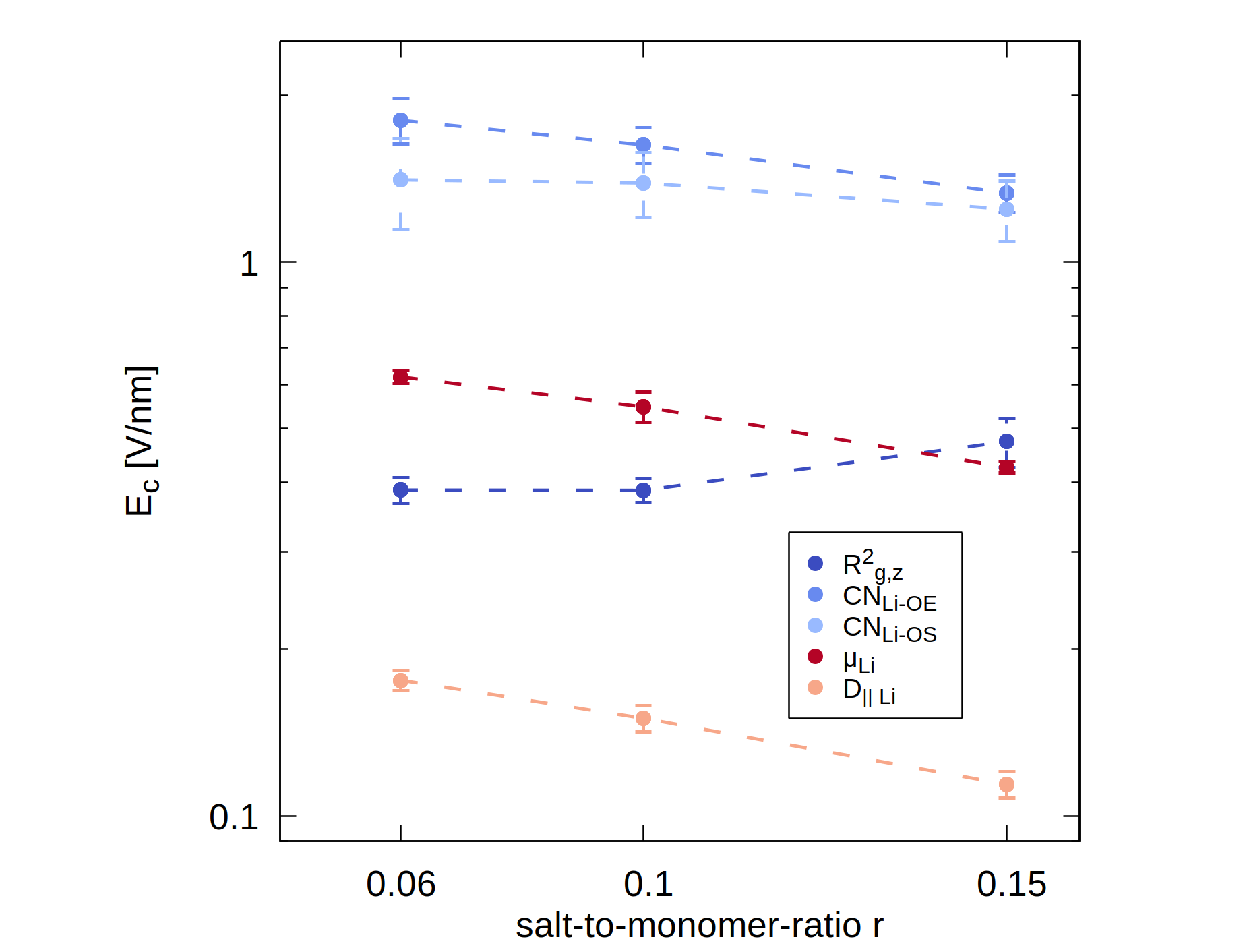}
\caption{Critical field strengths $\text{E}_c$ extracted from fitting the field dependence of dynamic transport (red) and structural (blue) properties via  $X(E) = X_0 + \chi\cdot E^2$.}
\label{fig:e_c_dynamics_structure}
\end{figure}

Figure \ref{fig:e_c_dynamics_structure} systematically summarizes the thus quantified degree of field-induced enhancement of the lithium dynamics and structural alterations, which we have touched before only in a qualitative manner. Note that none of the subsequent conclusions depend on the specific choice of the factor of 2 in equation \eqref{eq:critical_field_strength}.

First, we find that the critical fields $E_c$ split into two domains, i.e. $E_c$ values above and below 1\,V/nm. It requires strong electric fields to alter the immediate lithium environment characterized in terms of oxygen coordination numbers. The coordination numbers display an increasing susceptibility to the field for increasing salt content.
The nonlinear response of the lithium dynamics as well as the polymer chain alignment, on the other hand, occurs at considerably smaller field strengths. As stated previously $\text{D}_{\parallel}$ features by far the strongest nonlinear effects.  
Whereas $\mu$  and $\text{D}_{\parallel}$ follow the concentration dependence of the coordination numbers, the gyration radius $\text{R}_{\text{g,z}}^2$ ranging in a field regime in between displays the opposite dependence on salt content. As we further recall that the plateauing trend of $\text{R}_{\text{g,z}}^2$ is not reflected in the field dependence of $\mu$  and $\text{D}_{\parallel}$, it seems implausible that $\text{R}_{\text{g,z}}^2$ conditions the transport enhancement.
These observations suggest that there is , if at all, only a weak impact of structural modifications on the field-enhanced dynamics. 
However, no strict conclusions can be drawn on the causality of structure and dynamics since we may have missed to study the decisive structural quantity.

\textbf{Field switching.} \quad Up to now our findings are based on the steady-state behavior of the electrolyte systems under influence of a static electric field, which has not yet yielded a firm conclusion about the possible causality between structural modifications and nonlinear dynamics.  
In order to discriminate the contributions to the nonlinear dynamics due to (1) tilting the energy landscape and, possibly, (2) structural alteration effecting a beneficial reshaping of the potential barriers, we investigate with temporal resolution the response of an equilibrium structure to perturbation via a strong electric field. 

\begin{figure}[H]
\centering
\includegraphics[width=0.5\textwidth]{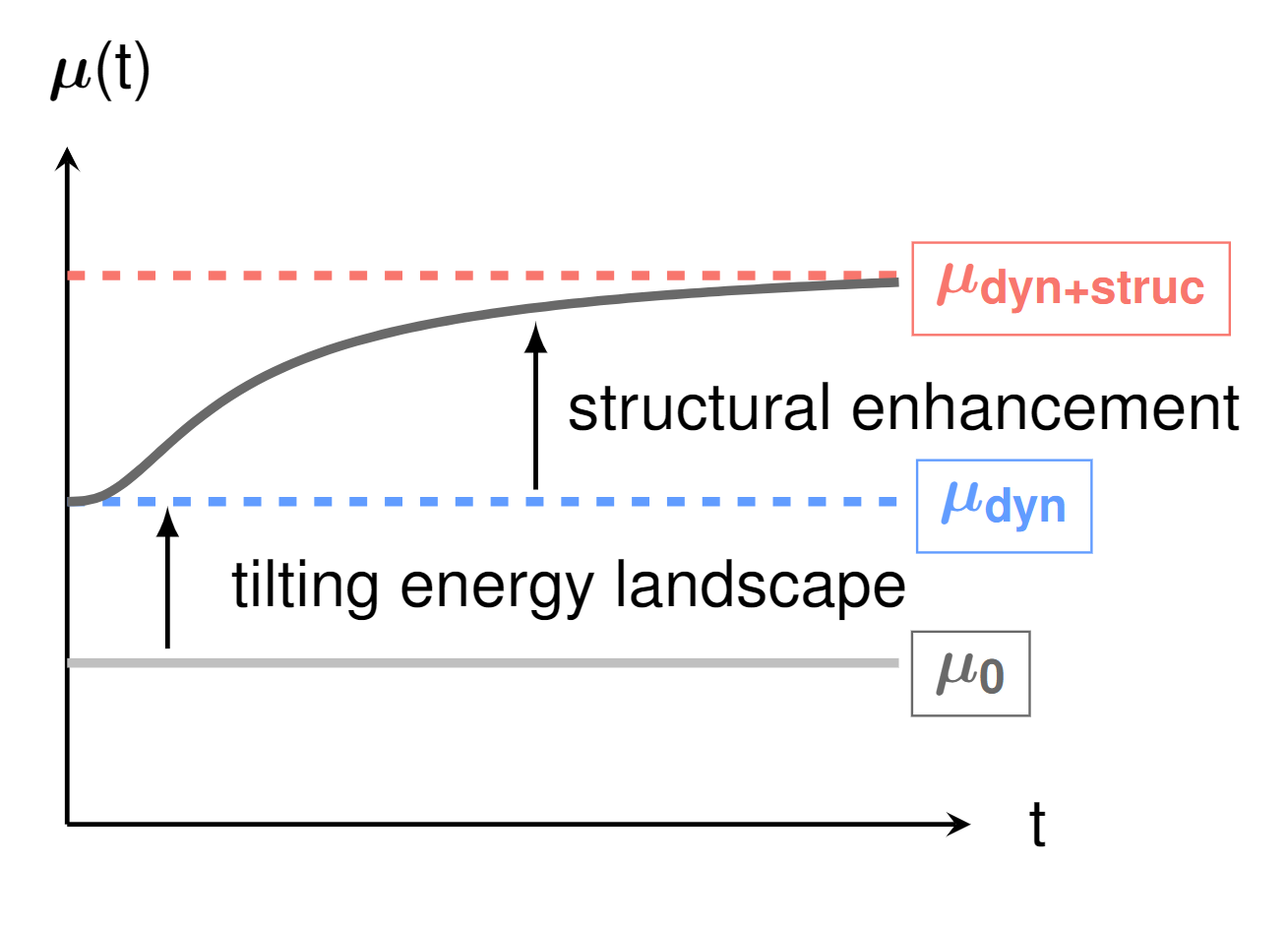}
\caption{Sketch of anticipated time dependence of $\mu$(t) for structurally conditioned enhancement of the dynamics.}
\label{fig:sketch_expectation_mu}
\end{figure}

Our expectations on the temporal evolution of $\mu(t)$, assuming that structural modifications contribute to the dynamic enhancement, are qualitatively depicted in Figure \ref{fig:sketch_expectation_mu}. Tilting the energy landscape instantaneously lowers the activation barriers and thereby raises the linear response mobility $\mu_0$ to $\mu_{\text{dyn}}$. The structural impact comes into effect after a finite time, i.e. when the structural observable responds to the external field and relaxes to its steady-state value. Hence, we expect $\mu(t)$ to approach its steady-state value $\mu_{\text{dyn+struc}}$ concurrently from below. \newline
Since our focus is now placed on the time dependence of lithium transport, $\mu$ can no longer be obtained from the long time limit of the drift velocity $\nu$ as in the stationary state analysis.
Therefore, we estimate the instantaneous mobility $\mu(t)$:
\begin{equation}
\mu(t) = \dfrac{\langle z(t+\Delta t/2)\rangle - \langle z(t-\Delta t/2)\rangle}{E\cdot \Delta t} ,
\end{equation}
the drift velocity is thus evaluated for a constant time lag $\Delta t$ between the reference positions. We show in \Cref{supp-fig:different_methods_of_extracting_mobility} that for a sufficiently short time lag $\mu(t)$ does not depend on the choice of $\Delta t$. As a compromise between statistical noise and temporal resolution, we choose $\Delta t$\,=\,10\,ps for our following analysis. \newline
The starting structures for this additional simulation set for the r\,=\,0.06 system are extracted from the $\text{E}_{\text{z}}\,=\,0.0$\,V/nm trajectory. The simulation procedure then consists of two steps: First, we freeze the polymer chains via harmonic restraints and simultaneously switch the field to $\text{E}_{\text{z}}\,=\,1.0$\,V/nm. In this setting the ions may adjust to their local energy minima in an otherwise unchanged environment. Referring to our introductory sketch in Figure \ref{fig:exemplary_energy_landscape_tilt_structure} this setting captures the plain tilting of the energy landscape.
In the second step, the chains are released from the restraints and data is collected over a set of 100 individual simulations of 5\,ns duration each. Further details on the simulation protocol are given in the Supplementary Information section S8.

In Figure \ref{fig:field_response_structure} we show the time evolution of  $\text{R}_{\text{g,x/z}}^2$ and the lithium coordination numbers after the polymer chains have been released from the restraints. 
We observe that for approximately 100\,ps the structural quantities remain unchanged by the external field until they relax monotonously to their steady-state values within the simulation time window. Particularly the coordination numbers are at equilibrium after 1\,ns. On that basis, we assume $\mu(t)$ to maintain $\mu_{\text{dyn}}$ until the structure starts transitioning and concurrently approach $\mu_{\text{dyn+struc}}$.
We further emphasize that while $\text{R}_{\text{g,x/z}}^2$ quantifies structural changes on a mesoscopic level, the coordination numbers are a very local measure. Since both quantities respond with approximately the same delay of 100\,ps to the external field, this suggests a typical time scale for other structural modifications.

\begin{figure}[H]
\centering
\includegraphics[width=1.0\textwidth]{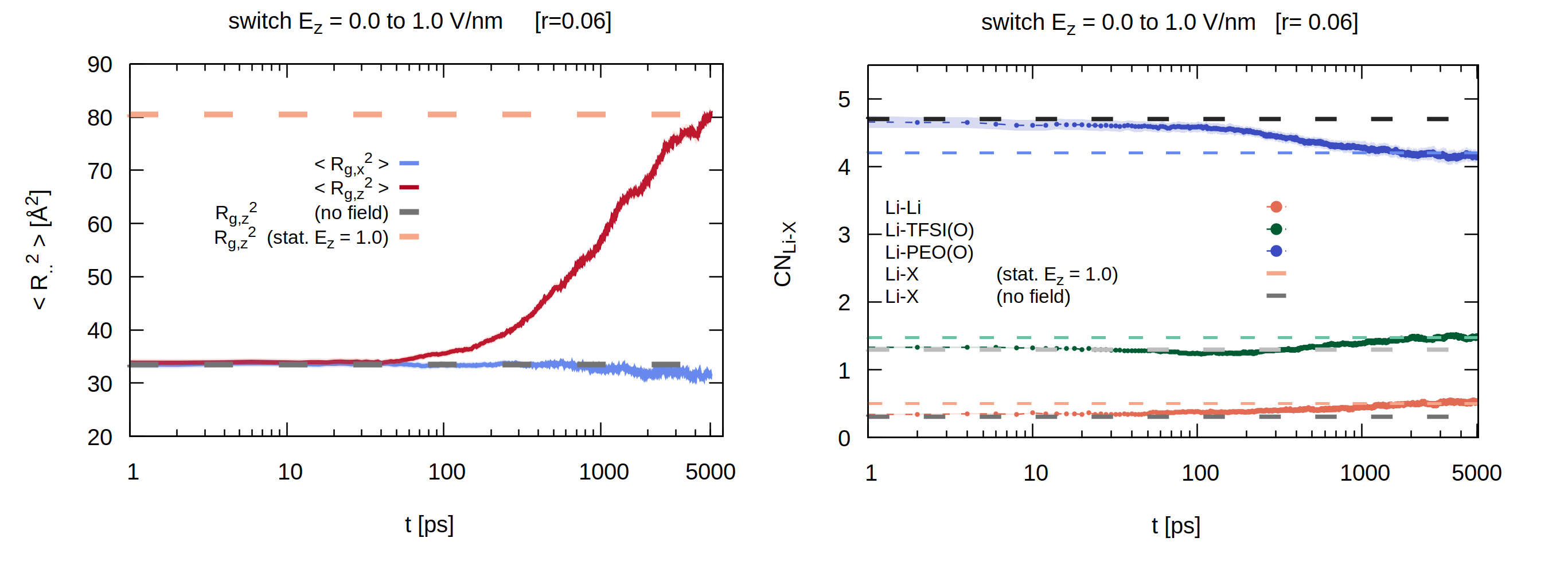}
\caption{Time evolution of $\text{R}_{\text{g,x}}^2$ and $\text{R}_{\text{g,z}}^2$ (left) and lithium coordination environment (right) for field switched from $\text{E}_{\text{z}}\,=\,0.0$ to 1.0\,V/nm. The dashed lines correspond to the respective stationary state values.}
\label{fig:field_response_structure}
\end{figure}

\begin{figure}[H]
\centering
\includegraphics[width=0.7\textwidth]{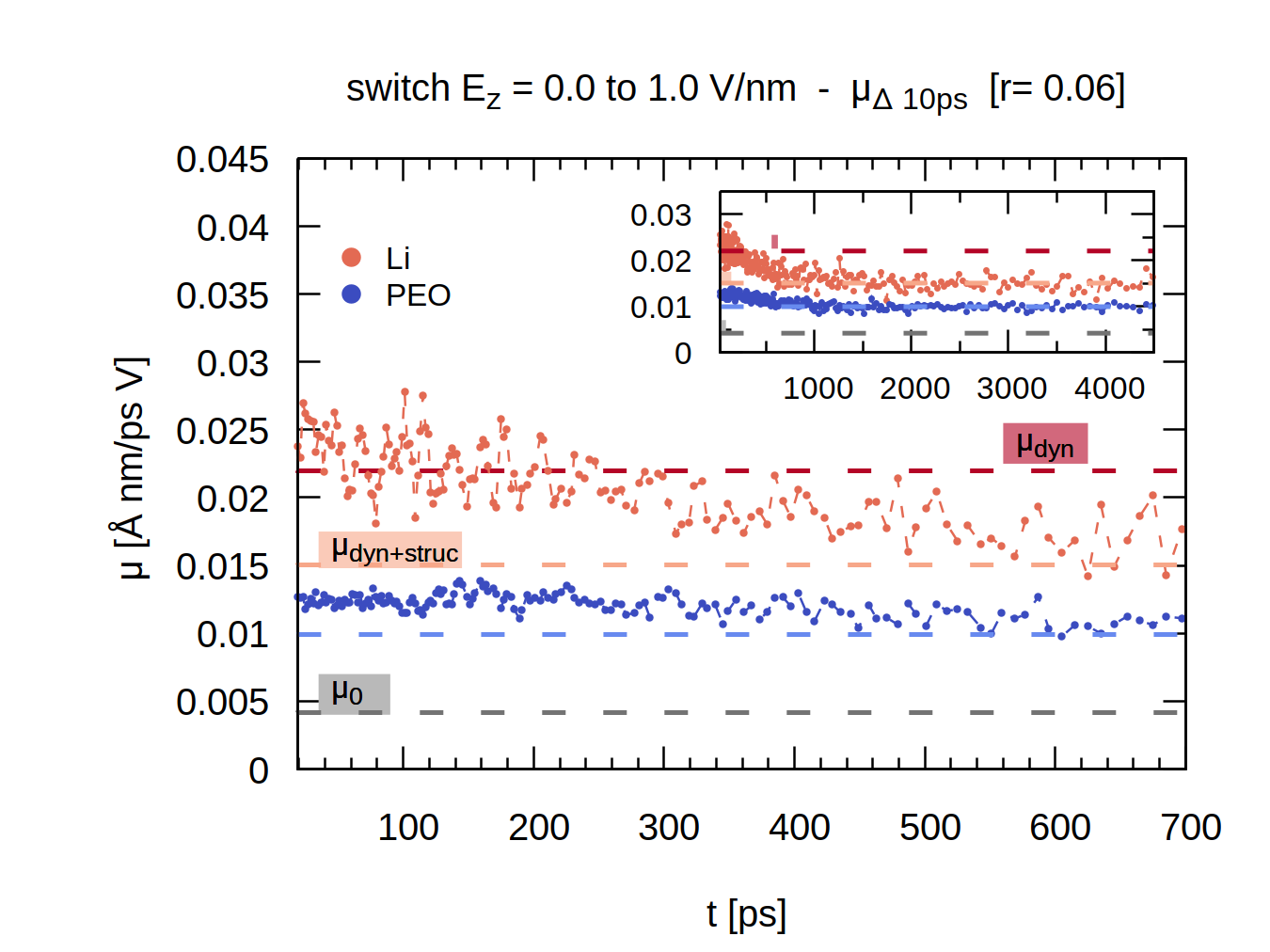}
\caption{Instantaneous lithium and polymer (center of mass) mobilities calculated for a time lag of 10\,ps (see Supplementary Information for further information) as a function of time since the field was switched from $\text{E}_{\text{z}}\,=\,0.0$ to 1.0\,V/nm. The dashed orange and blue horizontal lines correspond to the stationary state mobilities of lithium and polymer for $\text{E}_{\text{z}}\,=\,1.0\,$V/nm. The inset shows the long time behavior of $\mu_{\text{Li}}$ and $\mu_{\text{PEO}}$.}
\label{fig:field_response_dynamics_mobility}
\end{figure}

\begin{figure}[H]
\centering
\includegraphics[width=0.7\textwidth]{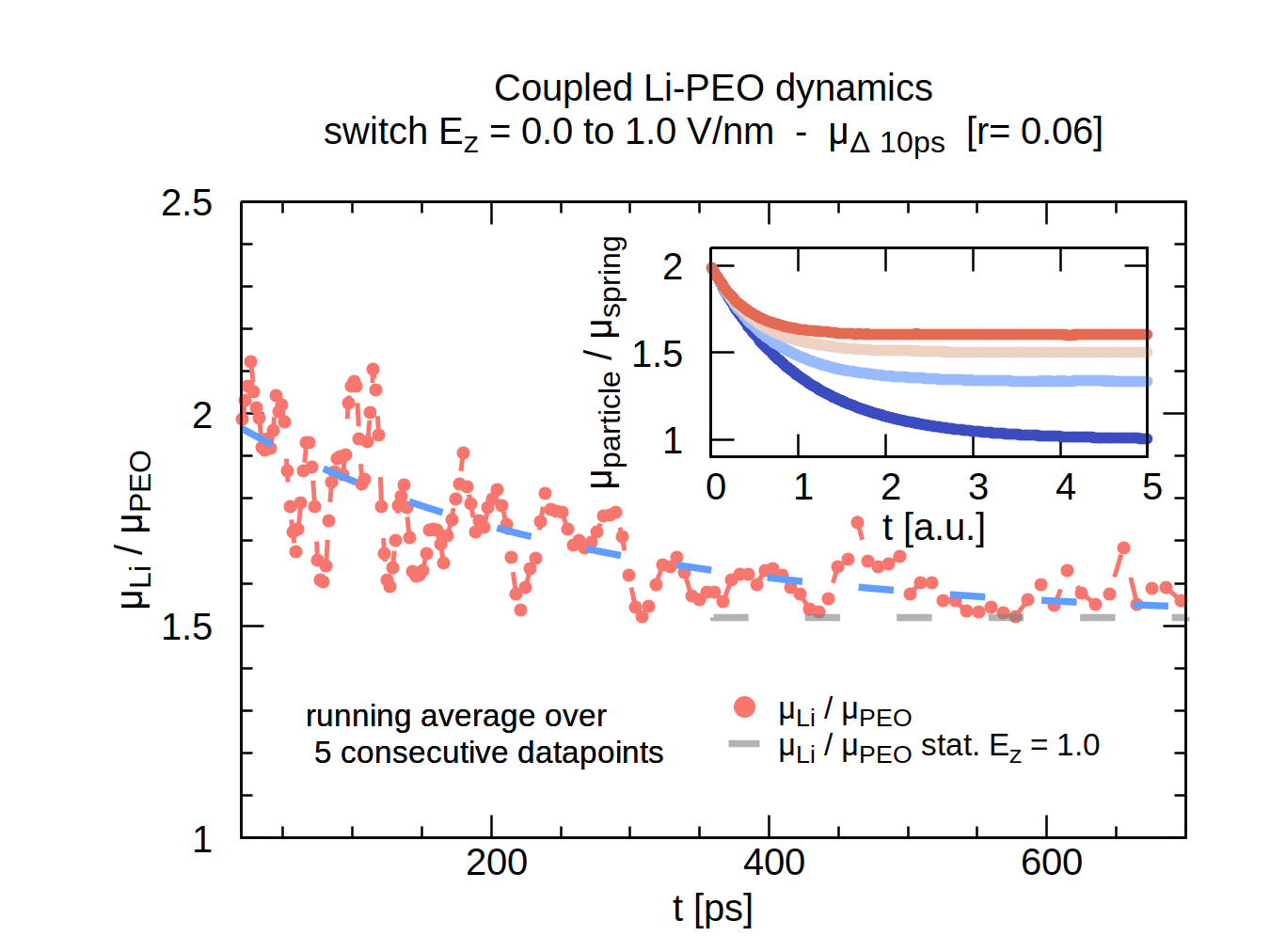}
\caption{Ratio of lithium and polymer mobilities as a function of time since the field was switched from $\text{E}_{\text{z}}\,=\,0.0$ to 1.0\,V/nm. The dashed line shows qualitatively the exponential decay ($\sfrac{\mu_{\text{Li}}(t)}{\mu_{\text{PEO}}(t)}\,=\, \alpha\cdot \exp\left(-t/\beta\right)+\gamma $ with $\alpha=0.49$, $\beta=249\,$ps and $\gamma=\lim_{t \to \infty} \,\sfrac{\mu_{\text{Li}}(t)}{\mu_{\text{PEO}}(t)}=1.52$) fitted as a guide to the eye. \newline The inset shows the mobility ratios for the simplified particle-spring-model set up by equations (8) and (9), which has been solved. The dark blue curve corresponds to the expected ratio of 1 if no particle-spring-exchange events are considered. For increasing exchange rates, i.e. shorter residence time of a particle on a spring, the condition $x=y$ is renewed with a rate $\Gamma$ and the long-time plateau shifts to higher values.}
\label{fig:field_response_dynamics_mobility_ratio}
\end{figure}

The numerical results for $\mu(t)$ are shown in Figure \ref{fig:field_response_dynamics_mobility}. 
Indeed, we find that tilting the potential energy landscape effectuates an increase of the linear response $\mu_0$ to $\mu_{\text{dyn}} \approx  4\cdot\mu_0$.  \newline
However, it is very unexpected that $\mu_{\text{dyn}} > \mu_{\text{dyn+struc}}$, thus $\mu(t)$ approaching its steady-state value from above. 
Interestingly, the polymer center of mass mobility $\mu_{\text{PEO}}$ is initially constant and then decreases to its steady-state value as well. 
We speculate the slowing $\mu_{\text{PEO}}$ to be a complex process beyond the scope of this work, possibly involving the observed structural alterations. 
Since lithium couples strongly to the polymer, we assume that the decrease of $\mu_{\text{PEO}}$ transfers on $\mu_{\text{Li}}$, which is why we now consider $\mu_{\text{Li}}/\mu_{\text{PEO}}$.
Figure \ref{fig:field_response_dynamics_mobility_ratio} shows that the velocity of the lithium ions is initially twice as fast compared to the center of mass of the polymer chains and then decays exponentially to the steady-state ratio of $\mu_{\text{Li}}/\mu_{\text{PEO}} \, \approx \,1.5$.

We attempt to understand the slowing down of the lithium dynamics qualitatively by means of a toy model. 
As discussed previously for the steady-state simulations under field application, the lithium ions pull the polymer chains into a stretched shape. Since such stretching reduces the number of possible chain conformations, i.e. the conformational entropy, the polymer responds to its decoiling with an elastic counter force comparable to an entropic spring \cite{Doi1988,Ortiz1999,Strobl1997,Liese2017}.

In the idealized model, the lithium ion is portrayed as a particle attached to a spring, with the latter being representative of the polymer chain. 
The particle experiences an external pull $F$ as well as an elastic back drag from the chain, scaling with the chain spring constant $k$. 
In the real case system the polymer chains are on average coordinated, and thus pulled, by multiple lithium ions, which is why we consider the tensile force acting on the polymer springs as an effective pull $\tilde{F}$.
The overdamped dynamics of the particle position $x(t)$ and the center of mass of the spring $y(t)$ are described by a set of two differential equations:
\begin{align} 
\gamma_p\,\,  \dot{x} \qquad &= \quad  -k \left(x-y\right) + F\\
\gamma_s\,\,  \dot{y} \qquad &= \quad \tilde{F},
\label{eq:toy_model}
\end{align}
where $\gamma_p$ and $\gamma_s$ are the friction coefficients of particle and spring. Note that equation \ref{eq:toy_model} is chosen in order to reproduce the basically constant PEO mobility directly after the switching.
Under the reasonable assumption of $x(0)\,=\,y(0)$, i.e. the particle and center of mass of the spring showing on average no initial displacement relative to each other, the solution yields:
\begin{align}
x(t) &= -\dfrac{\gamma_p}{k}\cdot\left(\dfrac{F}{\gamma_p}+\dfrac{\tilde{F}}{\gamma_s}\right)\exp\left(-\dfrac{k}{\gamma_p}t\right) + \dfrac{\tilde{F}}{\gamma_s}\cdot t + \dfrac{\gamma_p}{k}\cdot\left(\dfrac{F}{\gamma_p}+\dfrac{\tilde{F}}{\gamma_s}\right)  \\
y(t) &= \dfrac{\tilde{F}}{\gamma_s}\cdot t.
\end{align}
Whereas the center of mass of the spring displays a uniform motion at a steady velocity of $v_{\infty}\,=\,\sfrac{\tilde{F}}{\gamma_s}$, the particle dynamics relax exponentially to $v_{\infty}$. In other words, the particle initially migrates faster than the spring until its motion is increasingly restrained by the thereby elongated spring:
\begin{equation}
\dfrac{\mu_{\text{Li}}(t)}{\mu_{\text{spring}}(t)} \quad \propto  \dfrac{\dot{x}(t)}{\dot{y}(t)} \quad = \quad 1 \quad +  A\cdot \exp\left(-Bt\right),
\end{equation}
where $A\,=\,\left(\dfrac{F\gamma_s}{\tilde{F}\gamma_p}+1\right)$ and $B\,=\,\dfrac{k}{\gamma_p}$.
However, in the real case system the lithium ion is bound to the polymer chain for a mean residence time $\tau$ only before it hops into a new coordination environment, for example a different polymer chain. 
Such events serve as a renewal process for the lithium dynamics, which then become uncorrelated to their past. While we know from the steady-state simulations at $\text{E}_{\text{z}}$\,=\,0.0\,V/nm and $\text{E}_{\text{z}}$\,=\,1.0\,V/nm that the residence times drop from over 260\,ns to 2.8\,ns, we observe a significant amount of exchange events before 3\,ns have passed after the field switch as shown in \Cref{supp-fig:residence_time_lithium_chain_no_rejumps}.

We now demonstrate how incorporation of such hopping events, which occur at a rate $\Gamma$, cause an upshift of the particle velocity at long times.
Assuming that the lithium particles hop onto a new polymer spring at a rate $\Gamma$, the condition $x\,=\,y$ is on average renewed at a rate $\Gamma$ as well. 
For an exponentially distributed residence time on the chain \newline $P_{\Gamma}\,=\,\Gamma\cdot\exp\left(-\Gamma t\right)$ \cite{Diddens2013}, the long-time plateau for the ratio of particle and spring velocities gets an additional contribution as displayed in the inset in Figure \ref{fig:field_response_dynamics_mobility_ratio}:
\begin{equation}
\lim_{t \to \infty} \,\, \dfrac{\mu_{\text{Li}}(t)}{\mu_{\text{spring}}(t)} \qquad = \qquad  1 +  \int_0^{\infty}dt A\cdot\exp\left(-Bt\right) \cdot \Gamma\exp\left(-\Gamma t\right)\qquad = \qquad 1  +  \dfrac{A\Gamma}{B+\Gamma}.
\end{equation}

\begin{figure}[H]
\centering
\includegraphics[width=0.9\textwidth]{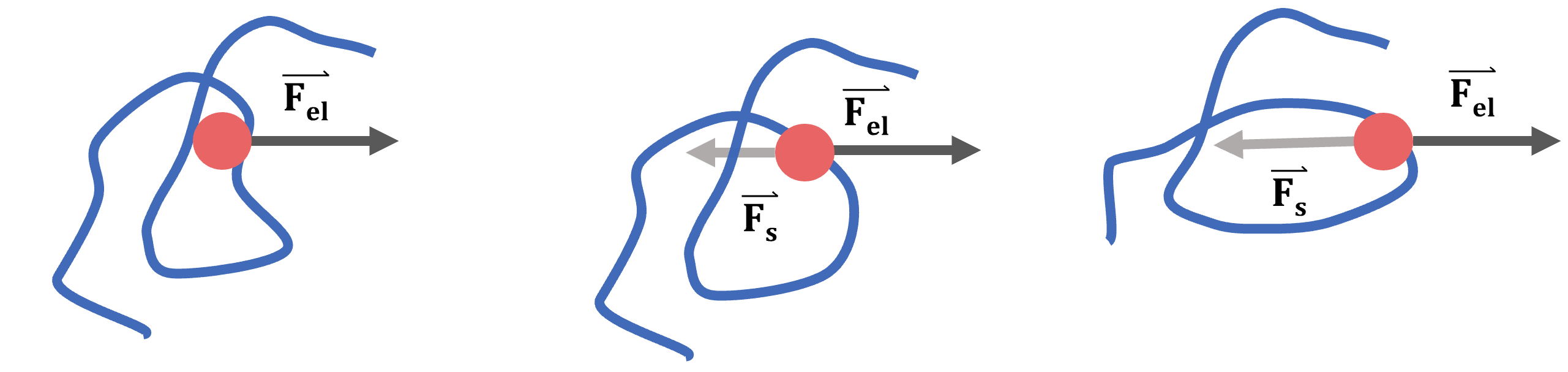}
\caption{Sketch of electric field and polymer chain exerting an electric force $\vec{\text{F}}_{\text{el}}$, respectively entropic force $\vec{\text{F}}_{\text{S}}$ on lithium.}
\label{fig:sketch_entropic_counterforce_lithium}
\end{figure}

Vividly speaking, the minimal model provides us with the qualitative conception that the motion of the individual lithium ions is held back by an entropic penalty arising from the chain stretching as sketched in Figure \ref{fig:sketch_entropic_counterforce_lithium}. Yet, the lithium ions exchange the polymer chains and may initially, i.e. after hopping onto a new chain, migrate at a speed faster than the center of mass of the chain they are bound to.

\begin{figure}[H]
\centering
\includegraphics[width=0.7\textwidth]{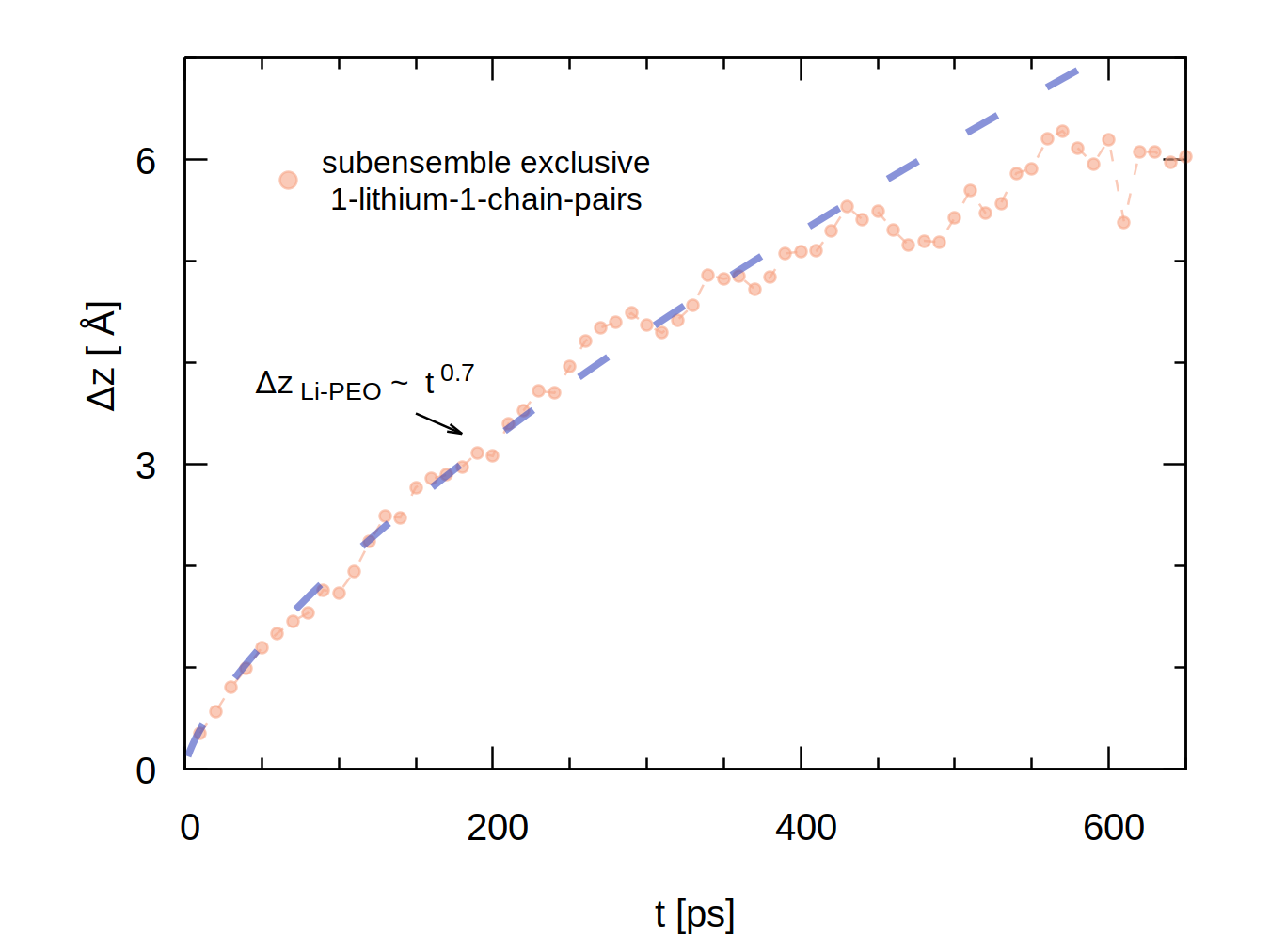}
\caption{Time evolution of lithium displacement relative to the center of mass of the polymer chain, that it is attached to, $\Delta z_{\text{Li-PEO}}$ for the subensemble of distinct lithium-chain-pairs existent at t\,=\,0\,ps, where lithium coordinates to a single chain and the chain is coordinated exclusively by this lithium ion.}
\label{fig:field_response_dynamics_dz}
\end{figure}

We gather further evidence supporting this interpretation from analyzing $\Delta z_{\text{Li-PEO}}$, the displacement of lithium relative to the center of mass of the chain that the lithium ion is attached to, as a function of the time since the field has been switched on. \newline
To meet the assumptions upon which the idealized toy model is based as close as possible, we discuss $\Delta z_{\text{Li-PEO}}(t)$ for the subensemble of special lithium-chain pairs that exist since t\,=\,0\,ps and, secondly, for which the chain is coordinated via a single lithium ion only and the lithium ion in return is bound exclusively to this chain as well. As shown in Figure \ref{fig:field_response_dynamics_dz} the drift of the lithium ions is sublinear in the reference frame of its coordinating polymer with $\Delta z_{\text{Li-PEO}}(t) \propto t^{\, 0.7}$ at short times, which is indicative of the restraining effect of the polymer chain. \newline
Whereas initially $\Delta z_{\text{Li-PEO}}(0)\,=\,0$ holds, the relative shift plateaus at long times which means that the forces acting on the lithium ion must balance, i.e. lithium pulling the polymer chain in field direction and concurrently experiencing a counter force originating from the thereby stretched chain. \newline
We note that according to Rouse theory, and due to the strong coupling of lithium and polymer segmental dynamics, a scaling of $\Delta z_{m\text{-com}}(t) \propto \sqrt t $ is predicted analytically for an arbitrary monomer $m$, that is being pulled by an external force, relative to the center of mass of the chain. We assume that the deviation of the scaling exponent may be related to the onset of structural modifications of the polymer, which is no longer covered by the Rouse model.\newline

\textbf{Conclusion} \quad We have performed an extensive all-atomistic molecular dynamics simulation study on the impact of external electric fields (0\,-\,1\,V/nm) on structural and transport properties in polymer electrolyte systems. 
Motivated by recent experimental findings on nonlinear dynamics in PEO/LiTFSI electrolytes and arising speculations on field-induced ordering of the polymer host matrix that facilitates ion motion \cite{Rosenwinkel2019}, our aim was to understand the correlation of field effects (if any) between structure and ion dynamics.

We found strong nonlinearities of the transport properties $\mu$ and $\text{D}_{\parallel}$ whose origin can be attributed to a tilt of the potential energy surface in field direction and, possibly, to an additional decrease of the activation barriers due to structural changes.

We observed significant structural alterations of the lithium coordination environment and polymer chain elongation in the presence of an electric field. We found that the lithium ions are progressively liberated from the polymer backbone, and furthermore that an asymmetric attachment of lithium to the chain causes a stretching of the polymer in direction of the field. This could be interpreted as a tentative indicator of the aforementioned channel structures.

To discuss the nonlinear effects for both structural and dynamical observables in a common framework, we defined a critical field strength $\text{E}_{\text{c}}$ that quantifies the field susceptibility of the observable. The results suggest a weak influence of the structural changes on the enhancement of the dynamics because the field strengths necessary to change the structure are either significantly higher than $\text{E}_{\text{c,dyn}}$, or they display the opposite dependence on salt concentration. Furthermore, we recall the observation that the structural field dependence of the chain elongation levels off for fields approaching $1\,$V/nm whereas no such saturation is observed for the dynamical properties. 

Since it cannot be excluded that already minor modification of a yet not discussed structural observable provokes a dramatic reduction of the activation barriers, the causality issue remained unresolved to this point. 
Therefore, we employed non-equilibrium simulations and monitored the temporal evolution of structure and dynamics in response to an external field. Starting from an $\text{E}_{\text{z}}\,=\,0.0$\,V/nm configuration we switched to $\text{E}_{\text{z}}\,=\,1.0$\,V/nm and observed that $\mu$ is, counterintuitively, slowed down.
We were able to present a microscopic understanding for this behavior and conclude on the question of causality: The nonlinear enhancement of the lithium dynamics owes to the tilting of the energy landscape in direction of the external field. Because the lithium ions pull the polymer into a stretched shape, thereby reducing the conformational entropy of the chains, the lithium dynamics decelerate.

Within this theoretical framework we might also give an answer to the opposed concentration dependencies for chain structure measured via $\text{R}_{\text{g,z}}^2$ on the one hand, and the lithium transport properties $\mu$ and $\text{D}_{\parallel}$ on the other. Because an increasing chain stretching implies a higher entropic penalty, the dynamics are increasingly impeded for decreasing salt content, which has been previously found to allow for a higher ordering of the polymer host.
Hence, we are able to give evidence, quite surprisingly, on an adverse net effect of the structural ordering on the dynamics, that means causing a slowing down.

Referring to the uncertain origin of the field-induced transport enhancement which was observed in eNMR measurements, performed at electric field strengths five orders of magnitude below the field strengths where our simulation study reveals nonlinear contributions \cite{Rosenwinkel2019}, we suggest that the eNMR can be rather explained by bulk flow effects, as already given as an alternative explanation in that work.

We hope that our findings on the rich dynamical behavior of lithium in a polymer electrolyte system draw attention to the nonlinear effects, which may emerge in the close environment of charged interfaces, i.e. the vicinity of electrodes where even minor drops of the electric potential can result in high fields.


\begin{acknowledgement}
Analysis and simulations have been performed on the computing cluster PALMA2 at the University of M\"unster. We thankfully acknowledge the financial support from MWIDE NRW as part of the "GrEEn" project (funding code: 313-W044A) and the Federal Ministry of Education and Research (BMBF) for funding within the FestBatt cluster (funding number 03XP0174B ).
\end{acknowledgement}

\begin{suppinfo}
Simulation procedure, velocity distributions, coordination numbers (Li-Li, Li-TFSI(O)), PEO-monomer dipole angle distributions, $\mu_{\text{TFSI}}$, $\text{R}_{\text{g,z}}^2$ for ions subjected to the external field only, pull-stretching-mechanism and $s_{\text{asym}}$ statistics, comparison of $\text{R}_{\text{g,z}}^2$, $\mu$ and $\text{D}_{\text{parallel}}$ for longer polymer chains, $\mu_{\text{Li}}(t)$ for switching $\text{E}_{\text{z}}$, $\tau_{\text{Li-PEO chain}}$, $\Delta \tilde{z}_{\text{Li-PEO}}$ analysis including all lithium-chain pairs.

\end{suppinfo}

\bibliography{literature}

\end{document}


\textbf{S1: Simulation procedure}

The simulation study presented in this work was carried out by means of the molecular dynamics (MD) software package GROMACS (version 2018.6) \cite{Berendsen1995,VanDerSpoel2005,Pall2015,Abraham2015}. The force field parameters for poly(ethylene oxide) (PEO) rely on the classic all-atom, non-polarizable OPLS-AA force field \cite{WilliamL.Jorgensen1996}, whereas the Li[TFSI] interactions were parameterized by the widely acknowledged OPLS-AA-based CL\&P force field developed by Canongia Lopes and Padua \cite{CanongiaLopes2012,Lopes*2004,Lopes2004,Shimizu2010}.
To account for the neglected polarizability, which entails effective charge screening and transfer \cite{Salanne2015,Nasrabadi2017,Dommert2012,Leontyev2014,Leontyev2011,Youngs2008}, in a mean field sense, all partial atom point charges were uniformly rescaled by a factor of 0.8 according to prevalent practice \cite{MullerPlathe1995,Costa2015,Mogurampelly2017,Mogurampelly2017a,Sunda2015,Mondal2014,Bhargava2007,Chaban2011}. While other simulation studies urgently promote to employ polarizable force field models \cite{And2000,OlegBorodin*2006b}, the gained accuracy in defining the atomistic local environment comes at the expense of computational effort and feasible simulation times, which are required to be sufficiently long to sample the diffusive regime of the particle dynamics for this research purpose.

If not stated otherwise, all molecular dynamics simulations were performed according to the following protocol. The initial configurations were generated in the gas phase at a temperature of $423\,$K with the Packmol package \cite{Martinez2009}, where both the lithium salt and coiled PEO chains, as they exist in the molten state, were randomly distributed in a cubic box. After an energy minimization the system was shrunk under NPT conditions for 2.5 ns using a high coupling constant of $2\,$ps to a Berendsen barostat with a reference pressure of 1 bar at a small integration time step of $0.5\,$fs (Berendsen thermostat, $\tau_T\,=\,1\,$ps). To ensure a well relaxed configuration the system was further equilibrated for $200\,$ns with a step size of $2\,$fs in a constant NPT ensemble, where pressure and temperature were controlled by a Berendsen barostat and a velocity-rescale thermostat \cite{Berendsen1984,Bussi2007}. Moreover, the linear constraint solver (LINCS) was used to constrain the hydrogen bonds \cite{Hess1997,Hess*2007}. Taking the hence equilibrated system as a starting structure, a constant external electric field oriented in $z$-direction was switched on and the system was further propagated for $90\,$ns, before beginning data acquisition in a  productive simulation run in a constant NPT ensemble, whereas here pressure and temperature couple to an extended Parrinello-Rahman and Nose-Hoover ensemble \cite{Parrinello1981,Nose1983,Nose1984,Hoover1985}. In order to reach the diffusive regime of the lithium ions and ensure sufficient sampling, the duration of the productive runs extends up to $2\,\mu$s, but covers at least $300\,$ns. The center of mass of the system was repositioned every $10$th simulation step.

\begin{figure}[H]
\centering
\includegraphics[width=1.0\textwidth]{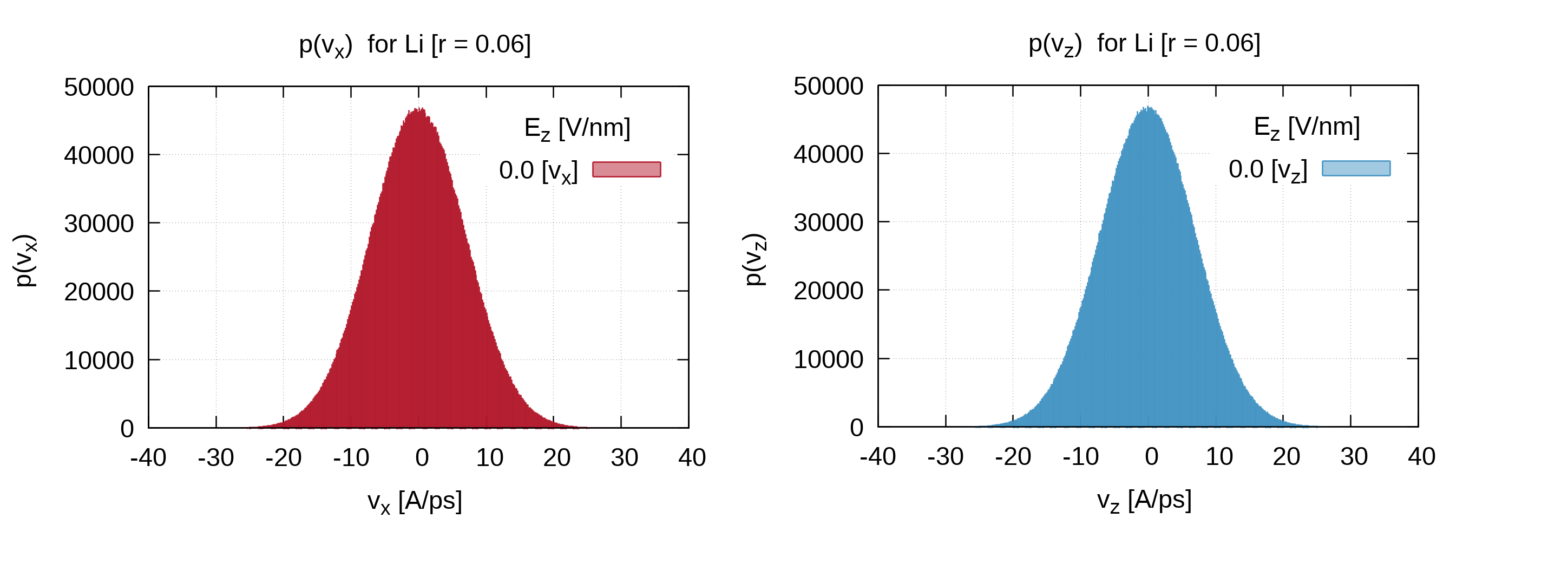}
\caption{Histogram of lithium velocities in arbitrary  directions in absence of an external field.}
\label{fig:histogram_velocities_E_z_0_0}
\end{figure}

\begin{figure}[H]
\centering
\includegraphics[width=1.0\textwidth]{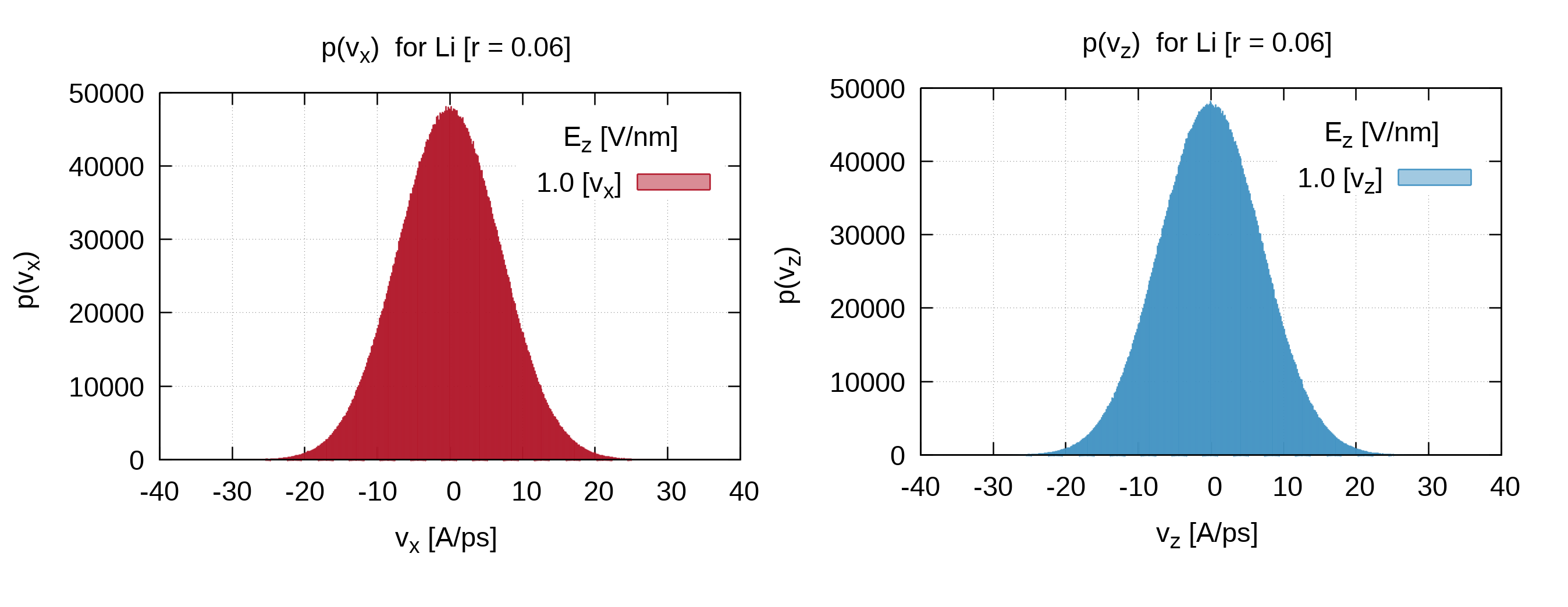}
\caption{Histogram of lithium velocities in $x$ and $z$-direction for application of an external field strength of 1.0 V/nm in $z$-direction.}
\label{fig:histogram_velocities_E_z_1_0}
\end{figure}

We further present exemplary velocity distribution profiles for the lithium ions parallel and perpendicular to the external electric field (see Figures \ref{fig:histogram_velocities_E_z_0_0} and  \ref{fig:histogram_velocities_E_z_1_0}) to ensure correct sampling of the Maxwell-Boltzmann distribution. Computing the non-gaussianity parameter $\alpha = 1/3 \cdot \langle  v_i^4 \rangle / \langle   v_i^2 \rangle ^2 - 1 $ , i.e. the ratio of second and fourth moments of the instant velocity distribution in each dimension $i$, which would vanish for a truly Gaussian distribution, $\alpha_{\text{para}}$ and $\alpha_{\text{ortho}}$  yields values in the order of $10^{-2} \approx 0$.

\newpage
\textbf{S2: Dynamic response TFSI and lithium-chain mean residence times }

\begin{figure}[H]
\centering
\includegraphics[width=0.8\textwidth]{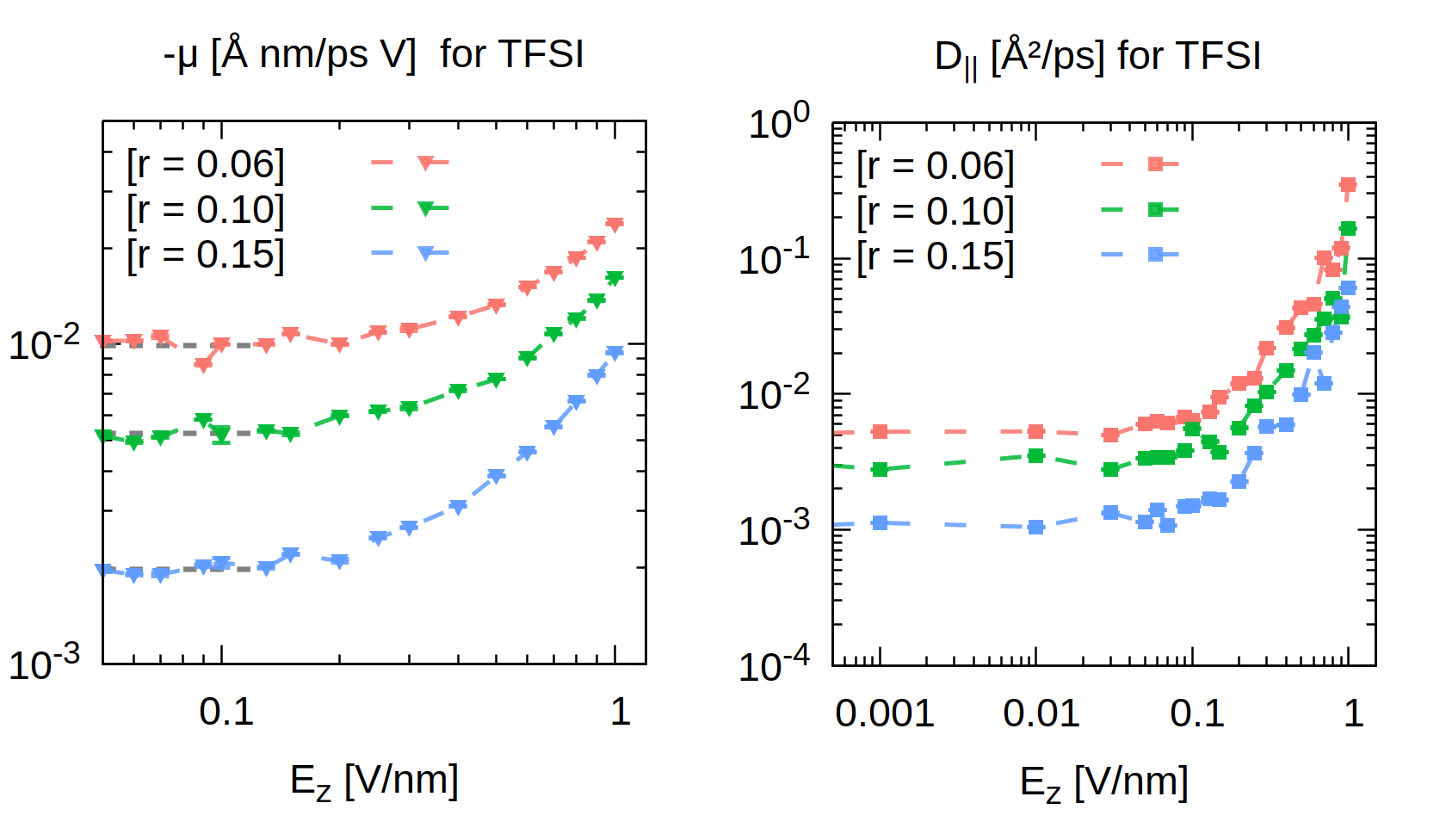}
\caption{TFSI mobilities $\mu$ and parallel diffusion constants $\text{D}_{\parallel}$ as a function of electric field strength for three different salt concentrations.}
\label{fig:nonlinear_dynamics_mu_d_para_tfsi}
\end{figure}

\begin{figure}[H]
\centering
\includegraphics[width=0.7\textwidth]{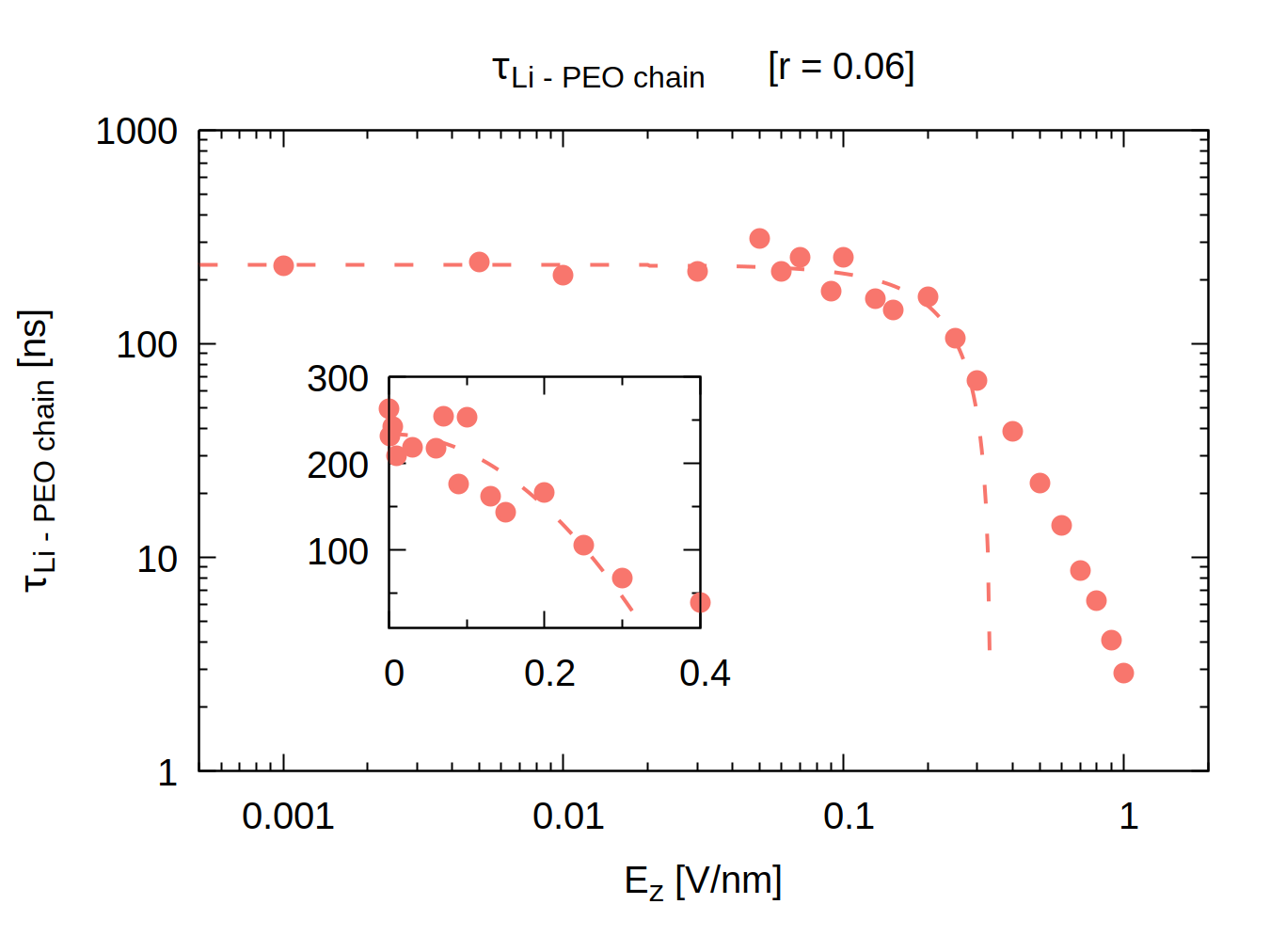}
\caption{Mean residence time of lithium on polymer chain as a function of $\text{E}_{\text{z}}$ for r\,=\,0.06 and example quadratic $\text{E}_{\text{c}}$ fit ($\text{E}_{\text{c}}$\,=\,0.33\,V/nm). }     
\label{fig:mean_residence_time_lithium_chain}
\end{figure}

\newpage


\textbf{S3: RDF and cumulative numbers}

For the highest salt content the average lithium coordination cage is composed of two TFSI anion oxygens and four ether oxygens in absence of an electric field, which is in excellent accordance with literature values for similar salt concentration \cite{Xu2004,Costa2015}. For lower salt contents the lithium-ether-oxygen coordination is higher, which may be attributed to a mere excess supply of vacant polymer segments \cite{Costa2015}.

\begin{figure}[H]
	\centering
	\includegraphics[width=0.7\textwidth]{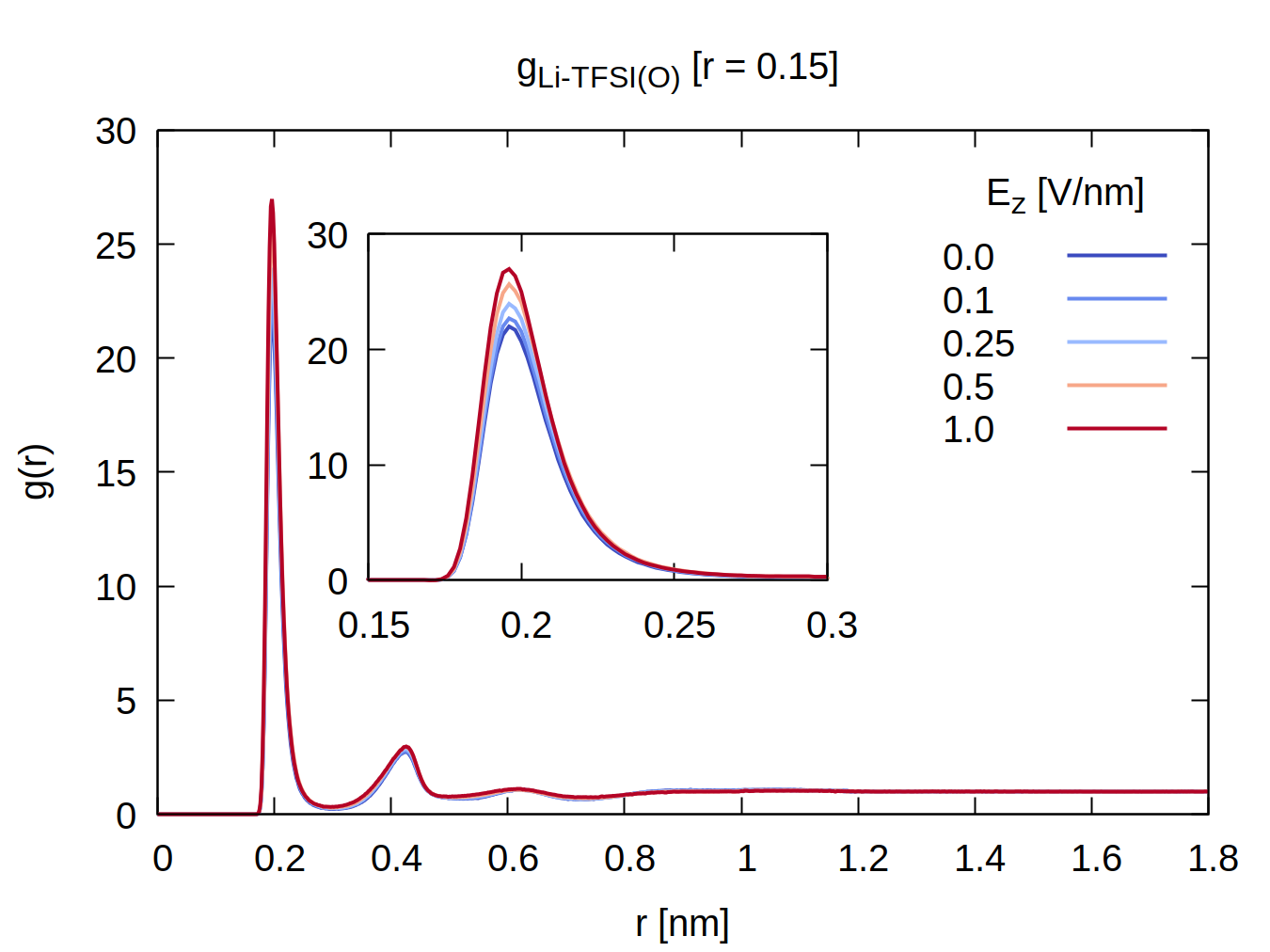}
	\caption{Exemplary lithium-TFSI(O) radial distribution functions as a function of electric field strength for the r = 0.15 electrolyte mixture.}
	\label{fig:RDF_LI_TFSI_OS}
\end{figure}

\begin{figure}[H]
	\centering
	\includegraphics[width=0.6\textwidth]{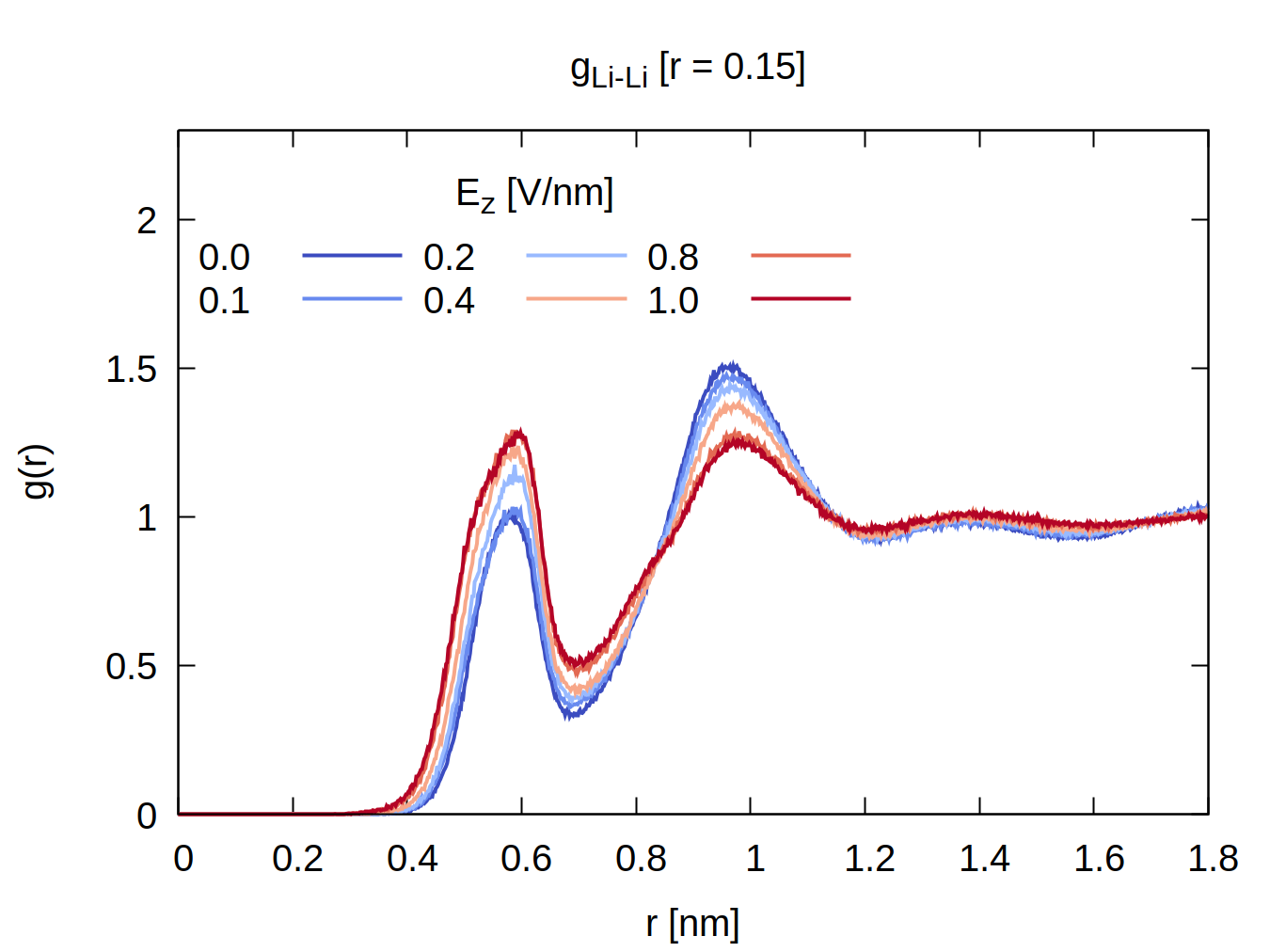}
	\caption{Exemplary lithium-lithium radial distribution functions as a function of electric field strength for the r = 0.15 electrolyte mixture.}
	\label{fig:RDF_LI_LI}
\end{figure}

\begin{figure}[H]
	\centering
	\includegraphics[width=0.7\textwidth]{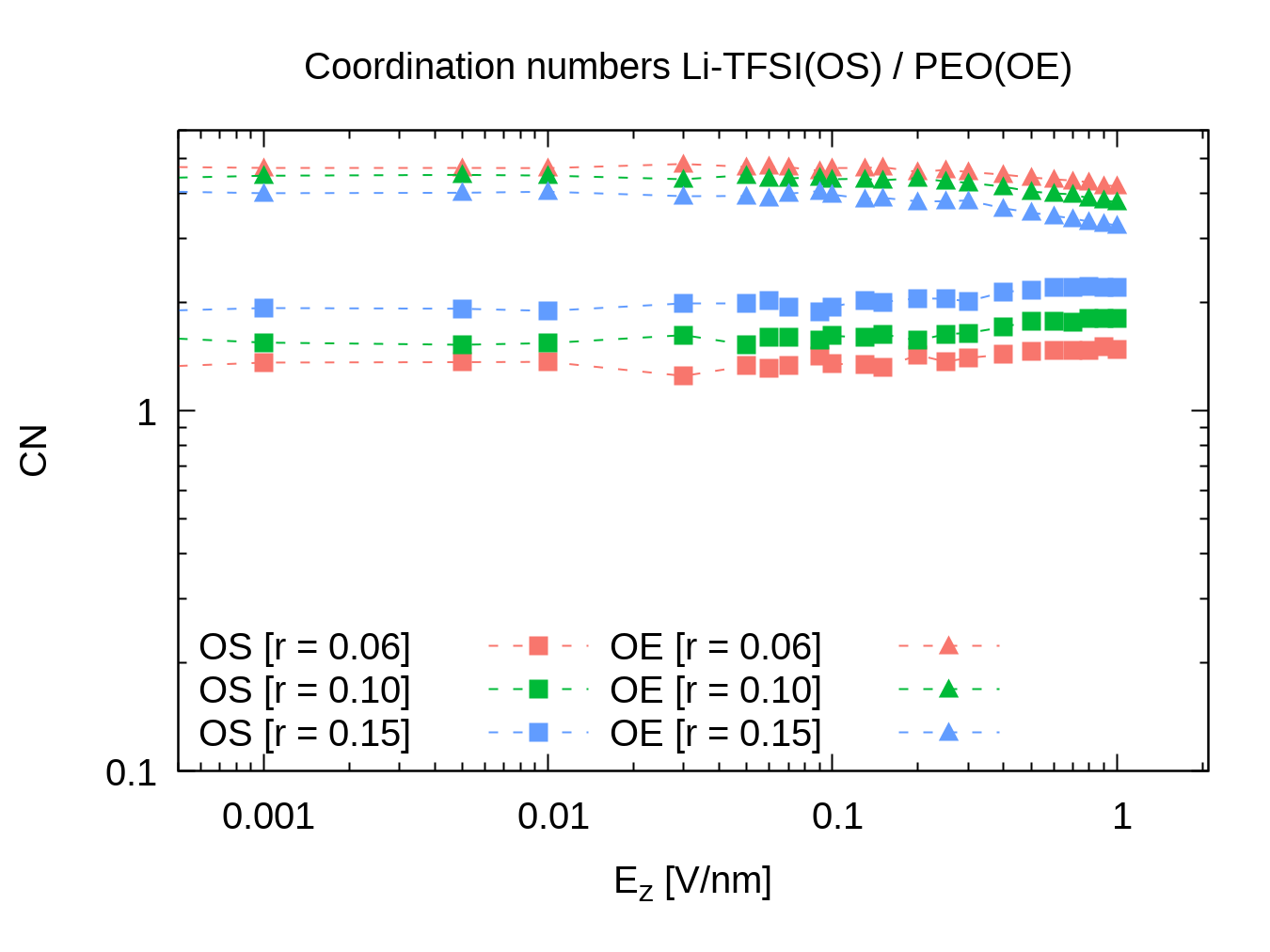}
	\caption{Electric field dependence of the Li - OS(TFSI) and Li - OE(PEO) coordination numbers for various salt concentrations.}
	\label{fig:cn_tfsi_o_peo_o_li_log}
\end{figure}

\begin{figure}[H]
	\centering
	\includegraphics[width=0.7\textwidth]{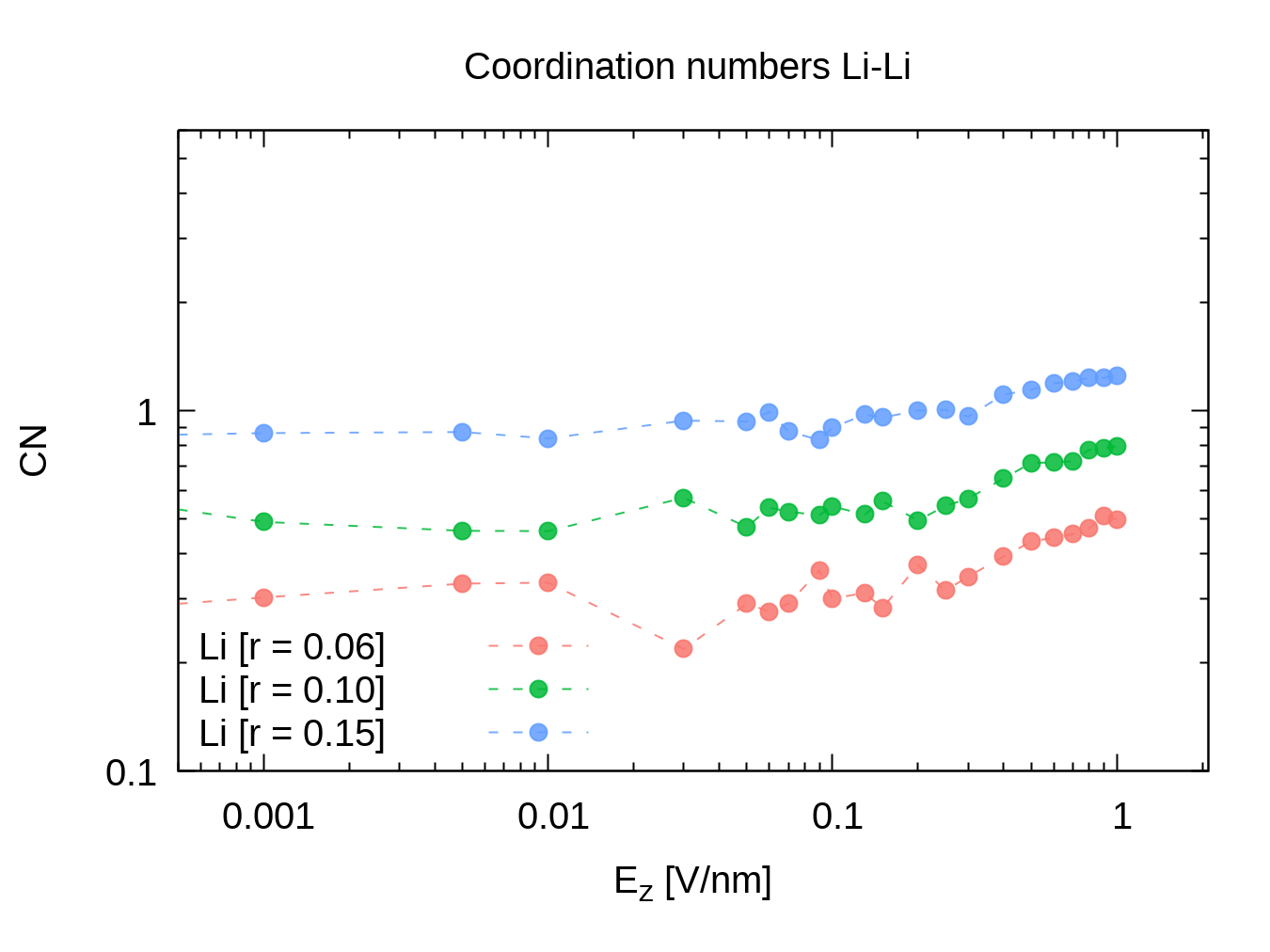}
	\caption{ Li-Li coordination numbers as a function of electric field strength for different salt contents.}
	\label{fig:cn_li_li_log}
\end{figure}

\begin{table} [hbtp]
	\caption{Relative frequencies of chain coordination motifs in absence of electric field and a maximum electric field strength of $\text{E}_{\text{z}}\,=\,1.0$~V/nm for various salt concentrations. The label 'free' means that lithium is not coordinated by the polymer.}\label{tab:ratio_coordination_chains}
	\centering
	\def\arraystretch{1.5}
	\begin{tabular*}{\textwidth}{c @{\extracolsep{\fill}} l l | l l|  l  l  l    }
		\hline
		\multirow{2}{3cm}{r =[Li]/[EO] }  &  \multicolumn{2}{c|}{free [\%] } &  \multicolumn{2}{c|}{1 chain [\%] }&  \multicolumn{2}{c}{2 chains [\%] } \\
		
		& $0.0$& $1.0$ [V/nm]& $0.0$& $1.0$ [V/nm]& $0.0$ & $1.0$ [V/nm]\\		
		\hline
		0.06 & 0.5 & 3.0 & 96.9 & 81.3  & 2.6 & 15.7  \\ 
		0.10 & 5.1 & 6.7 & 93.7 & 	82.6	& 1.2 & 10.7  \\
		0.15 & 5.7 & 13.6 & 93.6 & 79.2 & 0.7 & 7.2 \\ 		
		\hline
	\end{tabular*}
\end{table}

\textbf{S4: Monomer dipole orientation}

\begin{figure}[H]
\centering
\includegraphics[width=0.7\textwidth]{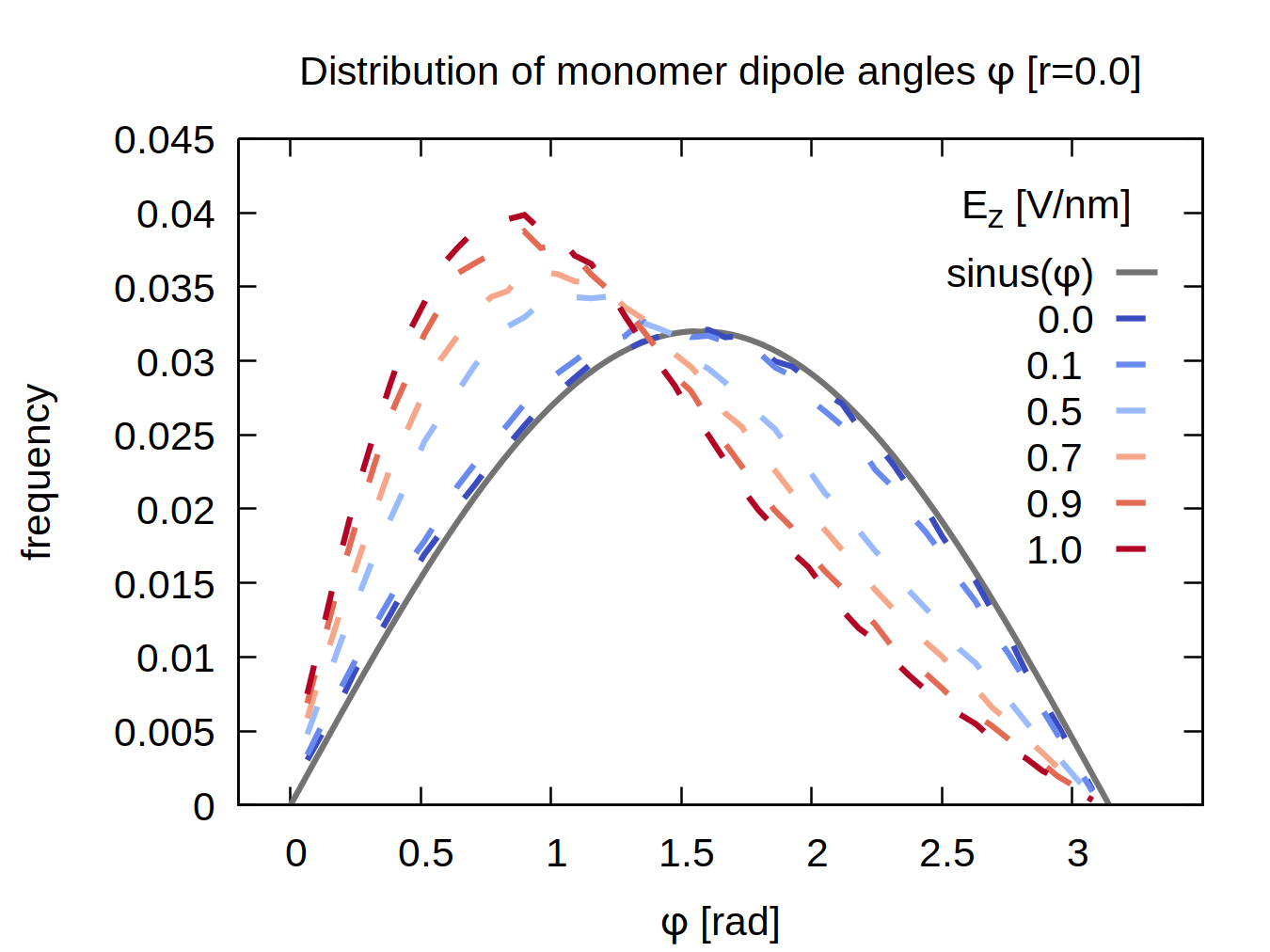}
\caption{Probability distribution of the angles of the individual monomer dipole C-O-C sections relative to field orientation in the plain polymer system for various electric field strengths.}
\label{fig:plain_polymer_dipoles}
\end{figure}

As mentioned in the main text we observe a deceleration of the polymer ordering captured via $\text{R}^2_{\text{g,z}}$ for electric field strengths above 0.6\,V/nm. To understand this behavior we study the orientation of the local monomer dipoles $\overrightarrow{p}_{\text{mon}}  $, constituted by the partially charged C-O-C segments, relative to the external field. The probability distribution of the angle between two randomly pointing vectors is proportional to $\sin(\varphi)$. 
We choose $\varphi$ as $\sphericalangle (\overrightarrow{\text{E}}_{\text{z}},\overrightarrow{p}_{\text{mon}} ) $.
As shown in Figures S\ref{fig:plain_polymer_dipoles} and S\ref{fig:r_0_06_polymer_dipoles} we recover the expected sinusoidal distribution in the plain polymer as well as the r\,=\,0.06 electrolyte mixture in absence of an external field when the structure of the polymer host has not received an orientation.
For application of an external field we observe a shift of $\varphi$ to smaller angles which means that individual monomer dipoles increasingly point in field direction. This local dipole $\overrightarrow{p}_{\text{mon}}$ alignment in field direction corresponds to an orientation of the C-O-C segment perpendicular to the field. Hence, the polymer chains contract and counteract the overall alignment in the field. To further assess the intensity of this effect, we normalize the distributions of $\varphi$ to  $\varphi_0$ when no field is applied. As shown in Figure S\ref{fig:r_0_06_polymer_dipoles_normed} the effect is less pronounced for the salt-in-polymer electrolyte.

\begin{figure}[H]
\centering
\includegraphics[width=0.7\textwidth]{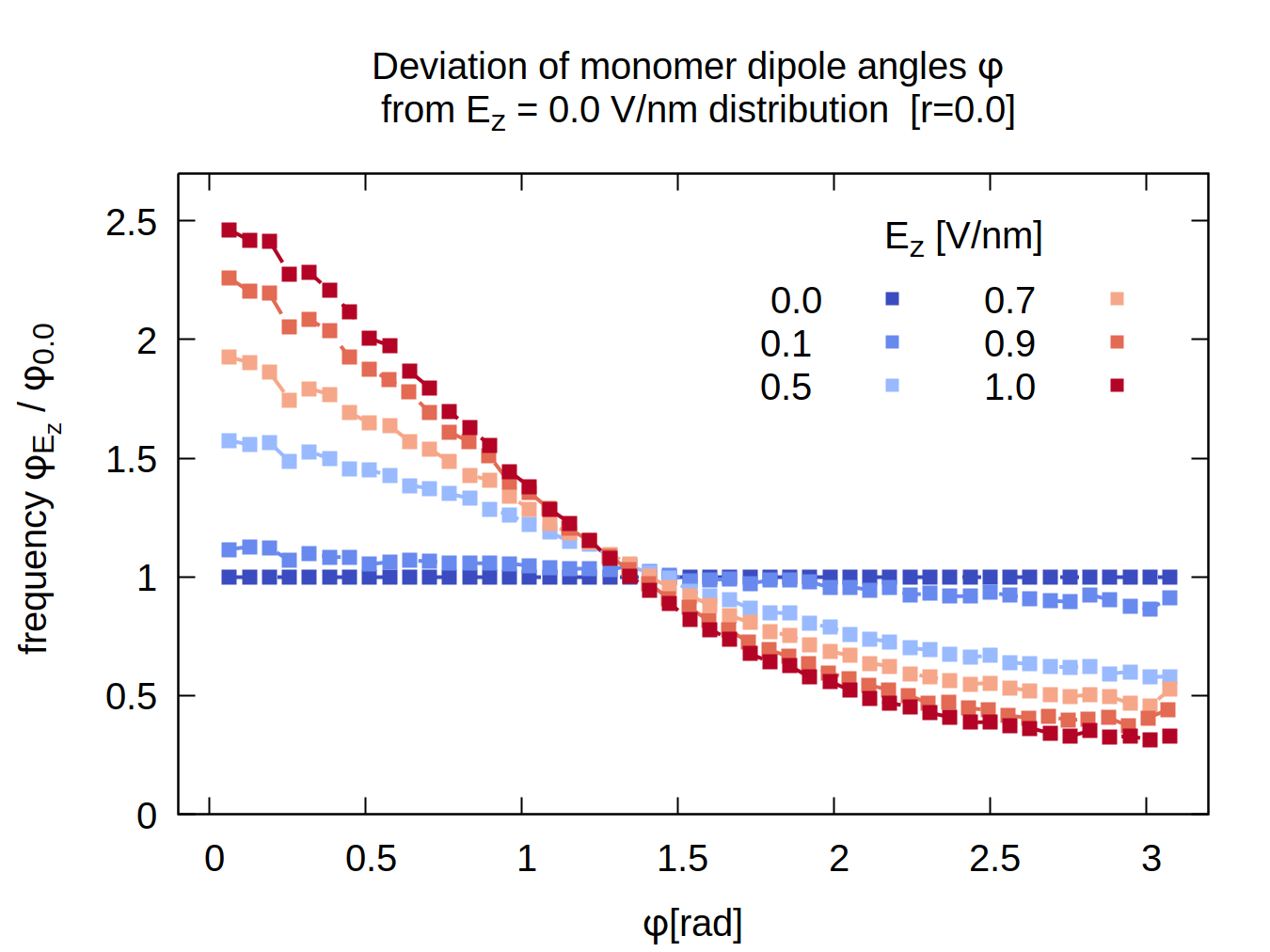}
\caption{Deviation of probability distribution in Figure S\ref{fig:plain_polymer_dipoles} from field-free reference $\varphi_{0,\text{r\,=\,0.0}}$.}
\label{fig:plain_polymer_dipoles_normed}
\end{figure}

\begin{figure}[H]
\centering
\includegraphics[width=0.7\textwidth]{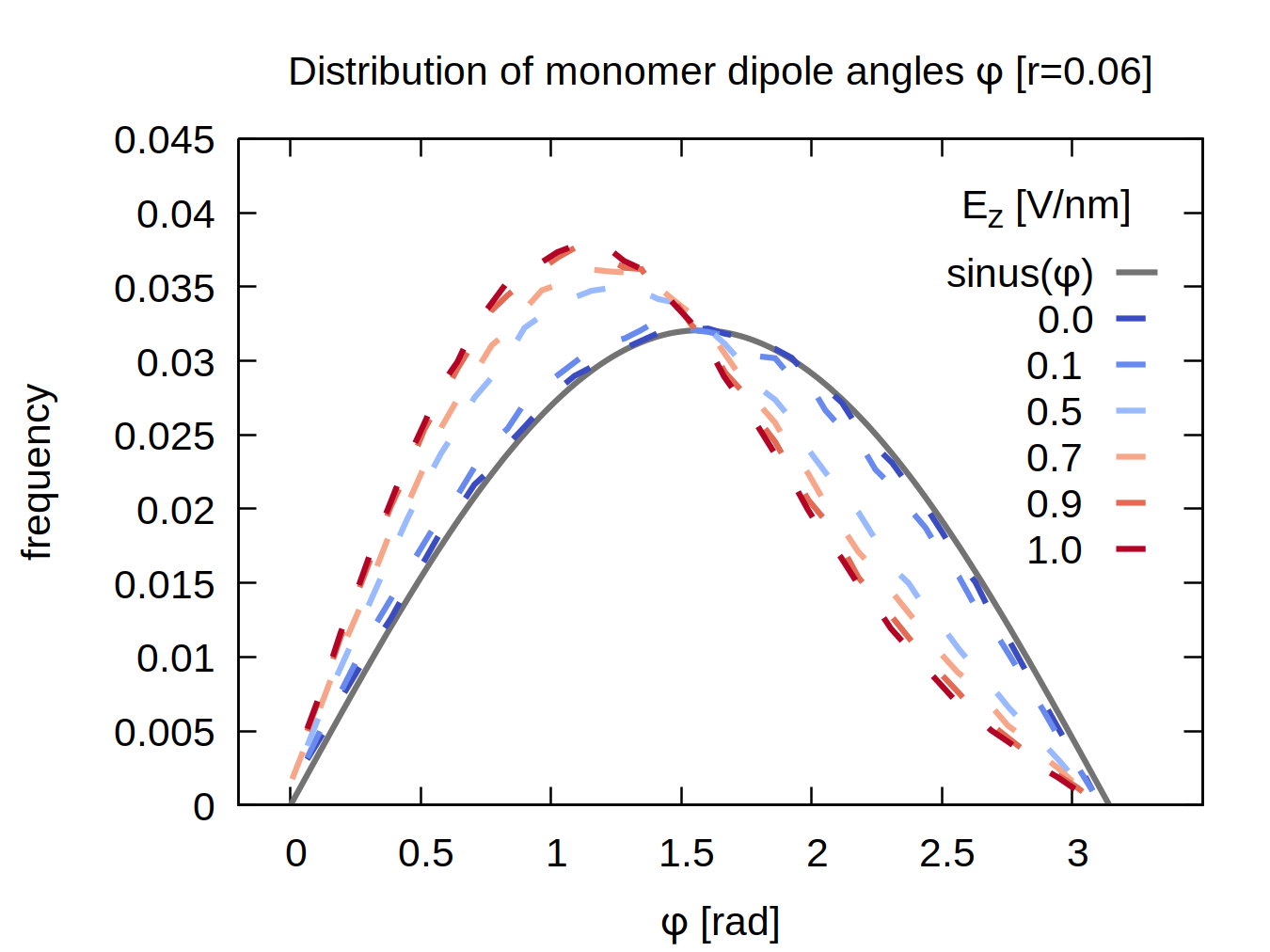}
\caption{Probability distribution of the angles of the individual monomer dipole C-O-C sections relative to field orientation in the $r\,=\,0.6$ salt-in-polymer mixture for various electric field strengths.}
\label{fig:r_0_06_polymer_dipoles}
\end{figure}

\begin{figure}[H]
\centering
\includegraphics[width=0.7\textwidth]{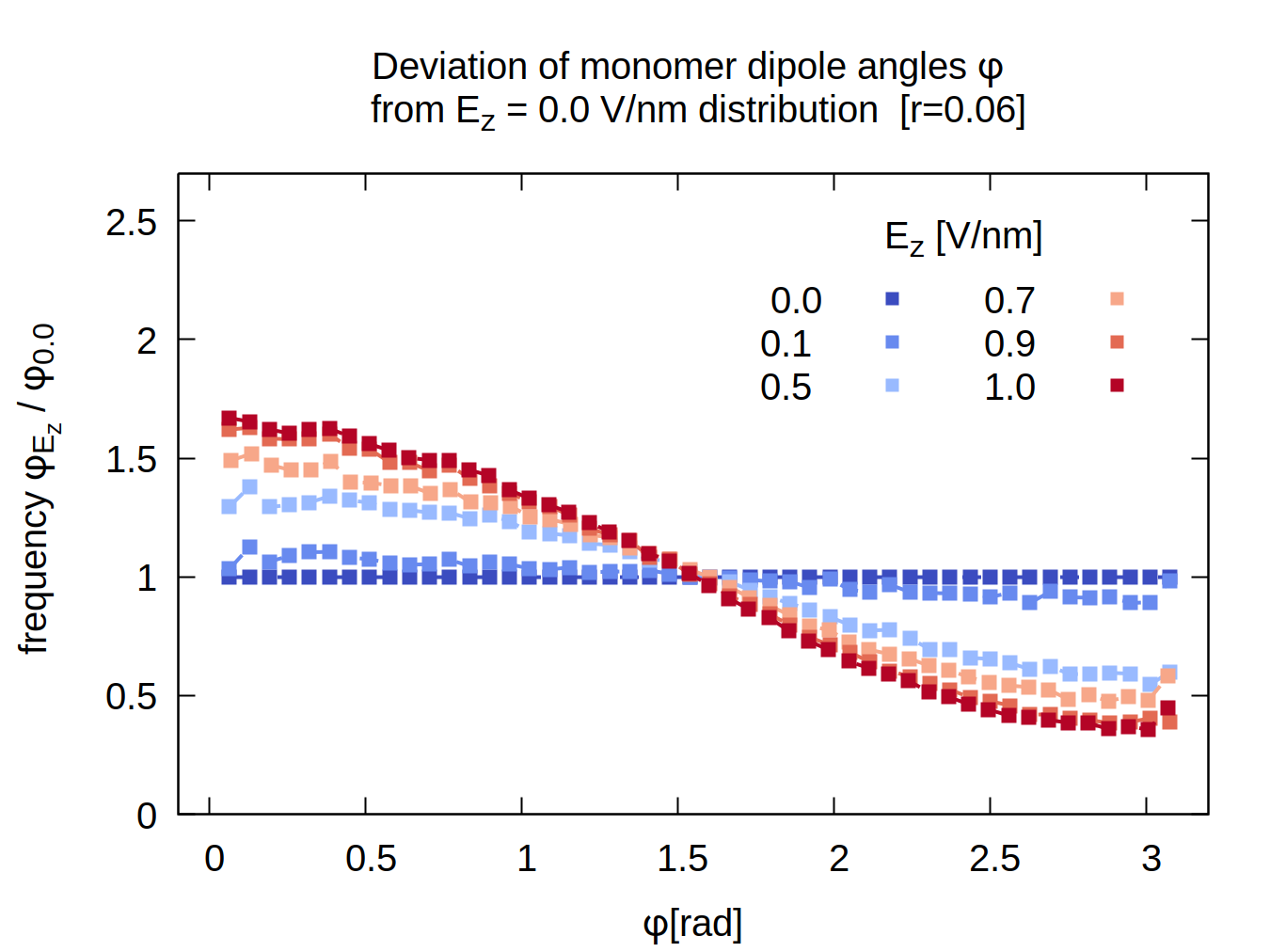}
\caption{Deviation of probability distribution in Figure S\ref{fig:r_0_06_polymer_dipoles} from field-free reference $\varphi_{0,\text{r\,=\,0.06}}$.}
\label{fig:r_0_06_polymer_dipoles_normed}
\end{figure}

\textbf{S5: Comparison $\text{R}^2_{\text{g,z}}$ for electric field applied on ions only}

\begin{figure}[H]
	\centering
	\includegraphics[width=0.7\textwidth]{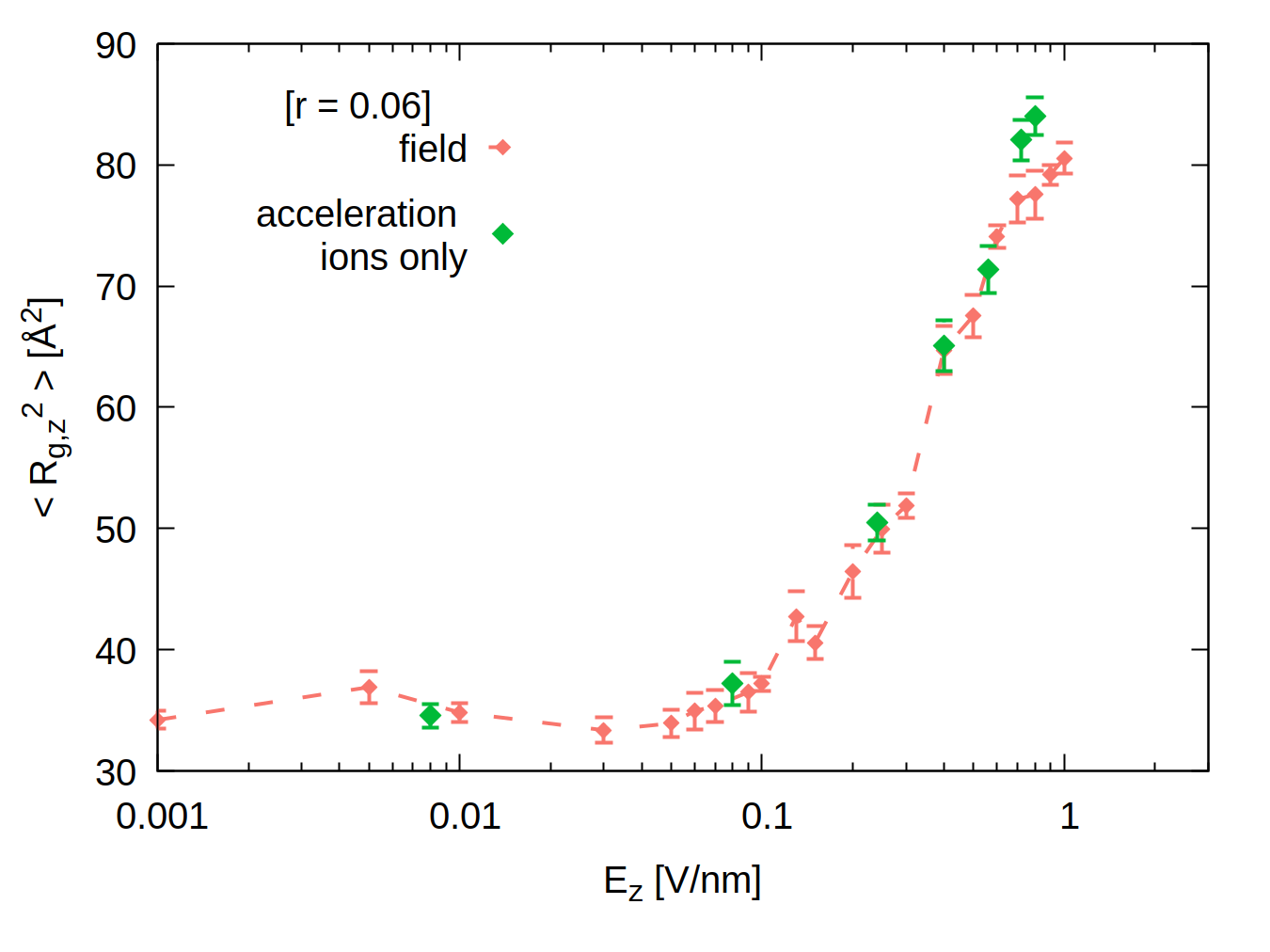}
	\caption{Comparison of gyration radii $\text{R}^2_{\text{g,z}}$ for the $r\,=\,0.06$ salt-in-PEO mixture for electrostatic force exerted on of all partial charges in the system ('field' , i.e. including the polymer atoms' partial charges) and ionic charge carriers only ('acceleration', i.e. lithium and TFSI only).}
	\label{fig:R_g_field_acceleration}
\end{figure}


\newpage
\textbf{S6: Structure formation principle - $\text{s}_{\text{asym}}$ analysis}

As mentioned in the main body of the paper we provide additional information on the chain ordering and coordination pattern for lithium on the polymer backbone. The observables used for this matter are the number of lithium ions coordinated to the backbone $\text{n}_{\text{backbone}}$, the asymmetry of their coordination sites as captured by $\text{s}_{\text{asym}}$ , the polymer chain's mean-squared gyration radius $\text{R}^2_{\text{g,z}}$ in field direction and the number of monomers coordinating lithium.

\begin{figure}[H]
\centering
\includegraphics[width=1.0\textwidth]{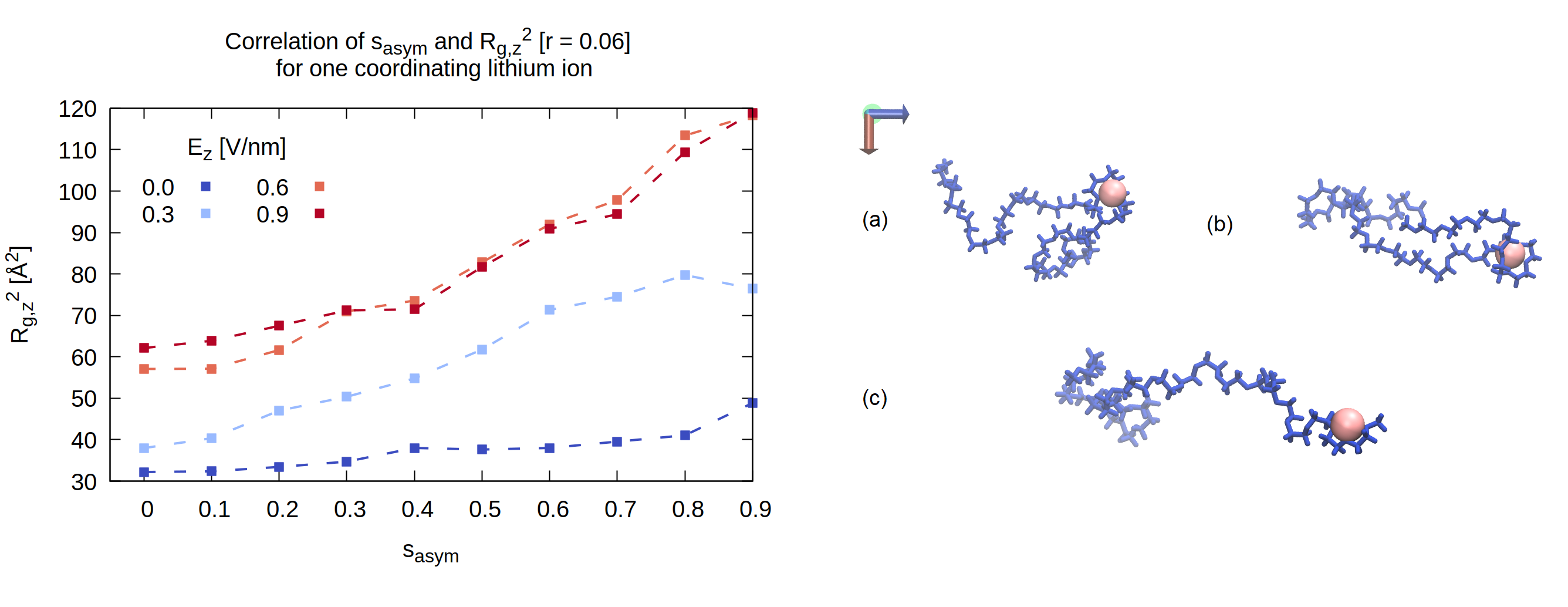}
\caption{Left: Relation between squared gyration radius in field direction $\text{R}_{\text{g,z}}^2$ and asymmetric chain coordination parameter $s_{\text{asym}}$ for various electric field strengths exemplary for the r = 0.06 mixture. Right: Illustration of lithium induced stretching of the polymer chain. The snapshots are taken from the $\text{E}_{\text{z}}$\,=\,1.0\,V/nm simulation of the r\,=\,0.06 mixture. For (a) $s_{\text{asym}}$\,=\,0.0 and $\text{R}_{\text{g,z}}^2$\,=\,58\,\angstrom$^2$, (b) $s_{\text{asym}}$\,=\,0.5 and $\text{R}_{\text{g,z}}^2$\,=\,85\,\angstrom$^2$ and (c) $s_{\text{asym}}$\,=\,0.9 and $\text{R}_{\text{g,z}}^2$=\,130\,\angstrom$^2$.}
\label{fig:MECHANISM}
\end{figure}

We propose an explanatory model for the coiled-to-stretched transformation of the polymer conformation upon increasing field strength with the underlying idea of lithium, closely attached to the ether oxygens, pulling the polymer chain into the elongated shape. 
Therefore, we define an order parameter $s_{\text{asym}}$ that quantifies the degree of symmetry of the lithium coordination of a chain:
\begin{equation}
s_{\text{asym}} = \dfrac{| n_{\text{Li}} - (N-1)/2 |}{ (N-1)/2},
\end{equation}
where $n_{\text{Li}}$ denotes the average index of the monomers coordinating the lithium ions on the individual chain and $N$ is the total number of monomers of a chain. For example, for a symmetric coordination of a single lithium ion to the chain center the definition yields $s_{\text{asym}}$\,=\,0, whereas coordination to the outermost monomer at the chain end $s_{\text{asym}}$\,=\,1. 
We assume that the stretching is most effective when lithium exerts such a pulling force at the chain ends. To  be most sensitive to this effect, we determine $\text{R}_{\text{g,z}}^2$ as a function of $s_{\text{asym}}$ for a subensemble where only a single lithium ion is attached to the polymer chain.  
The correlations between $s_{\text{asym}}$ and $\text{R}_{\text{g,z}}^2$ are shown in Figure S\ref{fig:MECHANISM}.\newline We make two independent observations:  \newline
First, we note that $\text{R}_{\text{g,z}}^2$ increases for increasing $s_{\text{asym}}$ in absence of an external field. For coordination of lithium at a center position the chain loops in the crown-ether fashion which thus results in a more coiled conformation. For a lithium - chain end contact, the ether oxygen coordination number drops (see Figure S\ref{fig:s_asym_r_0_06_coordination_number_lithium_ether_oxygens}) and the coiling effect due to lithium becomes smaller. 
Second, we observe a dramatic stretching effect for increasing $\text{E}_{\text{z}}$ for increasing $s_{\text{asym}}$, which resonates well with our assumption of lithium-induced stretching. 
The fact that also $\text{R}_{\text{g,z}}^2$ for $s_{\text{asym}}\,=\,0$ shifts to higher values, we attribute to a local stretching as visually illustrated in the simulation snapshots in Figure S\ref{fig:MECHANISM} as well as the ordering effect of adjacent stretched chains.  

On that basis, the previous observation of more effective chain alignment in field direction for decreasing salt content can be understood as a mere concentration effect. As can be seen in Figure S\ref{fig:s_asym_statistics_r_0_15} the probability for an asymmetric distribution of the lithium ions on the polymer backbone decreases with increasing salt content and so does consequently the ability of the lithium ions to stretch the chains.
 
\begin{figure}[H]
\centering
\includegraphics[width=0.7\textwidth]{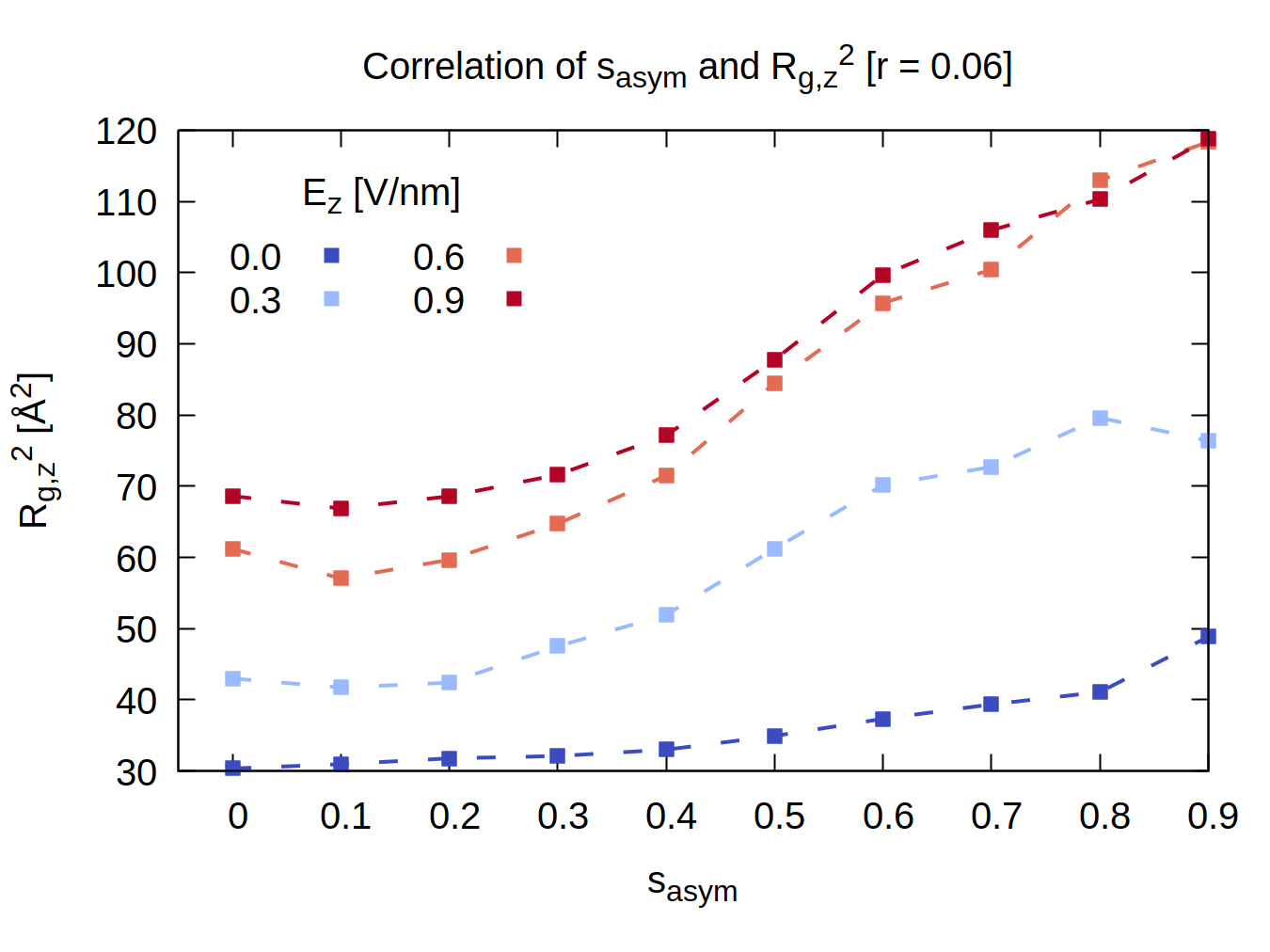}
\caption{Effect of $\text{E}_{\text{z}}$ on $\text{R}^2_{\text{g,z}}$ binned according to the asymmetry $\text{s}_{\text{asym}}$ of lithium coordination to backbone for taking into account all possible lithium coordination motifs, i.e. not only the subset of a single lithium coordinating to the backbone.}
\label{fig:s_asym_r_0_06_Rgz2_all_lithium_configurations}
\end{figure}

Figure S\ref{fig:s_asym_r_0_06_Rgz2_all_lithium_configurations} shows the correlation of $\text{R}^2_{\text{g,z}}$ and $\text{s}_{\text{asym}}$ in consideration of all possible lithium coordination motifs to the polymer backbone, i.e. all possible  $\text{n}_{\text{backbone}}$, and not only the subensemble of $\text{n}_{\text{backbone}}$\,=\,1 as discussed previously. Our underlying intention was to fairly compare the lithium-chain-dragging mechanism across the field range via this correlation while we remain aware of the significant change of the local polymer environment of lithium. Having said this, the relation between $\text{R}^2_{\text{g,z}}$ and $\text{s}_{\text{asym}}$ are recovered qualitatively for the complete ensemble of  $\text{n}_{\text{backbone}}$.

\begin{figure}[H]
\centering
\includegraphics[width=0.7\textwidth]{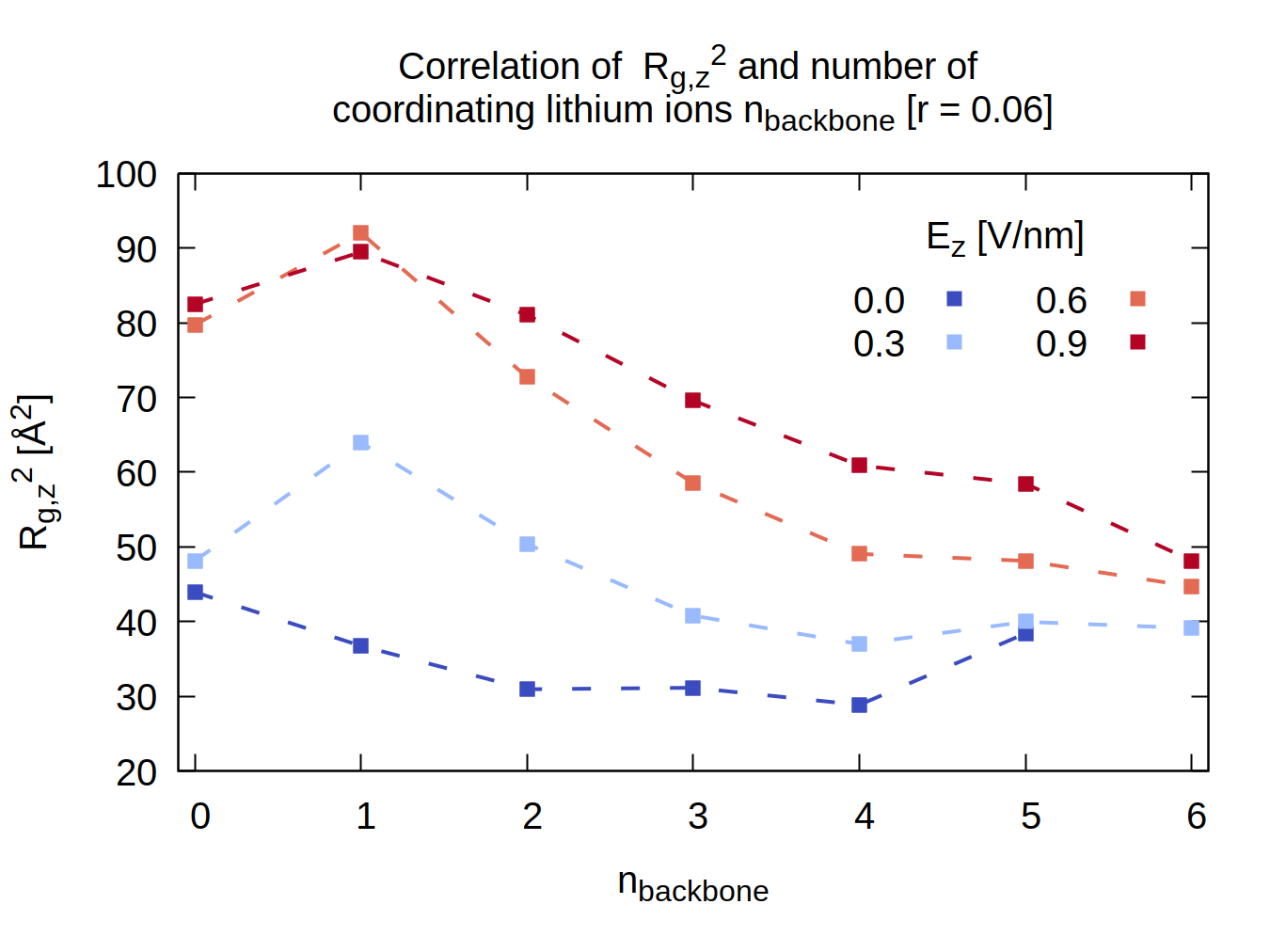}
\caption{Average $\text{R}^2_{\text{g,z}}$ as a function of the number of lithium coordinated to the polymer chain for various field strengths.}
\label{fig:Rgz2_number_lithium_on_backbone_r_0_06}
\end{figure}

\begin{figure}[H]
\centering
\includegraphics[width=0.7\textwidth]{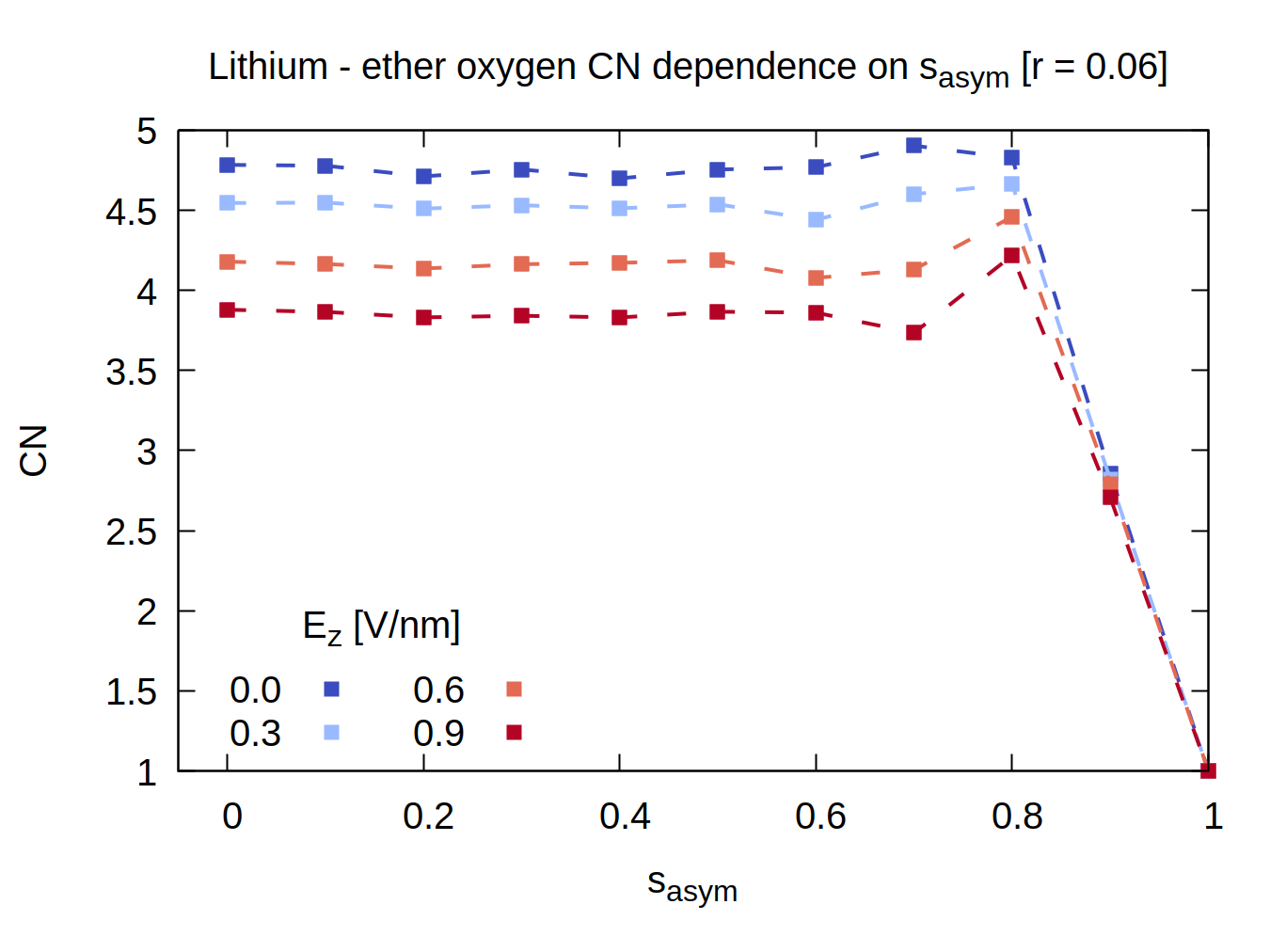}
\caption{Average lithium coordination via ether oxygens as a function of asymmetry of lithium coordination to the backbone for various field strengths.}
\label{fig:s_asym_r_0_06_coordination_number_lithium_ether_oxygens}
\end{figure}

In Figure S\ref{fig:Rgz2_number_lithium_on_backbone_r_0_06} we show the correlation between  $\text{R}^2_{\text{g,z}}$ and $\text{n}_{\text{backbone}}$ for various field strengths. We observe a shift of $\text{R}^2_{\text{g,z}}$  to higher values even when there is no lithium coordinating to the chain, i.e. $\text{n}_{\text{backbone}}$\,=\,0. As mentioned in before, we interpret this as indicative of the ordering effect of stretched chains.
Note that in absence of an external field, the equilibrium  $\text{R}^2_{\text{g,z}}$ of the plain polymer melt is recovered for $\text{n}_{\text{backbone}}$\,=\,0.

In Figure S\ref{fig:s_asym_r_0_06_coordination_number_lithium_ether_oxygens} we show how the average number of ether oxygens CN, provided by one chain, coordinating lithium changes as a function of $\text{s}_{\text{asym}}$ and electric field. Firstly, we observe that CN shifts to lower values for increasing field strength which can be explained by the progressing detachment of lithium from the polymer backbone. Secondly, we note a slight increase of CN for  $\text{s}_{\text{asym}}$\,=\,0.8 followed by a steep drop to CN\,=\,1 for $\text{s}_{\text{asym}}$\,=\,1. Whereas for  $\text{s}_{\text{asym}}$\,=\,0.8 the loose end of the polymer chain can wrap lithium efficiently, a maximum asymmetric position of lithium on the polymer backbone naturally corresponds to a single monomer contact.

\begin{figure}[H]
\centering
\includegraphics[width=0.7\textwidth]{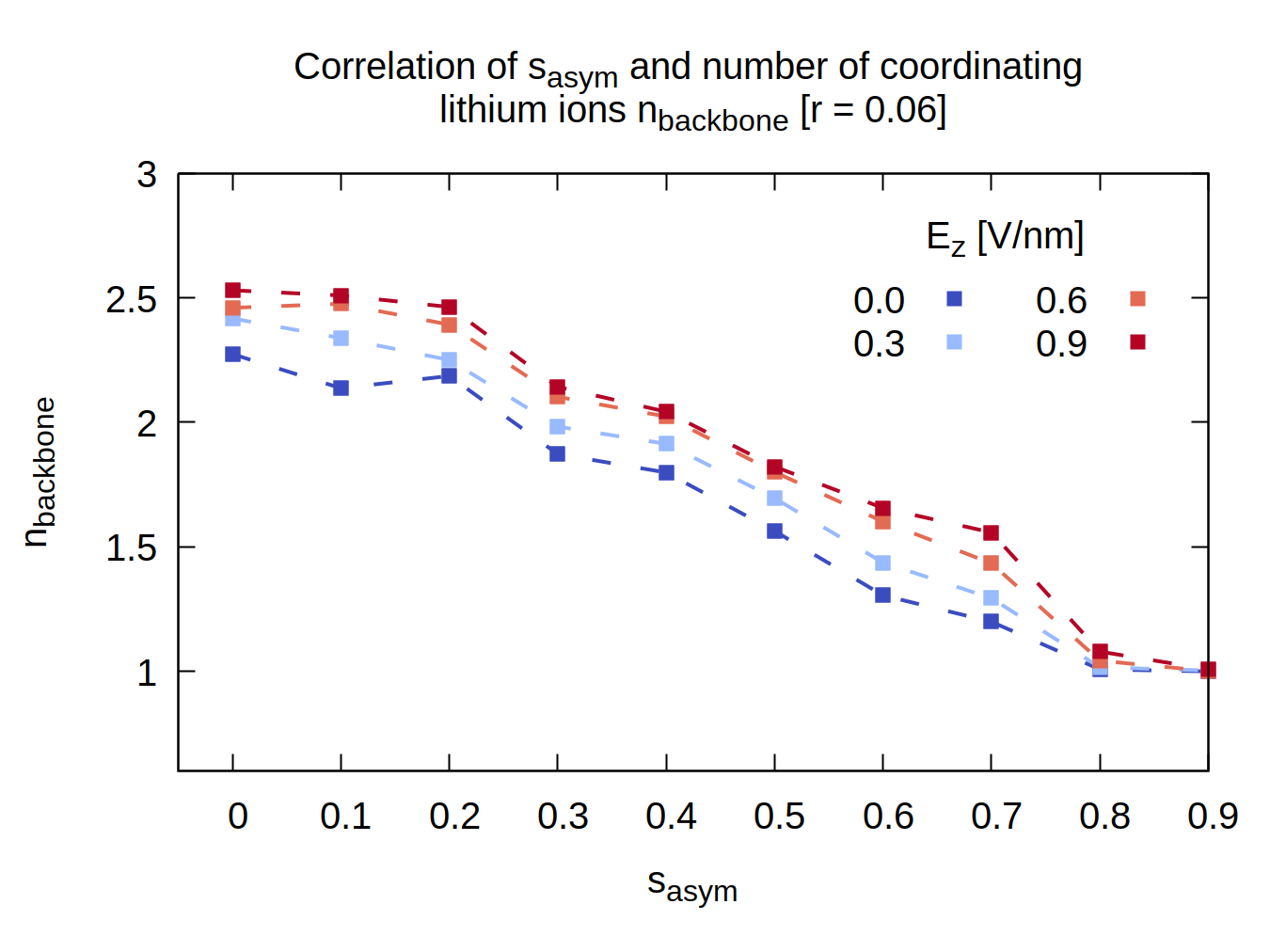}
\caption{Effect of $\text{E}_{\text{z}}$ on the average number of lithium ions on the backbone constituting the respective  $\text{s}_{\text{asym}}$.}
\label{fig:s_asym_number_lithium_on_backbone_r_0_06}
\end{figure}

Figure S\ref{fig:s_asym_number_lithium_on_backbone_r_0_06} provides information on the composition of the individual $\text{s}_{\text{asym}}$ values. While a very asymmetric lithium coordination of the backbone is usually provided by a single lithium ion attached at an outer position, a symmetric coordination of the chain is on average constituted via 2 or 3 lithium ions positioned on the chain. We further note an upshift of the number of lithium ions coordinated to the polymer chain $\text{n}_{\text{backbone}}$ for increasing field strength. We attribute this effect to the more frequent coordination of lithium via 2 chains as discussed in the main text. Consequently, polymer chains may increasingly share the same lithium ion.

\begin{figure}[H]
\centering
\includegraphics[width=0.7\textwidth]{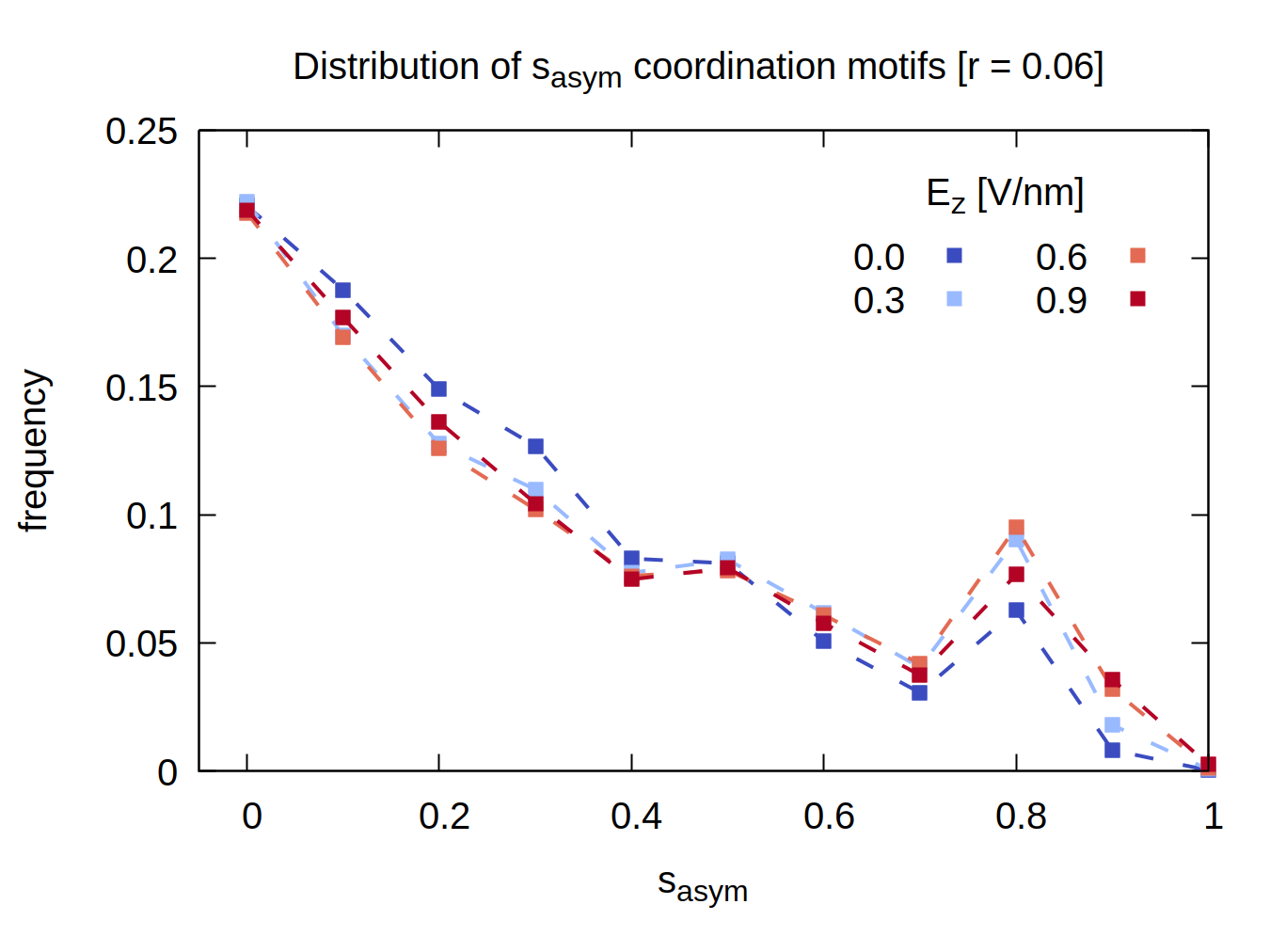}
\caption{Effect of $\text{E}_{\text{z}}$ on the probability distribution of  $\text{s}_{\text{asym}}$. An entirely asymmetric coordination, i.e.$\text{s}_{\text{asym}}$\,=\,1.0 is observed in less than 0.25\% of the samples. }
\label{fig:s_asym_statistics_r_0_06}
\end{figure}

\begin{figure}[H]
\centering
\includegraphics[width=0.7\textwidth]{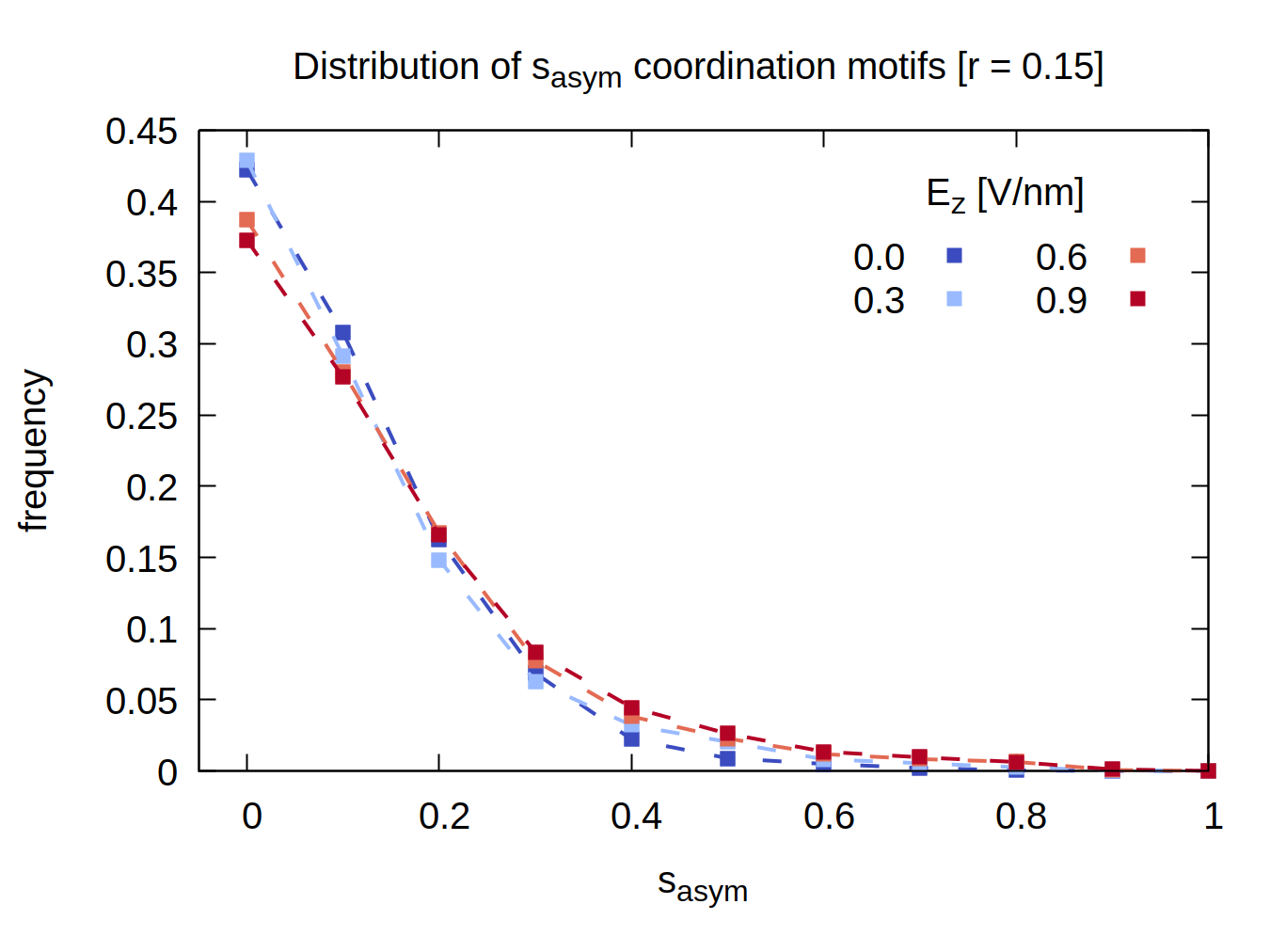}
\caption{Effect of $\text{E}_{\text{z}}$ on the probability distribution of  $\text{s}_{\text{asym}}$ for the salt content r\,=\,0.15. }
\label{fig:s_asym_statistics_r_0_15}
\end{figure}

Figure S\ref{fig:s_asym_statistics_r_0_06} shows the probability distribution of $\text{s}_{\text{asym}}$. The favorable wrapping of the polymer chain tail around lithium at $\text{s}_{\text{asym}}$\,=\,0.8 emerges as more frequent. The lithium-chain-dragging in field direction is further reflected in a slight increase of asymmetric coordination for increasing $\text{E}_{\text{z}}$.

\newpage
\textbf{S7: Structural and dynamic response for chain length N = 54}

Simulation of polymeric electrolytes by means of all-atomistic molecular dynamics is a challenging task due to the slow dynamics of the lithium ions, which strongly interact with the ether oxygens of the polymer backbone, and the long relaxation times of the polymer. Hence, atomistic-level simulation of such systems comes along at extensive computational cost. 
To assess within the frame of computational capabilities if the response of the polymer electrolyte system is qualitatively affected by the chain length, we performed additional simulations of the r\,=\,0.06 mixture but employing chains of double the length N\,=\,54. 
We extended the equilibration run to 500\,ns followed by simulation for data acquisition with a duration of 2\,$\mu$s for $\text{E}_{\text{z}}$\,<\,0.6\,V/nm and 1\,$\mu$s for the remaining field strengths.

\begin{figure}[H]
\centering
\includegraphics[width=0.7\textwidth]{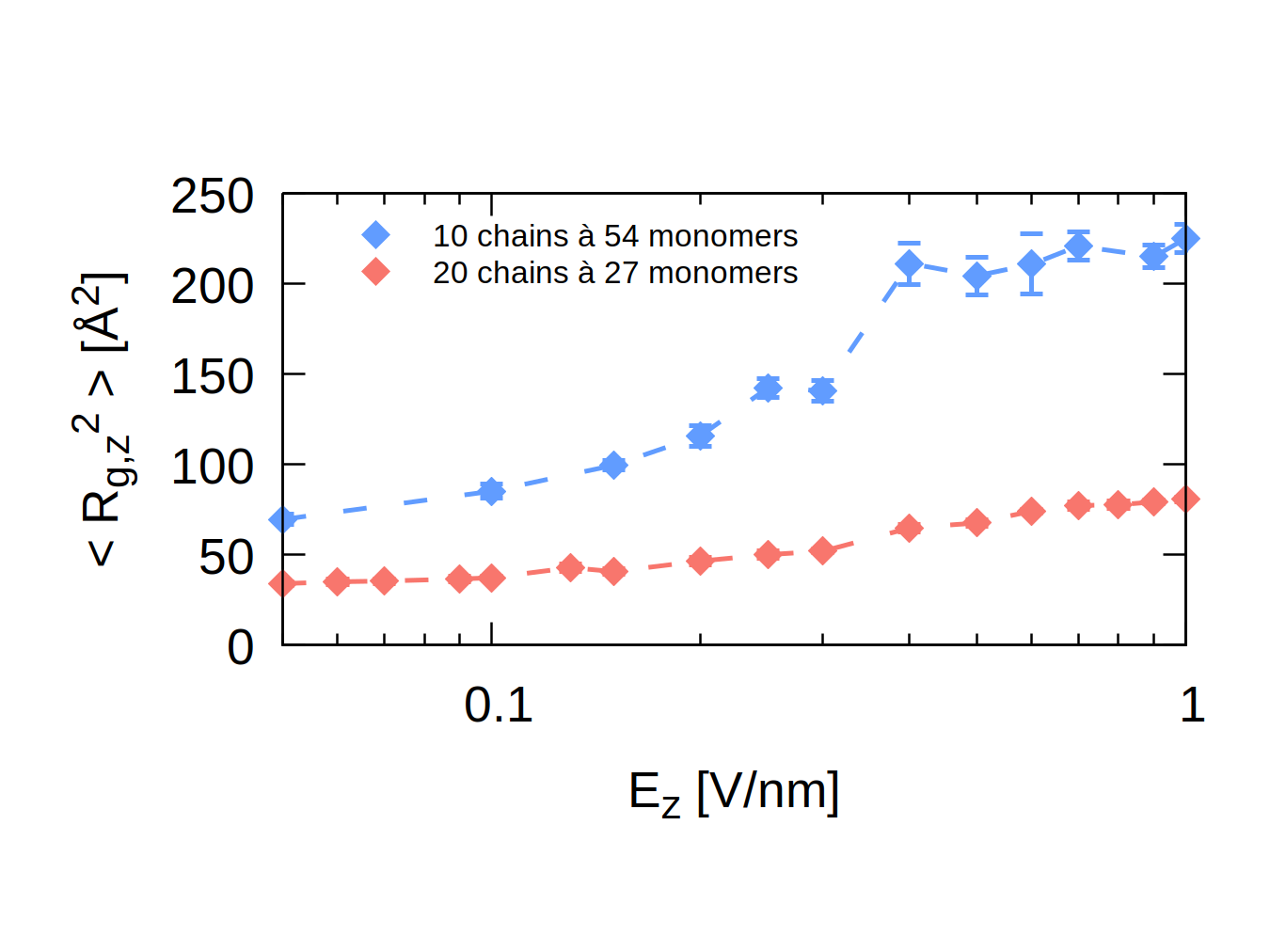}
\caption{Comparison of $\text{R}_{\text{g,z}}^2$ as a function of electric field strength for two different polymer chain lengths.}
\label{fig:chain_length_comparison_Rgz2}
\end{figure}

\begin{figure}[H]
\centering
\includegraphics[width=0.7\textwidth]{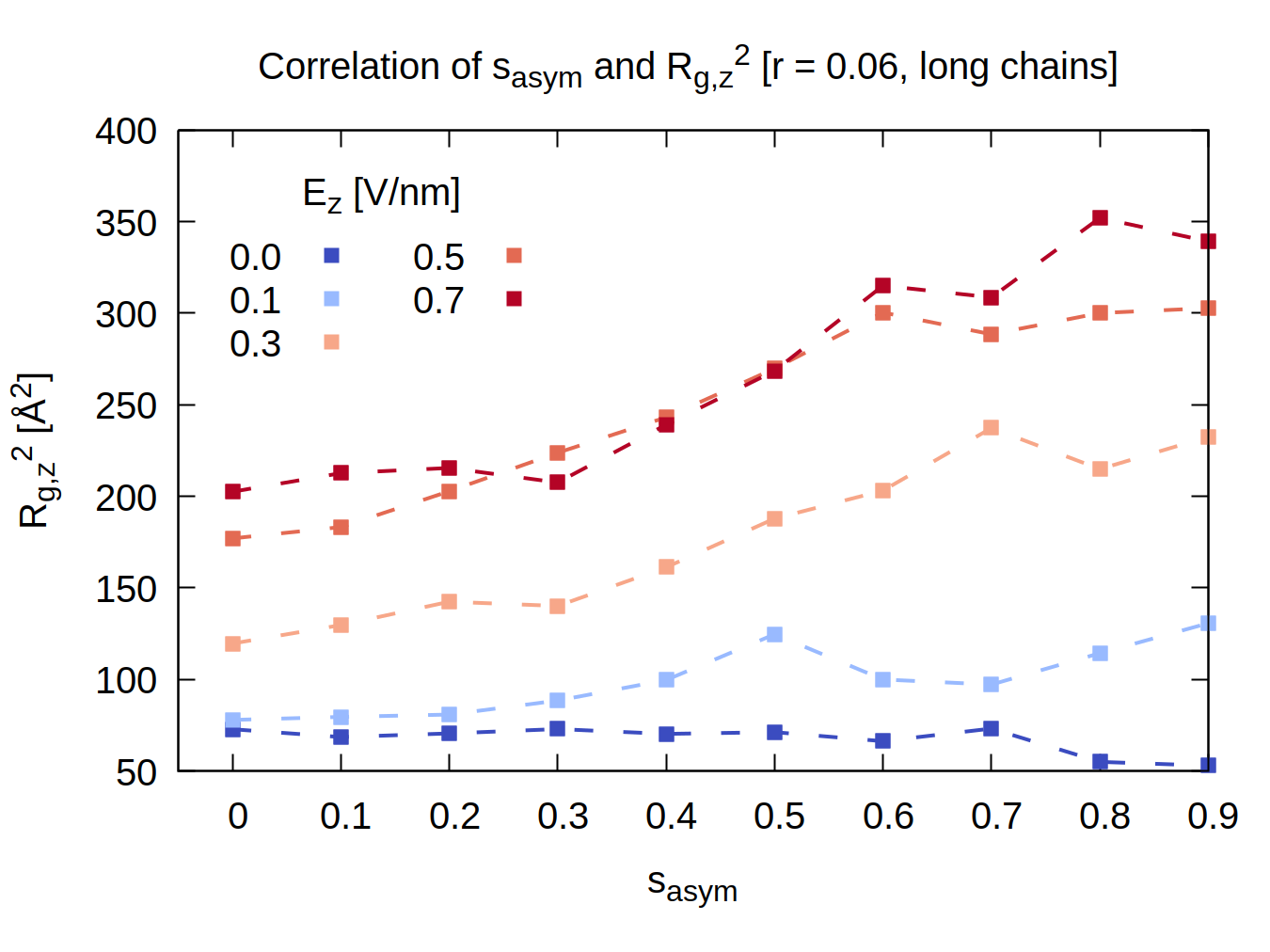}
\caption{Effect of $\text{E}_{\text{z}}$ on $\text{R}^2_{\text{g,z}}$ binned according to the asymmetry $\text{s}_{\text{asym}}$ of lithium coordination to backbone for the r\,=\,0.06 system containing chains of the length N\,=\,54.}
\label{fig:s_asym_Rgz2_long_chains}
\end{figure}

In Figure S\ref{fig:chain_length_comparison_Rgz2} we show the response of $\text{R}_{\text{g,z}}^2$  for both chain lengths. For the chain length N\,=\,54 $\text{R}_{\text{g,z}}^2$ plateaus at $\text{E}_{\text{z}}$\,> 0.4\,V/nm.
We note that the correlation between stretching of the polymer chains in direction of the electric field and asymmetric lithium coordination holds also for N\,=\,54 (see Figure S\ref{fig:s_asym_Rgz2_long_chains}). 
We assume that the coiled-to-stretched transformation is more pronounced for  N\,=\,54, because the stretching is more efficient for lithium exerting a pulling force on the chain end.

\begin{figure}[H]
\centering
\includegraphics[width=1.0\textwidth]{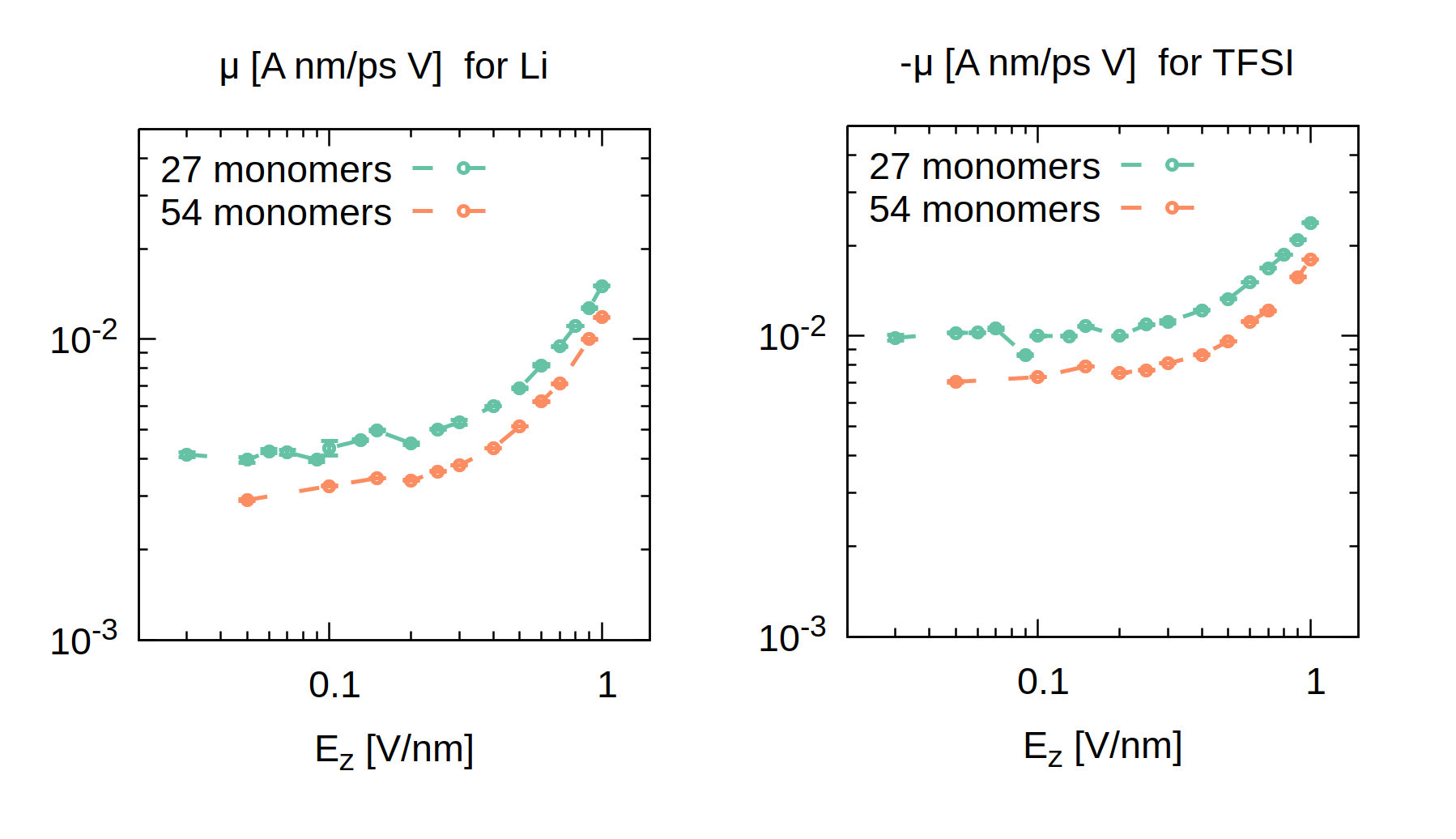}
\caption{Comparison of lithium and TFSI mobilities $\mu$ as a function of electric field strength for two different polymer chain lengths.}
\label{fig:chain_length_comparison_mobility}
\end{figure}

\begin{figure}[H]
\centering
\includegraphics[width=1.0\textwidth]{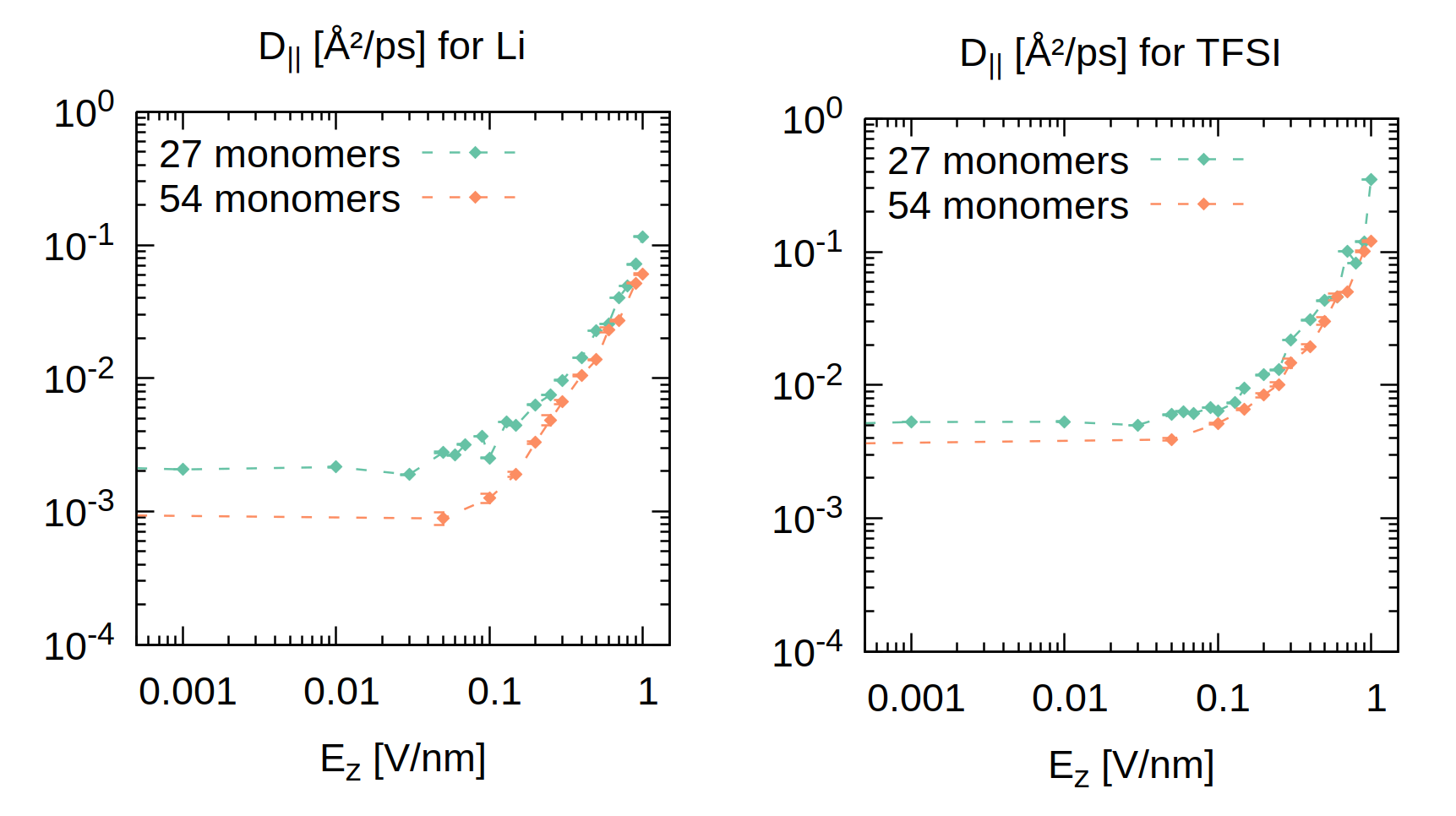}
\caption{Comparison of lithium and TFSI diffusion coefficients $\text{D}_{\parallel}$ as a function of electric field strength for two different polymer chain lengths.}
\label{fig:chain_length_comparison_diffusion}
\end{figure}

In Figures S\ref{fig:chain_length_comparison_mobility} and S\ref{fig:chain_length_comparison_diffusion} it is shown that the nonlinear effects for $\mu$ and $\text{D}_{\parallel}$ occur for longer chains as well. As known from experimental and theoretical literature \cite{Timachova2015,
Brooks2018,Chattoraj2015} transport properties decrease in magnitude for increasing chain length.

\newpage
\textbf{S8: Switching field from $\text{E}_{\text{z}}$\,=\,0.0 to 1.0\,V/nm}

We investigated the process of equilibration of structural and dynamic observables in response to field application for the r\,=\,0.06 electrolyte system.
To track the temporal evolution with sufficient statistics, we conducted a simulation series consisting of 100 short individual runs.
The simulation protocol for the individual runs splits up in three steps:

(1) The starting configurations were extracted from the $\text{E}_{\text{z}}\,=\,0.0\,$V/nm trajectory 18\,ns apart from each other.

\begin{figure}[H]
\centering
\includegraphics[width=0.7\textwidth]{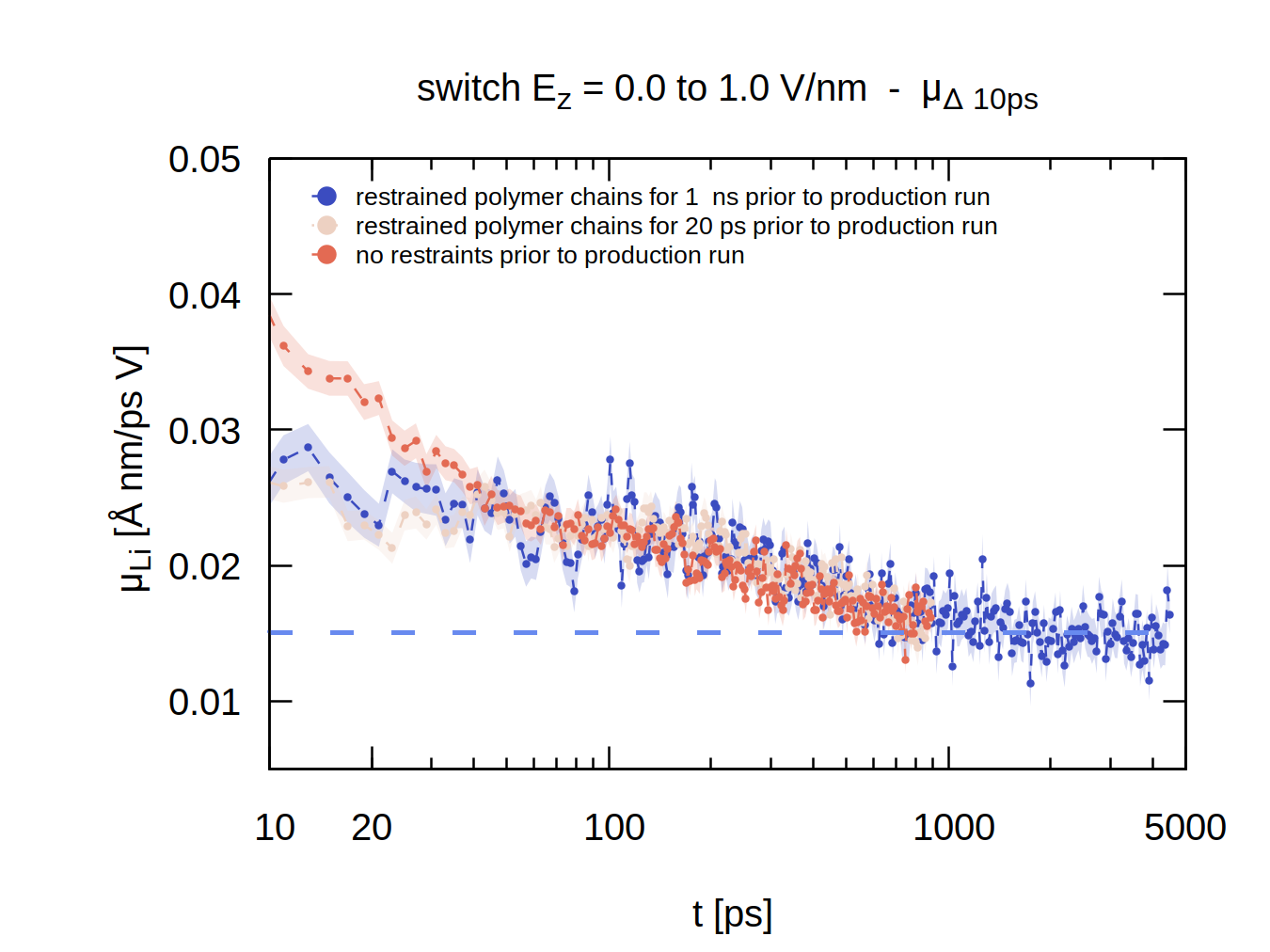}
\caption{Lithium mobility as a function of simulation time for different equilibration procedures prior to the production run. The dashed line represents the stationary state $\mu_{\text{Li}}$ for $\text{E}_{\text{z}}\,=\,1.0\,$V/nm.}
\label{fig:different_restrained_equilibration_setups}
\end{figure}

(2) A sudden tilting of the potential energy surface may cause ions to immediately slip into their  accordingly shifted local energy minima, thus provoking high instantaneous velocities of the particles. To separate this only initially effective increase of ionic motion from an enhancement of steady-state transport properties due to tilting or a structurally conditioned enhancement (see introductory sketch Figure 1 as well as sketch Figure 8), we generated a set of simulations preceded by three different equilibration protocols. 
Firstly, we performed production runs for direct application of $\text{E}_{\text{z}}\,=\,1.0\,$V/nm on the unperturbed $\text{E}_{\text{z}}\,=\,0.0\,$V/nm structures ('no restraints prior to production run' in Figure S\ref{fig:different_restrained_equilibration_setups}). Secondly, to permit the ions' adjustment to the newly shifted local energy minima due to $\text{E}_{\text{z}}\,=\,1.0\,$V/nm in an otherwise unchanged structural environment, we interposed a restrained equilibration run. The positions of the polymer backbone atoms were restrained in all directions by means of a harmonic potential with a force constant of 400.00\,kJ/mol. This restrained equilibration was carried out for either 20\,ps or 1\,ns at an integration time step of 0.5\,fs ('restrained polymer chains  for 1\,ns / 20\,ps prior to production run' in Figure S\ref{fig:different_restrained_equilibration_setups}).

(3) The production run for data acquisition was carried out for 5\,ns (respectively 1\,ns for the unrestrained / 20\,ps long restrained equilibration setup) using the same simulation parameters as for the previous steady-state simulations. The data was then averaged over all individual simulations.


In Figure S\ref{fig:different_restrained_equilibration_setups}, we show the impact of freezing the chain structure prior to the production run on $\mu_{\text{Li}}$(t). As discussed before the sudden tilting of the energy landscape causes the ions to quickly slip into their new local energy minima and thus translates into an elevated velocity in field direction (see 'no restraints') for short times after the field has been switched on. 
We observe that the increase of the initial $\mu_{\text{Li}}$(t) is smaller for prior restraints on the polymer chains and an equilibration duration of both 20\,ps and 1\,ns quantitatively yields the same behavior. We rationalize this by a fast local adjustment of the ions to the local energy minima during the restrained pre-equilibration.

\begin{figure}[H]
\centering
\includegraphics[width=0.7\textwidth]{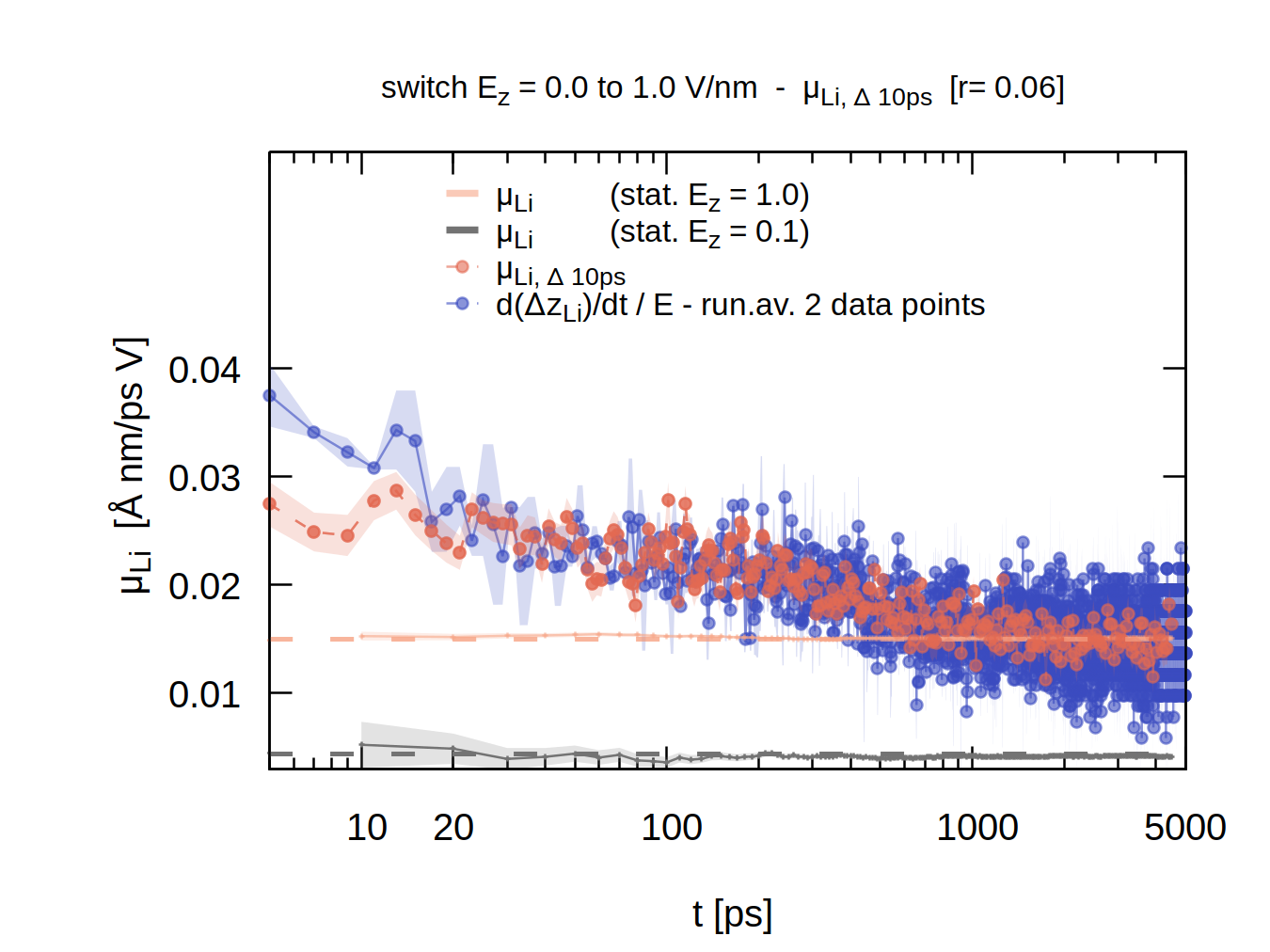}
\caption{Lithium mobility as a function of simulation time since the polymer chains are released from positional restraints for either obtaining $\mu_{\text{Li}}$ from the numerical derivative of d$(\Delta\text{z}_{\text{Li}})/\text{dt}$ or at a constant time lag of $\Delta\,10$ps. The dashed lines represent the stationary state $\mu_{\text{Li}}$ for $\text{E}_{\text{z}}\,=\,0.1\,$V/nm in the linear response regime, respectively $\text{E}_{\text{z}}\,=\,1.0\,$V/nm. The lag time dependence of $\mu_{\text{Li}}$ in the stationary state simulations is shown in the corresponding colors as well ($\mu_{\text{stat}}$ not shown as a function of simulation time but lag time to compute $\mu_{\text{stat}}$).  }
\label{fig:different_methods_of_extracting_mobility}
\end{figure}

As introduced in the main text, we measure the instantaneous mobility $\mu$(t) as:
\begin{equation}
\mu(t) = \dfrac{\langle \nu(t) \rangle}{E} = \dfrac{1}{E\cdot \Delta t} \left(\langle z(t+\dfrac{\Delta t}{2})\rangle - \langle z(t-\dfrac{\Delta t}{2} )\rangle \right) ,
\label{eq:current_mobility}
\end{equation}
the drift velocity is thus evaluated for a constant time lag $\Delta$t between the reference positions. 
For short time lags $\Delta t$ it is crucial to gather data with sufficient statistics because the drift motion is overlaid by thermal noise. We compare extracting $\mu(t)$ via equation \eqref{eq:current_mobility} for $\Delta t$\,=\,10\,ps to extracting the current velocity $\nu$ from the numerical derivative of the lithium displacement in field direction $\Delta z_{\text{Li}}(t)$ in Figure S\ref{fig:different_methods_of_extracting_mobility}. We show that the noise can be reduced for increasing $\Delta t$.


\begin{figure}[H]
\centering
\includegraphics[width=0.7\textwidth]{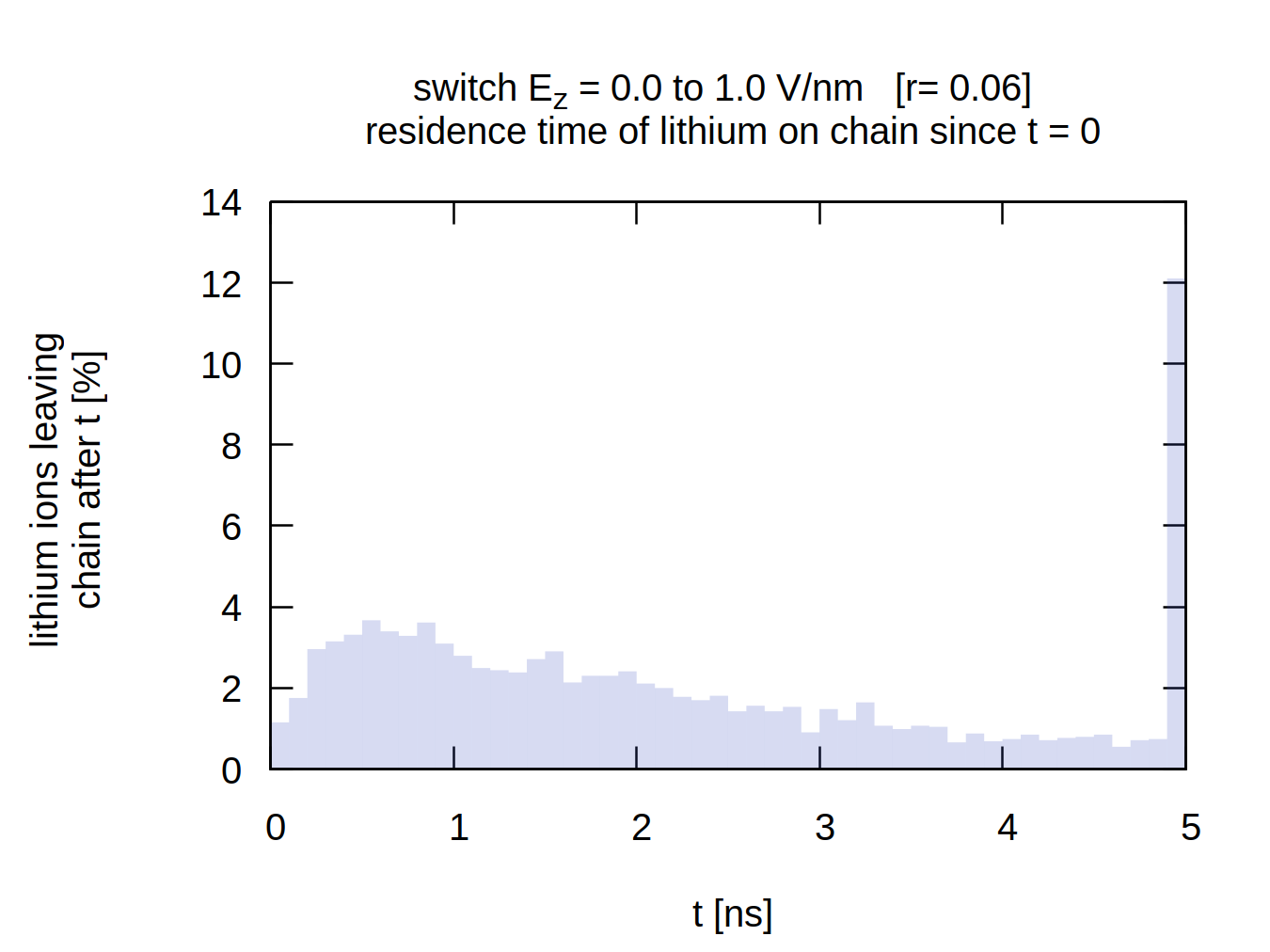}
\caption{Distribution of residence times of lithium on the polymer chains that lithium is attached to at t\,=\,0\,ps. Since the simulations are 5\,ns long, the last data bar must be understood to mean that after 5\,ns 12\% of initial lithium-chain-pairs are still preserved. }     
\label{fig:residence_time_lithium_chain_no_rejumps}
\end{figure}

\begin{figure}[H]
\centering
\includegraphics[width=0.7\textwidth]{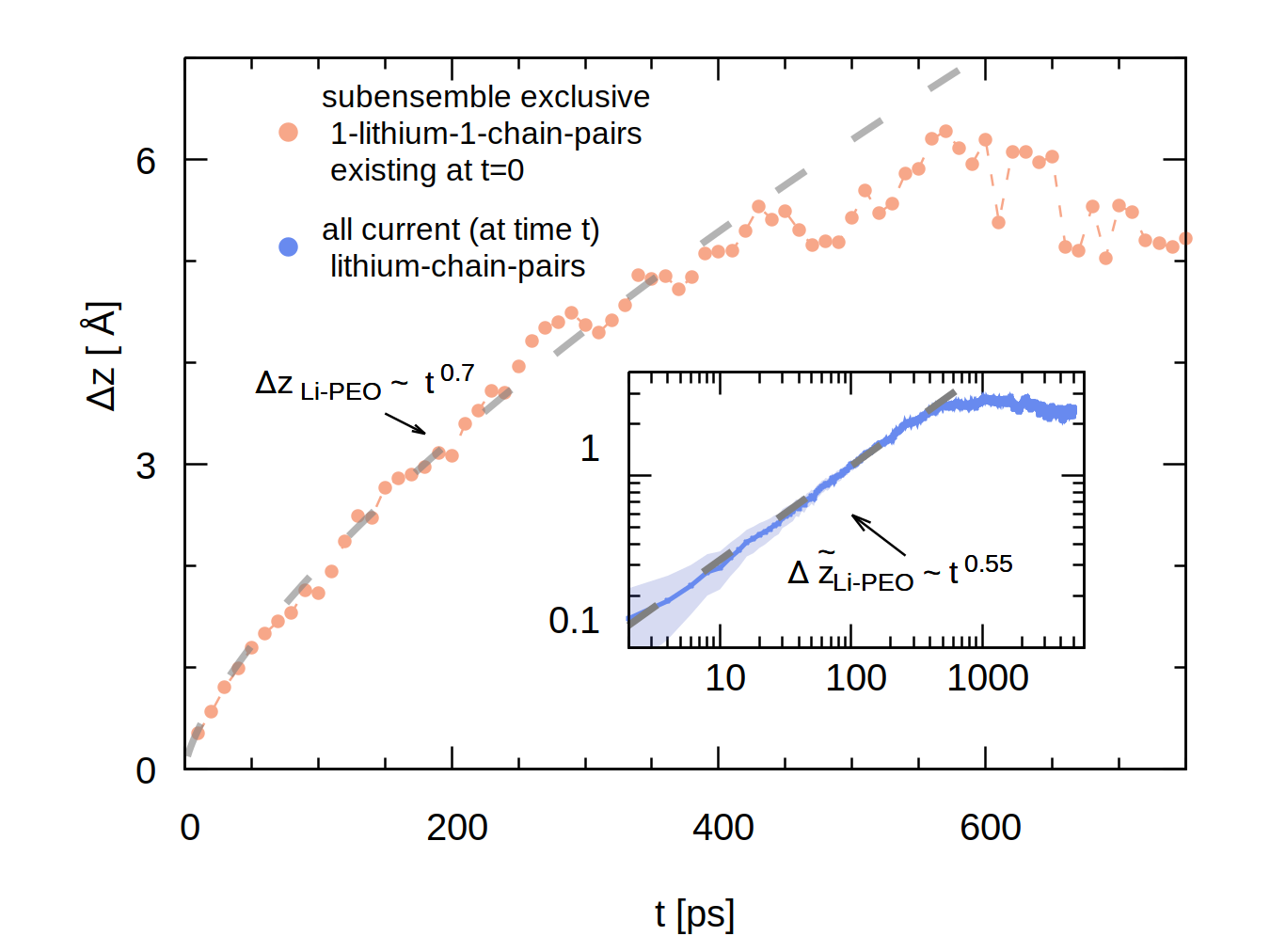}

\caption{As displayed in the main Figure 14: Time evolution of lithium displacement relative to the center of mass of the polymer chain, that it is attached to, $\Delta z_{\text{Li-PEO}}$ for the subensemble of distinct lithium-chain-pairs existent at t\,=\,0\,ps, where lithium coordinates to a single chain and the chain is coordinated exclusively by this lithium ion. The inset shows the same analysis for $\Delta \tilde{z}_{\text{Li-PEO}}$ including all lithium-chain pairs that are present at time t.}

\label{fig:compare_total_subensemble_delta_z_li_peo}
\end{figure}

The long time limit of the overall $\Delta \tilde{z}_{\text{Li-PEO}}$ effectively is a complex interplay of different effects: lithium transfers between chains and further the system comprises diverse coordination scenarios, for example lithium binding to multiple polymer chains or a polymer chain hosting multiple lithium ions (see Figure \ref{fig:compare_total_subensemble_delta_z_li_peo} for taking all lithium-chain-pairs into consideration). \newline

\newpage

\bibliography{literature}